\documentclass[useAMS, usenatbib]{mn2e}
\usepackage{color}
\usepackage{epsfig}
\usepackage{epstopdf}
\usepackage{amsmath, amssymb}
\usepackage{hyperref}
\usepackage{wasysym} 
\usepackage{subfigure}
\usepackage{multirow}
\usepackage{lipsum}

\usepackage{setspace}  

\newlength{\plotwidth}
\newlength{\fullwidth}
\renewcommand{\topmargin}{-1cm} 

\pagerange{\pageref{firstpage}--\pageref{lastpage}} \pubyear{2013}

\def\LaTeX{L\kern-.36em\raise.3ex\hbox{a}\kern-.15em
    T\kern-.1667em\lower.7ex\hbox{E}\kern-.125emX}

\title[Power spectrum multipole analysis of BOSS]{The clustering of galaxies in the SDSS-III Baryon Oscillation Spectroscopic Survey: Testing gravity with redshift-space distortions using the power spectrum multipoles}
\author[Florian Beutler et al.]
{\parbox{\textwidth}{Florian Beutler$^{1}$\thanks{E-mail: \texttt{fbeutler@lbl.gov}},
Shun Saito$^{1,2,3,4}$, Hee-Jong Seo$^{5,6}$, Jon Brinkmann$^{7}$, Kyle S. Dawson$^8$, Daniel J. Eisenstein$^{9}$, Andreu Font-Ribera$^1$, Shirley Ho$^{10,11}$, Cameron K. McBride$^9$, Francesco Montesano$^{12}$, Will J. Percival$^{13}$, Ashley J. Ross$^{13}$, Nicholas P. Ross$^{1,14}$, Lado Samushia$^{13}$, David J. Schlegel$^1$, Ariel G. S\'anchez$^{12}$, Jeremy L. Tinker$^{15}$, Benjamin A. Weaver$^{15}$}\vspace{0.4cm}\\
\parbox{\textwidth}{
$^{1}$Lawrence Berkeley National Lab, 1 Cyclotron Rd, Berkeley CA 94720, USA,\\
$^{2}$Kavli Institute for the Physics and Mathematics of the Universe (WPI),Todai Institues for Advanced Study, The University of Tokyo, Chiba 277-8582, Japan,\\
$^{3}$Department of Astronomy, University of California Berkeley, CA 94720, USA,\\
$^{4}$Department of Physics, University of California Berkeley, CA 94720, USA,\\
$^{5}$Berkeley Center for Cosmological Physics, LBL and Department of Physics, University of California, Berkeley, CA 94720, USA,\\
$^{6}$Center for Cosmology and Astroparticle Physics, Department of Physics, The Ohio State University, OH 43210, USA,\\
$^{7}$Apache Point Observatory, P.O. Box 59, Sunspot, NM 88349-0059,USA,\\
$^{8}$Department of Physics and Astronomy,  University of Utah, Salt Lake City, UT 84112, USA,\\
$^{9}$Harvard-Smithsonian Center for Astrophysics, 60 Garden St., Cambridge, MA 02138, USA,\\
$^{10}$Department of Physics, Carnegie Mellon University, 5000 Forbes Avenue, Pittsburgh, PA 15213, USA,\\ 
$^{11}$McWilliams Center for Cosmology, Carnegie Mellon University, 5000 Forbes Avenue, Pittsburgh, PA 15213, USA,\\
$^{12}$Max-Planck-Institut f\"ur Extraterrestrische Physik, Giessenbachstrasse, 85748 Garching, Germany,\\
$^{13}$Institute of Cosmology \& Gravitation, Dennis Sciama Building, University of Portsmouth, Portsmouth, PO1 3FX, UK,\\
$^{14}$Department of Physics, Drexel University, 3141 Chestnut Street, Philadelphia, PA 19104, USA,\\
$^{15}$Center for Cosmology and Particle Physics, New York University, New York, NY 10003 USA.}}

\begin{document}

\twocolumn[
\begin{@twocolumnfalse}
 
\label{firstpage}
\maketitle

\begin{abstract}
We analyse the anisotropic clustering of the Baryon Oscillation Spectroscopic Survey (BOSS) CMASS Data Release 11 (DR11) sample, which consists of $690\,827$ galaxies in the redshift range $0.43 < z < 0.7$ and has a sky coverage of $8\,498\,\deg^2$. We perform our analysis in Fourier space using a power spectrum estimator suggested by~\citet{Yamamoto:2005dz}. 
We measure the multipole power spectra in a self-consistent manner for the first time in the sense that we provide a proper way to treat the survey window function and the integral constraint, without the commonly used assumption of an isotropic power spectrum and without the need to split the survey into sub-regions. The main cosmological signals exploited in our analysis are the Baryon Acoustic Oscillations and the signal of redshift space distortions, both of which are distorted by the Alcock-Paczynski effect. Together, these signals allow us to constrain the distance ratio $D_V(z_{\rm eff})/r_s(z_d) = 13.89\pm 0.18$, the Alcock-Paczynski parameter $F_{\rm AP}(z_{\rm eff}) = 0.679\pm0.031$ and the growth rate of structure $f(z_{\rm eff})\sigma_8(z_{\rm eff}) = 0.419\pm0.044$ at the effective redshift $z_{\rm eff}=0.57$. We emphasise that our constraints are robust against possible systematic uncertainties. In order to ensure this, we perform a detailed systematics study against CMASS mock galaxy catalogues and N-body simulations. We find that such systematics will lead to $3.1\%$ uncertainty for $f\sigma_8$ if we limit our fitting range to $k=0.01$ - $0.20h/$Mpc, where the statistical uncertainty is expected to be three times larger. We did not find significant systematic uncertainties for $D_V/r_s$ or $F_{\rm AP}$. Combining our dataset with Planck to test General Relativity (GR) through the simple $\gamma$-parameterisation, where the growth rate is given by $f(z) = \Omega^{\gamma}_m(z)$, reveals a $\sim 2\sigma$ tension between the data and the prediction by GR. The tension between our result and GR can be traced back to a tension in the clustering amplitude $\sigma_8$ between CMASS and Planck.\vspace{1cm}
\end{abstract}

\end{@twocolumnfalse}
]


\section{introduction}

\begin{figure*}
\begin{center}
\epsfig{file=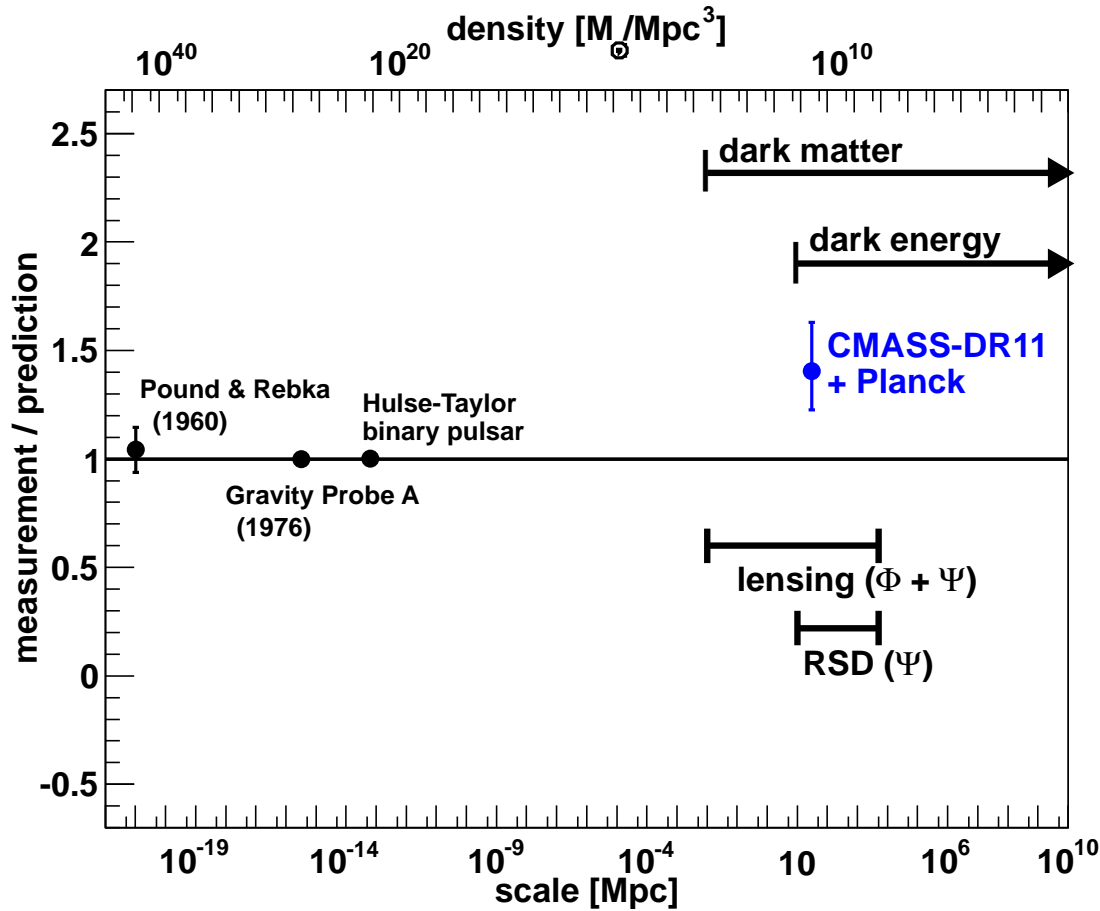,width=16cm}
\caption{Summary of different tests of General Relativity (GR) as a function of distance scale (bottom axis) and densities (top axis). The standard model of cosmology seems to run into problems (dark matter, dark energy) at large scales. Because these problems could indicate a breakdown of GR we need to test GR on large scales. Two probes which can do this are redshift space distortions (RSD) and lensing. While RSD measures the Newtonian potential $\Psi$, lensing measures the sum of the metric potentials $\Phi + \Psi$. However, any modification of gravity needs to pass the very precise tests on smaller scales (Pound \& Rebka experiment~\citealt{Pound:1960zz}, Gravity Probe A,~\citealt{Vessot:1980zz}, Hulse-Taylor binary pulsar~\citealt{Hulse:1974eb}, see~\citealt{Will:2005va} for a complete list). Note that the error bars for Gravity Probe A and the Hulse-Taylor binary pulsar are smaller than the data points in this plot. In this analysis we perform a $\Lambda$CDM consistency test (blue data point), where we use the CMASS-DR11 power spectrum multipoles together with Planck~\citep{Ade:2013zuv} to tests GR on scales of $\sim 30\,$Mpc (see section~\ref{sec:consistency}).}
\label{fig:keff}
\end{center}
\end{figure*}

The key to understand the dynamical properties of the Universe, its past and its future, is the understanding of gravity. Today's dominant theory of the origin of the Universe, the Big Bang model, is based on Albert Einstein's General Relativity (GR). The crucial idea behind GR, the connection between space and time into space-time first allowed us to talk about curved space and expanding space, terms which do not exist in Newton's gravity.

GR is a very powerful theory, which makes many testable predictions, like the deflection of light or gravitational waves. Despite the successes of our current understanding of gravity, there are several problems, which motivated scientists to search for alternative formulations or expansions of GR. One problem, which we will not pursue any further in this paper, is that GR cannot be combined with the other fundamental forces, since GR is not formulated as a quantum field theory. Another problem is that the motions of galaxies and galaxy clusters cannot be explained by GR and baryonic matter alone, but require the introduction of a new form of matter, so-called dark matter~\citep{Zwicky:1937zza,Kahn1959,Freeman:1970mx,Rubin:1970zza}, which nobody has yet observed directly. While the issue of dark matter has existed since the 1930s, in the late 1990s Type Ia supernova surveys made the intriguing discovery that the expansion of the Universe is accelerating~\citep{Riess:1998cb, Perlmutter:1998np}. This required the introduction of yet another ÓdarkÓ component, so-called dark energy, which would counteract the gravitational force leading to an accelerated expansion. The question now is whether these problems indicate a breakdown of GR or whether there are additional unknown components of the Universe. While the problems of GR on cosmological scales (dark matter and dark energy) gave birth to many new models of gravity (see e.g.~\citealt{Clifton:2011jh,Capozziello:2013fxa,Jain:2010ka}), so far none of these models has been able to convince scientists that it is time to abandon GR. 

Given that it is on cosmological scales where GR runs into trouble, it is on cosmological scales where we have to test GR. Figure~\ref{fig:keff} shows different tests of GR at different scales (see e.g.~\citealt{Will:2005va}). One interesting observable, which allows us to test GR on cosmic scales, is redshift space distortions (RSD)~\citep{Sargent:1977,Kaiser:1987qv,Hamilton:1997zq}. 
RSD are peculiar velocities of galaxies due to gravitational interaction. The line-of-sight component of this additional velocity cannot be easily separated from the Hubble flow and contaminates our measurement of the cosmic expansion. This makes the observed galaxy clustering anisotropically distorted, since the line-of-sight direction becomes ``special''. This is what we call RSD. The anisotropic pattern of RSDs in galaxy clustering allows us to extract information on the peculiar velocities which are directly related to the Newton potential through the Euler equation. Given the amount of matter in the Universe, GR makes a clear and testable prediction for the amplitude of this anisotropic signal. In the last decade, galaxy redshift surveys became large enough to test this prediction~\citep{Peacock:2001gs,Hawkins:2002sg,Tegmark:2006az,Guzzo:2008ac,Yamamoto:2008gr,Blake:2011rj,Beutler:2012px,Reid:2012sw,Samushia:2012iq,Chuang:2013hya,Nishimichi:2013aba}. 

In addition to the RSD signal, the galaxy power spectrum and correlation function carry geometric information. The measurement of the Baryon Acoustic Oscillation scale in the distribution of galaxies has become one of the most powerful probes of cosmology, together with the Cosmic Microwave Background~\citep{Ade:2013zuv}. The BAO scale has now been detected at several different redshifts~\citep{Eisenstein:2005su,Beutler:2011hx,Blake:2011en,Padmanabhan:2012hf,Anderson:2012sa,Slosar:2013fi,Anderson2.0}. Most notably the ongoing BOSS survey~\citep{Schlegel:2009hj} reduced the measurement uncertainty on the BAO scale to $1\%$~\citep{Anderson2.0}, which is still considerably larger than the expected systematic bias~\citep{Eisenstein:2004an,Padmanabhan:2009yr,Mehta:2011xf}. Measuring the galaxy clustering along the line-of-sight and perpendicular to the line-of-sight allows us to perform an Alcock-Paczynski (AP) test~\citep{Alcock:1979mp,Matsubara:1996nf,Ballinger:1996cd} with both the RSD and BAO signals. The Alcock-Paczynski test describes a distortion in an otherwise isotropic feature in the galaxy clustering when the assumed fiducial cosmological model used to transfer the measured redshifts into distances deviates from the true cosmology. This anisotropic signal may appear degenerate with the RSD signal in a featureless power spectrum. Using the BAO signal we can break this degeneracy and exploit all three signals, RSD, BAO and the Alcock-Paczynski effect for cosmological parameter constraints.

In this analysis we are going to use the CMASS sample of BOSS galaxies that will be included in the Sloan Digital Sky Survey (SDSS) Data Release 11 (DR11), which will become publicly available together with the final data (DR12) at the end of 2014. We use this dataset to constrain the growth of structure and the geometry of the Universe simultaneously. We measure the growth rate via the parameter combination $f(z)\sigma_8(z)$ and the geometry of the Universe via $D_V(z)/r_s(z_d)$ and $F_{\rm AP}(z)=(1+z)D_A(z)H(z)/c$ at an effective redshift of $z_{\rm eff} = 0.57$. The BAO signal and the AP effect constrain the geometry, i.e., $D_V(z)/r_s(z_d)$ and $F_{\rm AP}(z)$, thereby isolating the anisotropy in the clustering amplitude due to the RSD. The growth rate, $f(z)\sigma_8(z)$ is constrained by this RSD signal. We will make our analysis in Fourier space using the power spectrum monopole and quadrupole. The power spectrum multipoles are measured using a new power spectrum estimator suggested by~\citet{Yamamoto:2005dz}. The popular power spectrum estimator suggested by~\citet{Feldman:1993ky} (from here on FKP estimator) cannot be used to make angle-dependent measurements in BOSS because of the plane parallel approximation that this estimator implicitly makes (see section~\ref{sec:Yamamoto} for details). 

Since the power spectrum quadrupole is more sensitive to window function effects than the more commonly used monopole, we suggest a new way of including the window function into the power spectrum analysis. In order to robustly constrain the RSD and AP-test parameters, we model the anisotropic galaxy power spectrum using perturbation theory (PT) which fairly reflects a series of recent theoretical progresses. Our PT model accurately describes non-linear issues such as gravitational evolution, mapping from real to redshift space, and local and non-local galaxy bias. We also perform a detailed study of possible systematic uncertainties and quantify a systematic error for our parameter constraints. Our analysis has been done $``$blind$"$, meaning that all model tests and the set-up of the fitting conditions are investigated using mock data and only at the final stage do we fit the actual CMASS-DR11 measurements. The CMASS-DR11 constraints on $D_V(z)/r_s(z_d)$, $F_{\rm AP}(z)$ and $f(z)\sigma_8(z)$ are the most precise constraints to date using this technique.

This paper is organised as follows. In section~\ref{sec:data} we describe the BOSS CMASS-DR11 dataset. In section~\ref{sec:Yamamoto} we describe the power spectrum estimator used in our analysis and in section~\ref{sec:cov} we describe the mock catalogues together with the derivation of the covariance matrix. We then discuss the measurement of window function effects including the integral constraint in section~\ref{sec:win}. In section~\ref{sec:model} we discuss our model for the power spectrum multipoles, together with the modelling of the Alcock-Paczynski effect. We perform a detailed study of possible systematic uncertainties in section~\ref{sec:sys}, followed by the data analyses in section~\ref{sec:analysis}. We use our data constraints for cosmological tests in section~\ref{sec:cos} and conclude in section~\ref{sec:conclusion}. The appendix gives detailed derivations of equations used in our analysis.

The fiducial cosmology used to turn redshifts into distances assumes a flat $\Lambda$CDM universe with $\Omega_m = 0.3$. The Hubble constant is set to $H_0 = 100h\,$km/s/Mpc, with our fiducial model using $h = 0.7$. 

\section{The BOSS CMASS-DR11 dataset}
\label{sec:data}

\begin{figure}
\begin{center}
\epsfig{file=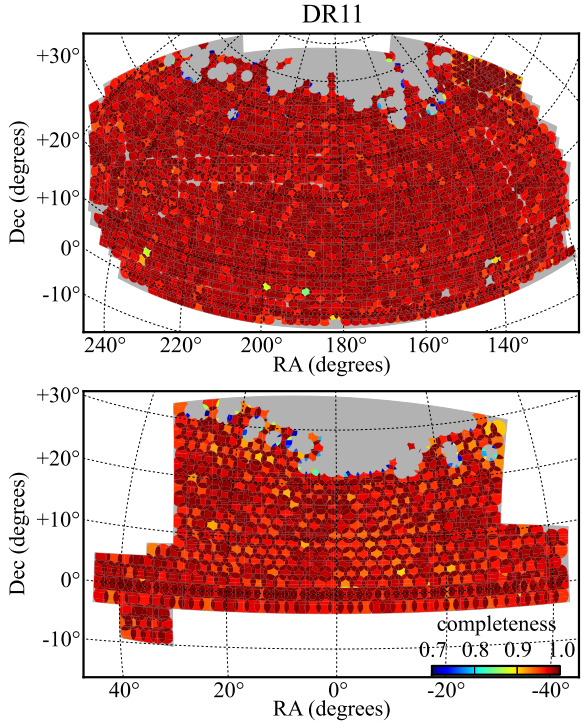,width=8cm}
\caption{The CMASS-DR11 North Galactic Cap (top) and South Galactic Cap (bottom) sky coverage. The grey region indicates the final footprint of  the survey (DR12). The colours indicate the completeness in the regions included in our analysis.}
\label{fig:footprint}
\end{center}
\end{figure}

BOSS, as part of SDSS-III~\citep{Eisenstein:2011sa,Dawson:2012va} is measuring spectroscopic redshifts of $\approx 1.5$ million galaxies (and $150\,000$ quasars) making use of the SDSS multi-fibre spectrographs~\citep{Bolton:2012hz,Smee:2012wd}. The galaxies are selected from multi-colour SDSS imaging~\citep{Fukugita:1996qt,Gunn:1998vh,Smith:2002pca,Gunn:2006tw,Doi:2010rf} and cover a redshift range of $z = 0.15$ - $0.7$, where the survey is split into two samples called LOWZ ($z=0.15$ - $0.43$) and CMASS ($z=0.43$ - $0.7$). In this analysis we are only using the CMASS sample. The survey is optimised for the measurement of the BAO scale and hence covers a large cosmic volume ($V_{\rm eff} = 2.31\times 10^9[$Mpc$/h]^3$) with a density of $\overline{n} \approx 3\times10^{-4}[h/$Mpc$]^{3}$, high enough to ensure that shot noise is not the dominant error contribution at the BAO scale~\citep{White:2010ed}. Most CMASS galaxies are red with a prominent $4000\,$\AA\; break in their spectral energy distribution. Halo Occupation studies have shown that galaxies selected like the CMASS galaxies are mainly central galaxies residing in dark matter halos of $10^{13}M_{\odot}/h$, with a $5$ - $10\%$ satellite fraction~\citep{White:2010ed}. CMASS galaxies are highly biased ($b\sim 2$), which boosts the clustering signal including BAO in respect to the shot noise level.

The CMASS-DR11 sample covers $6\,391\deg^2$ in the North Galactic Cap (NGC) and $2\,107\deg^2$ in the South Galactic Cap (SGC); the total area of $8\,498\deg^2$ represents a significant increase from CMASS-DR9, which covered $3\,265\deg^2$ in total. The sample used in our analysis includes $520\,806 $ galaxies in the NGC and $170\,021$ galaxies in the SGC. Figure~\ref{fig:footprint} shows the footprint of the survey in the two regions, where the grey area indicates the expected footprint of DR12.

We include three different incompleteness weights to  account for shortcomings of the CMASS dataset (see~\citealt{Ross:2012qm} and~\citealt{Anderson2.0} for details): A redshift failure weight, $w_{\rm rf}$, a fibre collision weight, $w_{\rm fc}$ and a systematics weight, $w_{\rm sys}$, which is a combination of a stellar density weight and a seeing condition weight. Each galaxy is thus counted as 
\begin{equation}
w_c = (w_{\rm rf} + w_{\rm fc} - 1)w_{\rm sys}.
\end{equation}
We will discuss these weights in more detail in section~\ref{sec:SN}. 

\section{The power spectrum estimator}
\label{sec:Yamamoto}

In this section we describe the power spectrum estimator we use to measure the multipole power spectrum from the CMASS-DR11 sample. We carefully address how to incorporate the incompleteness weights. Before explaining the estimator itself, we summarise different approximations commonly used in galaxy clustering analysis.

\subsection{Commonly used approximations}
\label{sec:approx}

Here we discuss different approximations used in galaxy clustering statistics, and if used in our analysis we discuss their impact on our measurement:
\begin{enumerate}
\item \textbf{Distant observer approximation:} Here one assumes that a displacement $\Delta x$ (e.g. caused by redshift space distortions) is much smaller than the distance, $|\vec{x}|$, to the galaxy itself. This approximation is commonly used for the volume element in the Jacobian mapping from real to redshift space. We assume the distant observer approximation when modelling the galaxy power spectrum in section~\ref{sec:taruya}.
\item \textbf{Local plane parallel approximation:} Here one assumes that the position vectors of a galaxy pair can be treated as parallel, meaning
\begin{equation}
\hat{\vec{k}}\cdot \hat{\vec{x}}_1 \approx \hat{\vec{k}}\cdot \hat{\vec{x}}_2 \approx \hat{\vec{k}}\cdot \hat{\vec{x}}_h,
\end{equation}
where $\hat{\vec{x}}_h = (\hat{\vec{x}}_1 + \hat{\vec{x}}_2)/2$ and $\hat{\vec{x}} = \vec{x}/|\vec{x}|$. This approximation is only valid for a galaxy pair with a small angular separation and hence will break down on large scales~\citep{Papai:2008bd}. It has been shown, however, that the local plane parallel approximation is a very good approximation for most galaxy samples even when they cover a large fraction of the sky~\citep{Samushia:2011cs,Beutler:2011hx,Yoo:2013zga}. Most of the anisotropic galaxy clustering measurements adopt this assumption including our analysis, where it is introduced in eq.~\ref{eq:Fx}.
\item \textbf{(Global) plane parallel approximation (or flat-sky approximation):} Here one assumes that the line-of-sight vector $\hat{\vec{x}}$ is the same for all galaxies in the survey, meaning
\begin{equation}
\hat{\vec{k}}\cdot\hat{\vec{x}} \approx \hat{\vec{k}}\cdot\hat{\vec{z}},
\end{equation}
where $\hat{\vec{z}}$ is the global line-of-sight vector. This approximation is included in the FKP estimator suggested by~\citet{Feldman:1993ky}. Since the line-of-sight vector only appears in the calculation of the cosine angle to the line-of-sight, $\mu$, the monopole power spectrum is not affected by this approximation. The higher order multipoles are strongly affected, except for very narrow angle surveys~\citep{Blake:2011rj}. The invalidity of the plane parallel approximation for the geometry of the CMASS sample~\citep{Yoo:2013zga} motivated the use of the power spectrum estimator suggested by~\citet{Yamamoto:2005dz} in our analysis. 
\end{enumerate}

\subsection{The~\citet{Yamamoto:2005dz} power spectrum estimator}

The multipole power spectrum of a galaxy distribution can be calculated as~\citep{Feldman:1993ky,Yamamoto:2005dz}
\begin{equation}
\begin{split}
P_{\ell}(\vec{k}) &= \frac{(2\ell + 1)}{2A}\bigg[\int d\vec{x}_1 \int d\vec{x}_2\, F(\vec{x}_1)F(\vec{x}_2)\times\\
&\;\;\;\;\;e^{i\vec{k}\cdot(\vec{x}_1-\vec{x}_2)}\mathcal{L}_{\ell}(\hat{\vec{k}}\cdot\hat{\vec{x}}_h) - S_{\ell}\bigg]
\end{split}
\label{eq:eq1}
\end{equation}
where $\mathcal{L}_{\ell}$ is the Legendre polynomial, $\vec{x}_h = (\vec{x}_1 + \vec{x}_2)/2$ and
\begin{align}
A &= \int d\vec{x}\,\left[n'_g(\vec{x})w_{\text{\tiny{FKP}}}(\vec{x})\right]^2,\\
F(\vec{x}) &= w_{\text{\tiny{FKP}}}(\vec{x})\left[n'_g(\vec{x}) -  \alpha'n_s(\vec{x})\right],
\end{align}
where $n'_g$ is the galaxy density, $n_s$ is the density of the random catalogue and $\alpha'$ is the ratio of real galaxies to random galaxies.
The shot noise term is given by
\begin{equation}
\begin{split}
S_{\ell} &= \int d\vec{x}\,n'_g(\vec{x})w_{\rm sys}(\vec{x})w^2_{\text{\tiny{FKP}}}(\vec{x})\mathcal{L}_{\ell}(\hat{\vec{k}}\cdot\hat{\vec{x}}) \\
&\;\;\;\;\;+\alpha'\int d\vec{x}\,n'_g(\vec{x})w^2_{\text{\tiny{FKP}}}(\vec{x})\mathcal{L}_{\ell}(\hat{\vec{k}}\cdot\hat{\vec{x}}).
\end{split}
\end{equation}
In our notation, quantities marked with a ($'$) include all completeness weights, like $\alpha' = N'_{\rm gal}/N_{\rm ran}$ where $N'_{\rm gal} = \sum^{N_{\rm gal}}_i (w_{\rm rf} + w_{\rm fc} - 1)w_{\rm sys}$. In CMASS-DR11, the completeness weights increase the average galaxy density by about $8\%$\footnote{In our analysis we have $N'_{\rm gal} = 558\,001$ for the NGC and $N'_{\rm gal} = 184\,145$ for the SGC, while the actually observed values are $N_{\rm gal} = 520\,806$ and $N_{\rm gal} = 170\,021$, respectively.}. Whenever we have to write the weighting explicitly, we use the completeness weight $w_c(\vec{x}) = (w_{\rm rf} + w_{\rm fc} - 1)w_{\rm sys}$. The random galaxies follow the redshift distribution of the weighted galaxy catalogue, $\langle \alpha'n_s\rangle = \langle n'_g\rangle$, which means that the randoms do not need a completeness weight. In addition to the completeness weight we employ a minimum variance weight, $w_{\rm FKP}(\vec{x})$, which applies to the data and random galaxies (see eq.~\ref{eq:wFKP1}).

Most power spectrum studies in the past employed a Fast Fourier Transform (FFT) to solve the double integral in eq.~\ref{eq:eq1}. 
Such an approach however, requires the (global) plane parallel approximation (see section~\ref{sec:Yamamoto} for the definition), which for wide-angle surveys like BOSS, introduces significant bias into the higher order multipoles of the power spectrum (see e.g.~\citealt{Yoo:2013zga}). The monopole of the power spectrum is unaffected by this assumption, because it does not require an explicit knowledge of the angle to the line-of-sight. \citet{Yamamoto:2005dz} suggested a power spectrum estimator which does not use the plane parallel approximation, for the price of significantly higher computation time. This is the estimator we employ in this analysis.

Using the relation $\int d\vec{x}\,n'_g(\vec{x})... \rightarrow \sum_{N_{\rm gal}}w_c(\vec{x})... \rightarrow \alpha'\sum_{N_{\rm ran}}...$, the integrals in eq.~\ref{eq:eq1} can be written as
\begin{align}
F_{\ell}(\vec{k}) &= \int d\vec{x}\; F(\vec{x}) e^{i\vec{k}\cdot\vec{x}}\mathcal{L}_{\ell}(\hat{\vec{k}}\cdot\hat{\vec{x}})\\
\begin{split}
&= \sum^{N_{\rm gal}}_{i} w_c(\vec{x}_i)w_{\text{\tiny{FKP}}}(\vec{x}_i)e^{i\vec{k}\cdot\vec{x}_i}\mathcal{L}_{\ell}(\hat{\vec{k}}\cdot\hat{\vec{x}}_i)\\
&\;\;\;\;\; -  \alpha'\sum^{N_{\rm ran}}_{i} w_{\text{\tiny{FKP}}}(\vec{x}_i)e^{i\vec{k}\cdot\vec{x}_i}\mathcal{L}_{\ell}(\hat{\vec{k}}\cdot\hat{\vec{x}}_i),
\end{split}
\label{eq:Fx}
\end{align}
where the local plane parallel approximate $\hat{\vec{k}}\cdot\hat{\vec{x}}_h \approx \hat{\vec{k}}\cdot\hat{\vec{x}}_i$ has been used. If we define
\begin{align}
D_{\ell}(\vec{k}) &= \sum^{N_{\rm gal}}_{i} w_c(\vec{x}_i)w_{\text{\tiny{FKP}}}(\vec{x}_i)e^{i\vec{k}\cdot\vec{x}_i}\mathcal{L}_{\ell}(\hat{\vec{k}}\cdot\hat{\vec{x}}_i),\label{eq:D}\\
R_{\ell}(\vec{k}) &= \sum^{N_{\rm ran}}_{i} w_{\text{\tiny{FKP}}}(\vec{x}_i)e^{i\vec{k}\cdot\vec{x}_i}\mathcal{L}_{\ell}(\hat{\vec{k}}\cdot\hat{\vec{x}}_i),
\label{eq:R}
\end{align}
the power spectrum estimate is given by~\citep{Yamamoto:2005dz, Blake:2011rj}
\begin{equation}
\begin{split}
P_{\ell}(\vec{k}) &= \frac{(2\ell + 1)}{2A}\bigg[\left(D_{\ell}(\vec{k}) - \alpha' R_{\ell}(\vec{k})\right)\times\\
&\;\;\;\;\;\left(D_0(\vec{k}) - \alpha' R_0(\vec{k})\right)^*  - S_{\ell}\bigg],
\end{split}
\end{equation}
where the $^*$ represents the complex conjugate. The normalisation is given by
\begin{align}
A &= \sum_i^{N_{\rm gal}}n'_g(\vec{x}_i)w_c(\vec{x}_i)w^2_{\rm FKP}(\vec{x}_i)\\
&= \alpha'\sum_i^{N_{\rm ran}}n'_g(\vec{x}_i)w^2_{\rm FKP}(\vec{x}_i)
\end{align}
and the shot noise for each multipole is defined as
\begin{equation}
\begin{split}
S_{\ell} &= \sum^{N_{\rm gal}}_i w_{c}(\vec{x}_i)w_{\rm sys}(\vec{x}_i)w_{\text{\tiny{FKP}}}^2(\vec{x}_i)\mathcal{L}_{\ell}(\hat{\vec{k}}\cdot\hat{\vec{x}}_i)\\
&\;\;\;\;\;+ \alpha'^2\sum^{N_{\rm ran}}_i w_{\text{\tiny{FKP}}}^2(\vec{x}_i)\mathcal{L}_{\ell}(\hat{\vec{k}}\cdot\hat{\vec{x}}_i).
\end{split}
\end{equation}
Note that because $\int^1_{-1}K\mathcal{L}_{\ell}(\mu)d\mu = 0$ for $\ell > 0$ and any constant $K$, the shot noise term will vanish for the quadrupole ($\ell = 2$) and hexadecapole ($\ell = 4$) if the window function is isotropic. In order to minimise the additional shot noise contribution from the random catalogue to the power spectrum and its error, we generate a very large (i.e., dense) random catalogue with $\alpha' = 0.036$.

\begin{figure*}
\begin{center}
\epsfig{file=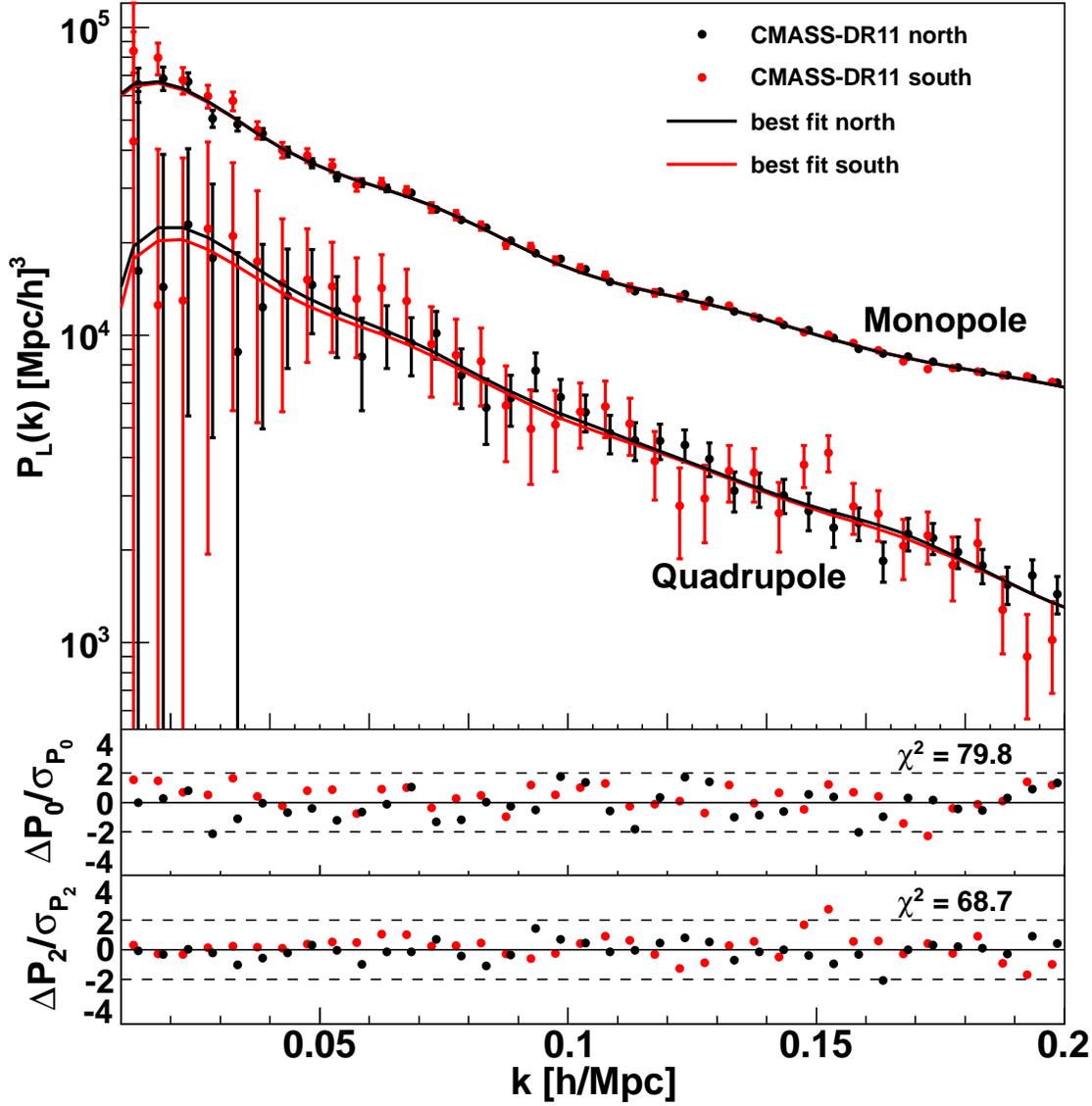,width=16cm}
\caption{The measured CMASS-DR11 monopole (top) and quadrupole (bottom) power spectra. The black data points are the measurement of the North Galactic Cap (NGC) and the red data points are the measurement of the South Galactic Cap (SGC) of CMASS-DR11. The black data points have been shifted by $\Delta k = 0.001h/$Mpc to the right for clarity. The error bars are the diagonal of the covariance matrix. Because of the smaller volume in the SGC the error bars are larger by a factor of $\sim 1.6$. The solid black and red lines represent the best fitting power spectra for the NGC (black) and SGC (red) respectively (fitting range $k = 0.01$ - $0.20h/$Mpc, see section~\ref{sec:fit}). The red and black lines are based on the same cosmology and only differ in the effect of the window function (see section~\ref{sec:win}). The lower two panels show the difference between the measured monopole and the best fitting monopole (middle panel) and the measured quadrupole and the best fitting quadrupole (bottom panel), both relative to the diagonal elements of the covariance matrix. We fit the monopole and quadrupole simultaneously. The best fitting $\chi^2$ is $66.6+73.9 = 140.5$ (NGC + SGC) for $152$ bins and $7$ free parameters (see Table~\ref{tab:para}). The contribution to $\chi^2$ from the monopole and quadrupole alone is given in the middle and lower panel, for comparison.}
\label{fig:ps1}
\end{center}
\end{figure*}

The final power spectrum is then calculated as the average over spherical k-space shells
\begin{align}
P_{\ell}(k) = \langle P_{\ell}(\vec{k})\rangle &= \frac{1}{V_k}\int_{\rm \text{k-shells}} d\vec{k}\;P_{\ell}(\vec{k})\\
&= \frac{1}{N_{\rm modes}}\sum_{k-\frac{\Delta k}{2} < |\vec{k}| < k+\frac{\Delta k}{2}}P_{\ell}(\vec{k}),
\label{eq:averaging}
\end{align}
where $V_k$ is the volume of the k-space shell and $N_{\rm modes}$ is the number of $\vec{k}$ modes in that shell. In our analysis we use $\Delta k = 0.005h/$Mpc. 

The method described above has a bias at larger scales arising from the discreteness of the gridding in k-space~\citep{Blake:2011rj}. The effect can be estimated by comparing a model power spectrum with a gridded model power spectrum, where the gridded model power spectrum is defined as
\begin{align}
P_{\ell}^{\rm gm}(\vec{k}) = \frac{(2\ell + 1)\alpha'}{2A}\sum_{i=1}^{N_{\rm ran}} n'_g(\vec{x}_i)w^2_{\rm FKP}(\vec{x}_i) P^{\rm m}(k,\mu)\mathcal{L}_{\ell}(\hat{\vec{k}}\cdot\hat{\vec{x}}_i).
\end{align}
This should be averaged following eq.~\ref{eq:averaging} and compared to a model power spectrum of the form
\begin{equation}
P_{\ell}^{\rm em}(k) = \frac{(2\ell + 1)}{2}\int^1_{-1}d\mu\; P^{\rm m}(k,\mu)\mathcal{L}_{\ell}(\mu).
\label{eq:multipole27}
\end{equation}
The final estimate of the power spectrum is then given by
\begin{equation}
P^{\rm final}_{\ell}(k) = P_{\ell}(k) + P_{\ell}^{\rm em}(k) - P_{\ell}^{\rm gm}(k),
\label{eq:cor}
\end{equation}
where $P_{\ell}(k)$ on the right hand side is the measured power spectrum and $P^{\rm final}_{\ell}(k)$ is the measured power spectrum after being  corrected for the discrete gridding in k space. In our case this correction is $0.08\%$ ($1.5\%$) at $k=0.04h/$Mpc and $0.03\%$ ($0.09\%$) at $k=0.10h/$Mpc for the monopole and quadrupole, respectively. We show the measurement of the power spectrum monopole and quadrupole for CMASS-DR11 NGC (black) and SGC (red) in Figure~\ref{fig:ps1}.

\subsection{The Poisson shot noise}
\label{sec:SN}

Here we are going to discuss the impact of the CMASS incompleteness weighting on the shot noise term. In principle, any arbitrary constant weight applied to observed galaxies should not change the shot noise term, since no information is added. For example if one decides to up-weight each galaxy by a constant factor, e.g. the average incompleteness of the survey, the shot noise term should not change. In CMASS we have several different kinds of weights, and here we argue that some of these weights use extra information, in a sense that they should reduce the shot noise:
\begin{enumerate}
\item \textbf{Fibre collision, $w_{\rm fc}$ and redshift failure, $w_{\rm rf}$ weight:} Galaxies which did not get a redshift due to fibre collision or redshift failure are still included in the galaxy catalogue by  double counting the nearest galaxy (see~\citealt{Ross:2012qm} for details). For each missing galaxy we know its angular position exactly. Even though the procedure to use the redshift of the closest galaxy is incorrect for some fraction of the missing galaxies~\citep{Guo:2011ai} it means we effectively put extra galaxies into the survey in a non-random fashion, which should reduce the shot noise term. We hence include the fibre collision as well as the redshift failure weights in the shot noise term.
\item \textbf{Systematic weights, $w_{\rm sys}$:} The CMASS sample shows correlations between the galaxy density and the proximity to a star as well as between the galaxy density and the seeing conditions for a particular observation. These correlations are removed using galaxy specific weights (systematic weights). Here we know only statistically that there were missed galaxies, but never know exactly where. To correct for these correlations we up-weight observed galaxies depending on their proximity to stars and the seeing condition for that particular observation. The correction is not random, but it is linked to a Poisson process (e.g. the existence of another galaxy around that star). Therefore we argue that the systematic weights should not reduce the shot noise. We also note that the systematic weights are much smaller than the fibre collision and redshift failure weight and hence the impact to the shot noise term is small.
\end{enumerate}
The shot noise term defines how the galaxy density field enters in the minimum variance weight, $w_{\rm FKP}$, and hence the arguments discussed above result in a minimum variance weight of the form: 
\begin{equation}
w_{\rm FKP}(\vec{x}) = \frac{1}{1 + \frac{n'_g(\vec{x})P_0}{w_{\rm sys}(\vec{x})}}.
\label{eq:wFKP1}
\end{equation}
A detailed derivation can be found in appendix~\ref{ap:minvar}. Since the systematic weights employed in our analysis are very small, our definition of $w_{\rm FKP}$ is almost identical to the commonly used
\begin{equation}
w_{\rm FKP}(\vec{x}) = \frac{1}{1 + n'_g(\vec{x})P_0}.
\label{eq:wFKP2}
\end{equation}
If we were to assume that the systematic weights, $w_{\rm sys}(\vec{x})$, reduce the shot noise, eq.~\ref{eq:wFKP1} and eq.~\ref{eq:wFKP2} would be identical. The value of $P_0$ defines the power spectrum amplitude at which the error is minimised. In this analysis we use $P_0 = 20\,000\,$Mpc$^3/h^3$, which corresponds to $k \sim 0.10h/$Mpc and evaluate the density in redshift bins.

Several studies in recent years reported deviations from the pure Poisson shot noise assumption~\citep{CasasMiranda:2001ym,Seljak:2009af,Manera:2009zu,Hamaus:2010im,Baldauf:2013hka}. Even though we discussed our definition of the shot noise term at length in this section, the parameter constraints we derive in this paper are fairly independent of the precise definition, since for all parameter constraints we are marginalising over a constant offset, $N$ (see section~\ref{sec:taruya}). 

\section{CMASS-DR11 mock catalogues}
\label{sec:cov}

\begin{figure*}
\begin{center}
\epsfig{file=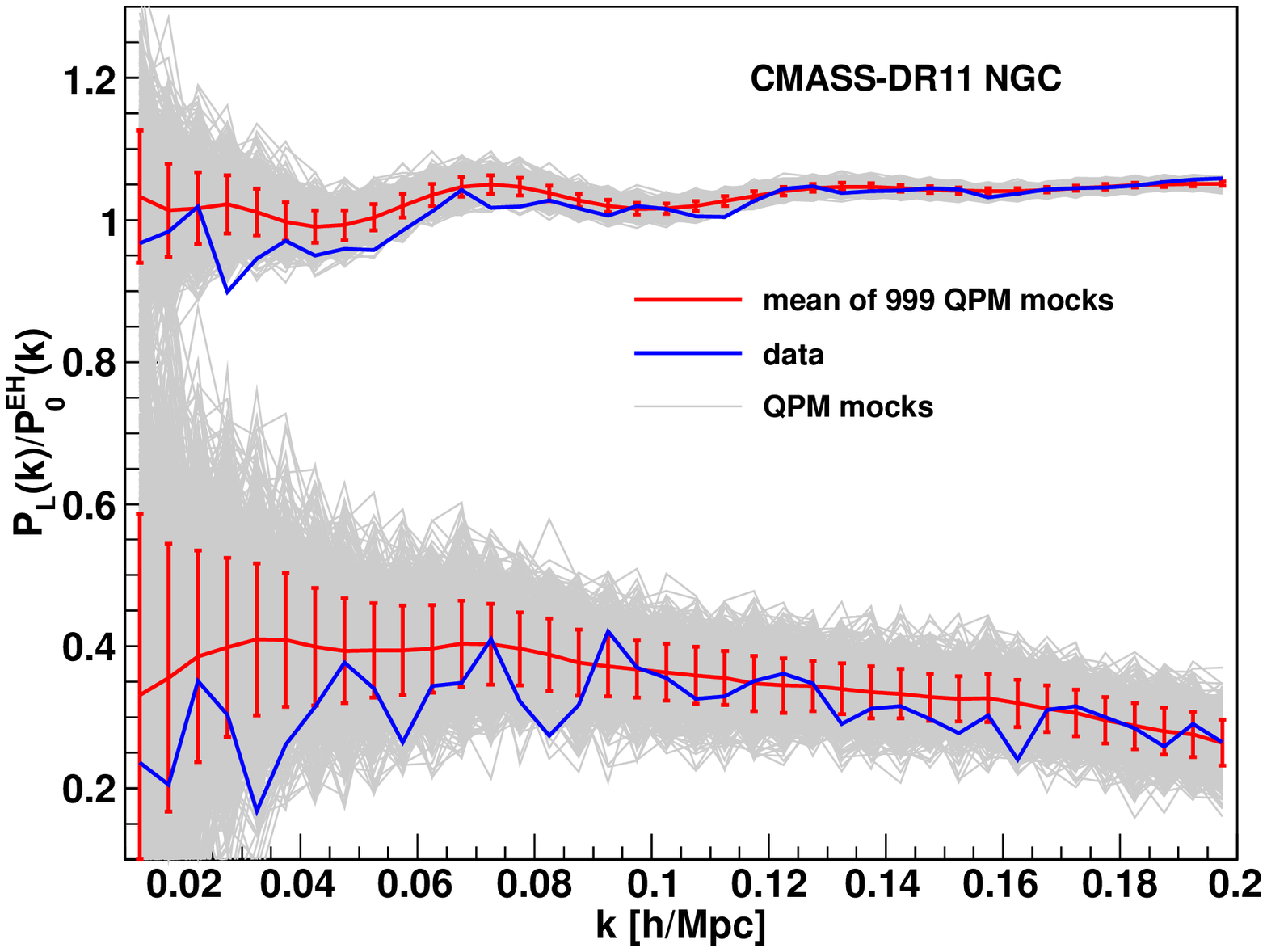,width=8.8cm}
\epsfig{file=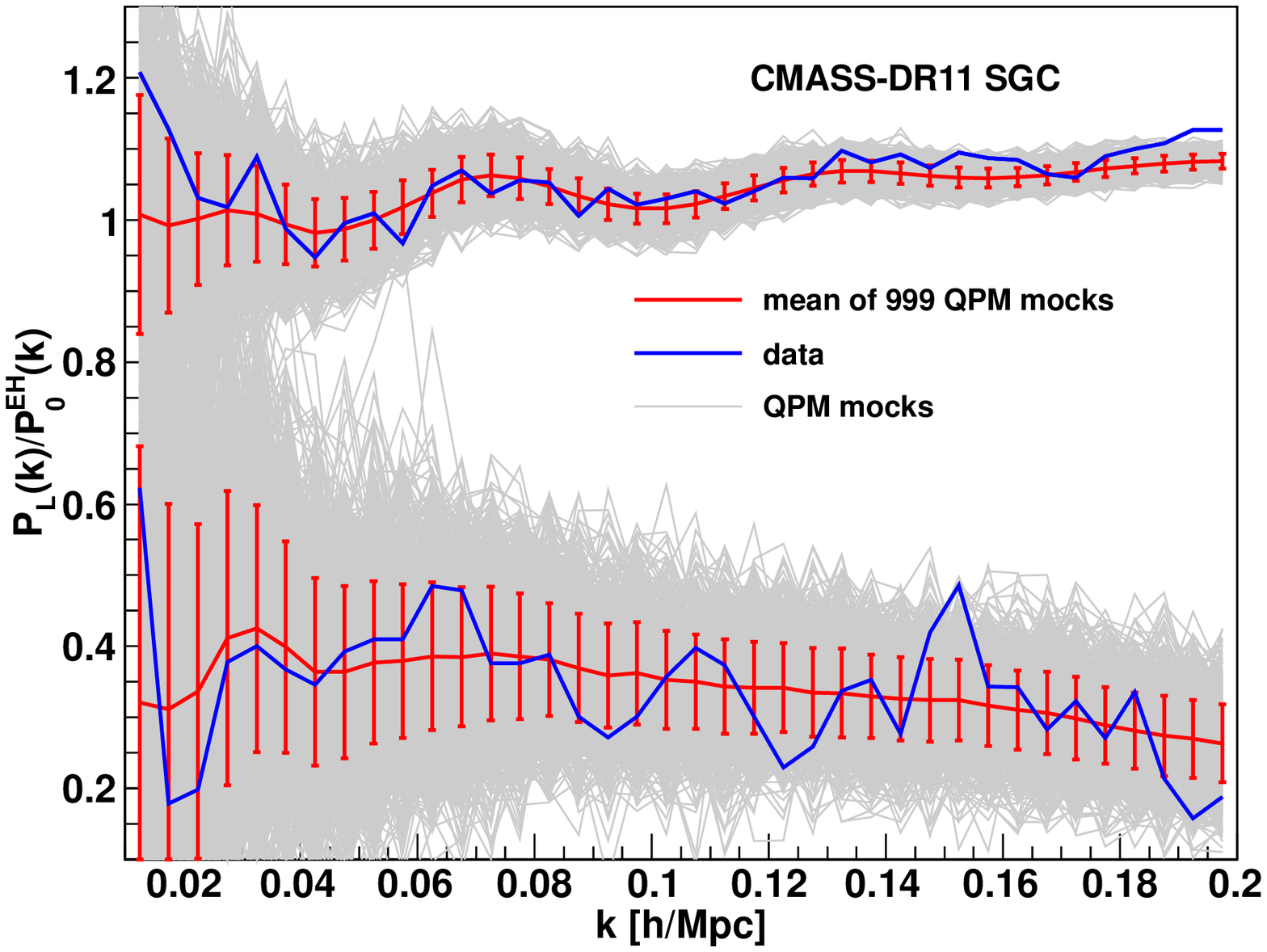,width=8.8cm}\\
\caption{The power spectrum monopole (top) and quadrupole (bottom) of the $999$ QPM mock catalogues (grey lines) for the North Galactic Cap (left) and the South Galactic Cap (right), relative to an~\citet{Eisenstein:1997ik} no-BAO monopole power spectrum. We plot the power spectrum without the shot noise subtraction, since this way, the scatter closely represents the diagonal of the covariance matrix. The red lines show the mean of all mock catalogues with the error representing the variance around the mean. The blue lines show the measured CMASS-DR11 power spectra.}
\label{fig:QPMmean}
\end{center}
\end{figure*}
In our analysis we use $999$ mock catalogues which follow the same selection function as the CMASS-DR11 sample. The catalogues are produced using quick particle-mesh (QPM) N-body simulations~\citep{White:2013psd} with $1280^3$ particles in a $[2560\,\text{Mpc}/h]^3$ box. These simulations have been found to better describe the clustering of CMASS galaxies compared to the previous version of CMASS mock catalogues~\citep{Manera:2012sc}, especially at small scales~\citep{McBride:2013}. Each simulation started from 2LPT initial conditions at $z=25$ and evolved to the present using time steps of $15\%$ in $\ln(a)$, where $a = (1 + z)^{-1}$ is the scale factor. The fiducial cosmology assumes flat $\Lambda$CDM with $\Omega_m=0.29$, $h=0.7$, $n_s=0.97$ and $\sigma_8=0.8$. We use the simulation output at $z=0.55$, where the simulation generated a sub-sample of the N-body particles and a halo catalogue using the friends-of-friends algorithm with a linking length of $0.2$ times the mean inter-particle spacing. The halo catalogue is then extended to lower masses by appointing a set of the sub-sampled particles as halos and assigning them a mass using the peak-background split mass function. The halos are then populated by galaxies using the Halo Occupation Distribution (HOD) formalism with the occupation functions (see e.g.~\citealt{Tinker:2011pv})
\begin{align}
\langle N_{\rm cen}\rangle_M &= \frac{1}{2}\left[1 + \text{erf}\left(\frac{\log M - \log M_{\rm min}}{\sigma_{\log M}}\right)\right],\\
\langle N_{\rm sat}\rangle_M &= \langle N_{\rm cen}\rangle_M\left(\frac{M}{M_{\rm sat}}\right)^{\alpha}\exp\left(\frac{-M_{\rm cut}}{M}\right),
\end{align}
where we use $M_{\rm min} = 9.319\times 10^{12}M_{\odot}/h$, $\sigma_{\log M} = 0.2$, $\alpha = 1.1$, $M_{\rm sat} = 6.729\times 10^{13}M_{\odot}/h$ and $M_{\rm cut} = 4.749\times 10^{13}M_{\odot}/h$ (Jeremy Tinker, private communication). In section~\ref{sec:sys} we will modify the HOD parameters to test possible systematic effects in our modelling of the power spectrum multipoles. For more details about the QPM mock catalogues see~\citet{McBride:2013} and~\citet{White:2013psd}.

\subsection{The covariance matrix}

\begin{figure*}
\begin{center}
\epsfig{file=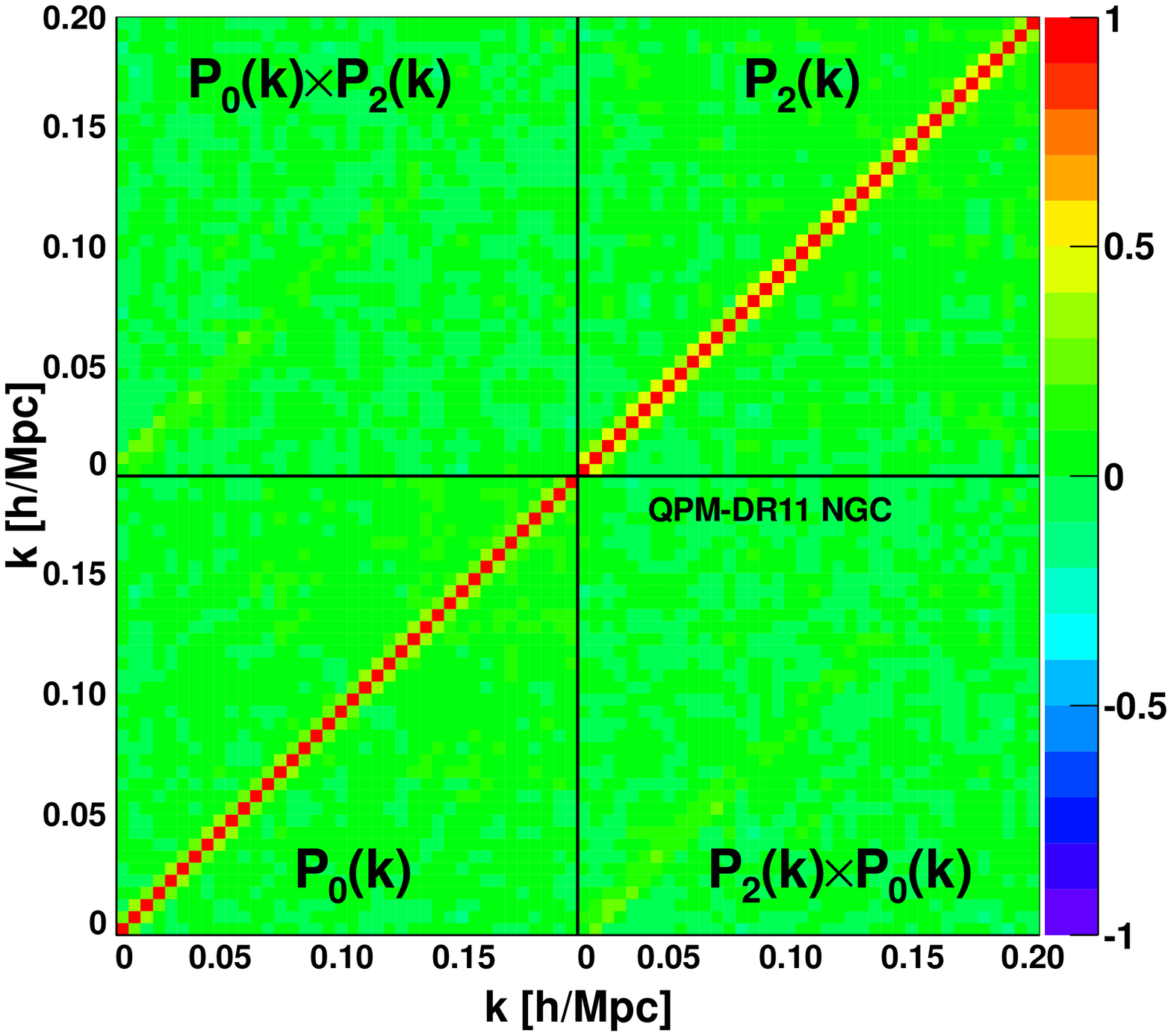,width=8cm}
\epsfig{file=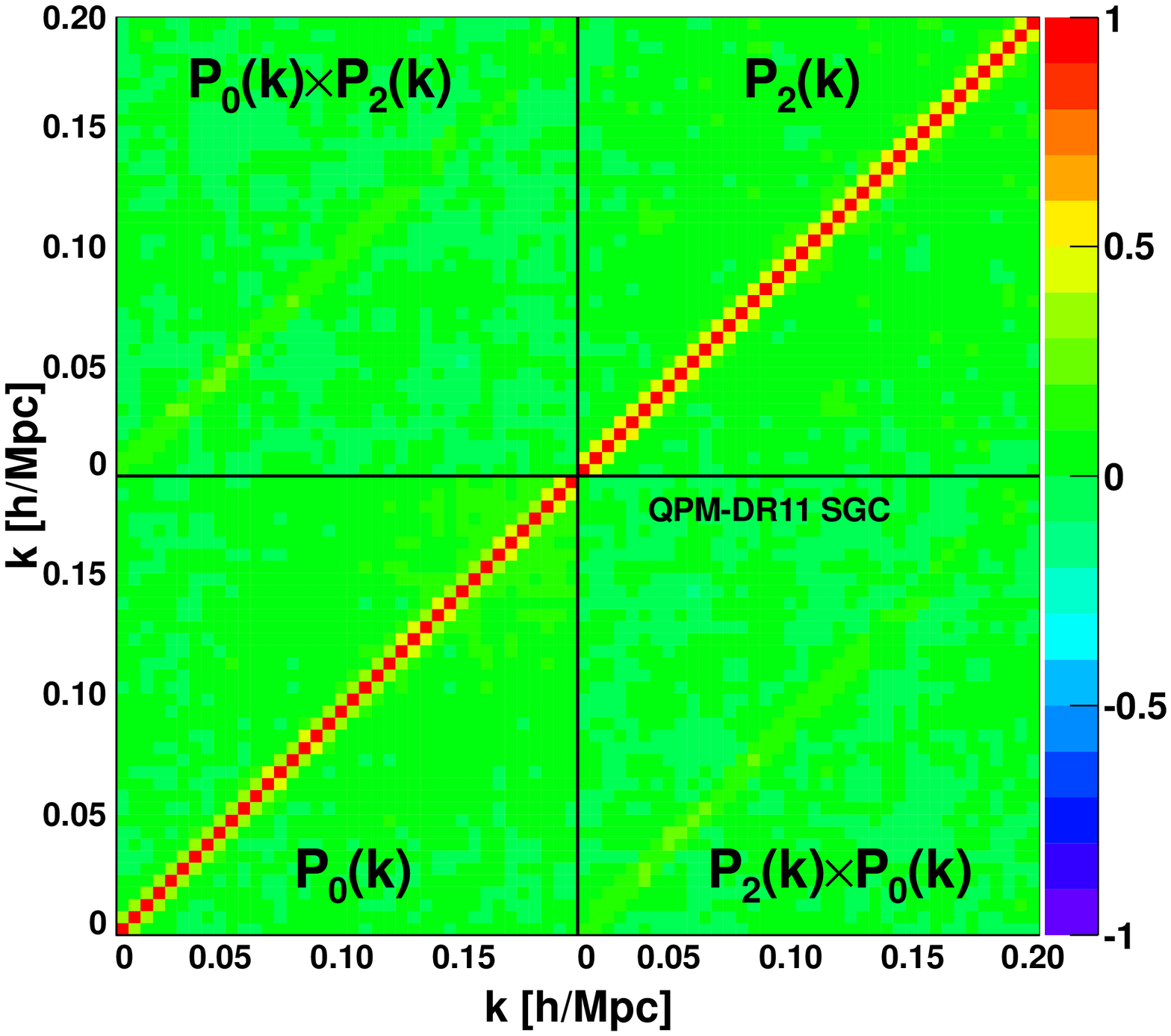,width=8cm}
\caption{The correlation matrix for the NGC (left) and SGC (right) of CMASS-DR11. The colour indicates the level of correlation, where red represents high correlation, blue represents high anti-correlation and green represents no-correlation. The correlation between the bins in the monopole is shown in the lower left hand corner, while the correlation between the $k$-bins in the quadrupole is shown in the upper right hand corner. The upper left hand corner and the lower right hand corner show the cross-correlations.}
\label{fig:covmatrixQPM}
\end{center}
\end{figure*}

We measure the power spectrum monopole and quadrupole for each of the $999$ QPM mocks, using the estimator introduced in section~\ref{sec:Yamamoto}. The $999$ power spectrum monopoles and quadrupoles are shown in Figure~\ref{fig:QPMmean} together with the mean (red) and the CMASS-DR11 measurements (blue). We can see that the mock catalogues closely reproduce the data power spectrum multipoles for the entire range of wavenumbers relevant for this analysis. 

The covariance matrix is then given by
\begin{equation}
C_{x,y} = \frac{1}{N_s-1}\sum^{N_s}_{n=1}\left[P_{\ell,n}(k_i) - \overline{P}_{\ell}(k_i)\right]\left[P_{\ell',n}(k_j) - \overline{P}_{\ell'}(k_j)\right],
\end{equation}
where $N_s = 999$ represents the number of mock realisations.
We estimate the covariance matrices for the NGC and SGC separately, i.e. treat them as statistically independent samples. This covariance matrix contains the monopole as well as the quadrupole, and the elements of the matrices are given by $(x,y) = (\frac{n_b\ell}{4}+i,\frac{n_b\ell'}{4}+j)$, where $n_b$ is the number of bins in each multipole power spectrum. Our $k$-binning yields $n_{b}=76$ $(56)$ for the fitting range $k_{\rm max}=0.01$ - $0.20$ $(0.01$ - $0.15)h/$Mpc, and hence the dimensions of the covariance matrices become $76\times 76$ ($56\times 56$) for the NGC and SGC. The mean of the power spectrum is defined as
\begin{equation}
\overline{P}_{\ell}(k_i) = \frac{1}{N_s}\sum^{N_s}_{n=1} P^n_{\ell}(k_i).
\end{equation}
The mock catalogues automatically incorporate the window function and integral constraint effect present in the data. Figure~\ref{fig:covmatrixQPM} shows the correlation matrix for CMASS-DR11 NGC (left) and SGC (right), where the correlation coefficient is defined as
\begin{equation}
r_{xy} = \frac{C_{xy}}{\sqrt{C_{xx}C_{yy}}}.
\end{equation}
The lower left hand corner shows the correlation between bins in the monopole, the upper right hand corner shows correlations between the bins in the quadrupole and the upper left hand corner and lower right hand corner show the correlation between the monopole and quadrupole. Most of the correlation matrix is coloured green, indicating no or a small level of correlation. This is expected for the linear power spectrum since each Fourier mode evolves independently. For larger wave-numbers non-linear effects will introduce correlations between bins, while for very small wave-numbers window function effects can introduce correlations. 

As the estimated covariance matrix $C$ is inferred from mock catalogues, its inverse, $C^{-1}$, provides a biased estimate of the true inverse covariance matrix, due to the skewed nature of the inverse Wishart distribution~\citep{Hartlap:2006kj}. To correct for this bias we rescale the inverse covariance matrix as 
\begin{equation}
C_{ij,\rm Hartlap}^{-1} = \frac{N_s - n_b - 2}{N_s - 1}C_{ij}^{-1},
\label{eq:covhartlap}
\end{equation} 
where $n_b$ is the number of power spectrum bins. With these covariance matrices we can then perform a standard $\chi^2$ minimisation to find the best fitting parameters.

\begin{figure}
\begin{center}
\epsfig{file=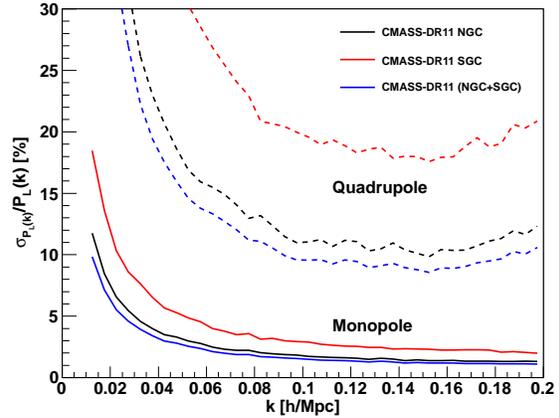,width=8cm}
\caption{Relative error using the diagonal elements of the covariance matrix of the power spectrum multipoles in CMASS-DR11. The upper three dashed lines show the quadrupole error and the lower three solid lines show the error in the monopole. Because of the larger volume, the error in the NGC of CMASS-DR11 (black lines) is about a factor of $1.6$ smaller than the error in the SGC (red lines). The power spectrum error for the entire CMASS-DR11 sample (blue lines) shows an error of $\sim 1.5\%$ in the monopole and $\sim 10\%$ in the quadrupole at $k=0.10h/$Mpc.}
\label{fig:cerror}
\end{center}
\end{figure}

In Figure~\ref{fig:cerror} we show the diagonal elements of the covariance matrix for the monopole and quadrupole power spectrum. We find an error of $\sim 1.5\%$ in the monopole and $\sim 10\%$ in the quadrupole at $k=0.10h/$Mpc. This represents the most precise measurement of the galaxy power spectrum multipoles ever obtained.

\section{The survey window function}
\label{sec:win}
 
The power spectrum estimator we discussed in section~\ref{sec:Yamamoto} is not actually estimating the true galaxy power spectrum, but rather the galaxy power spectrum convolved with the survey window function:
\begin{equation}
\begin{split}
P^{\rm conv}(\vec{k}) &= \int d\vec{k}' P^{\rm true}(\vec{k}')|W(\vec{k}-\vec{k}')|^2\\
&\;\;\;\;\; - \frac{|W(\vec{k})|^2}{|W(0)|^2}\int d\vec{k}' P^{\rm true}(\vec{k}')|W(\vec{k}')|^2.
\end{split}
\label{eq:surveywindow}
\end{equation}
The window function, $W(\vec{k})$ has the following two effects: (1) It mixes the modes with different wave-numbers and introduces correlations and (2) it changes the amplitude of the power spectrum at small $k$. First we discuss the first term of eq.~\ref{eq:surveywindow}, the convolution of the true power spectrum with the window function. The second term of eq.~\ref{eq:surveywindow}, the so-called integral constraint, will be discussed in the next subsection. We present the full derivation of the equations of this section in Appendix~\ref{ap:window} and restrict the discussion here to the main results.\vspace{1cm}

\subsection{The convolution of the power spectrum with the window function}

\begin{figure*}
\begin{center}
\epsfig{file=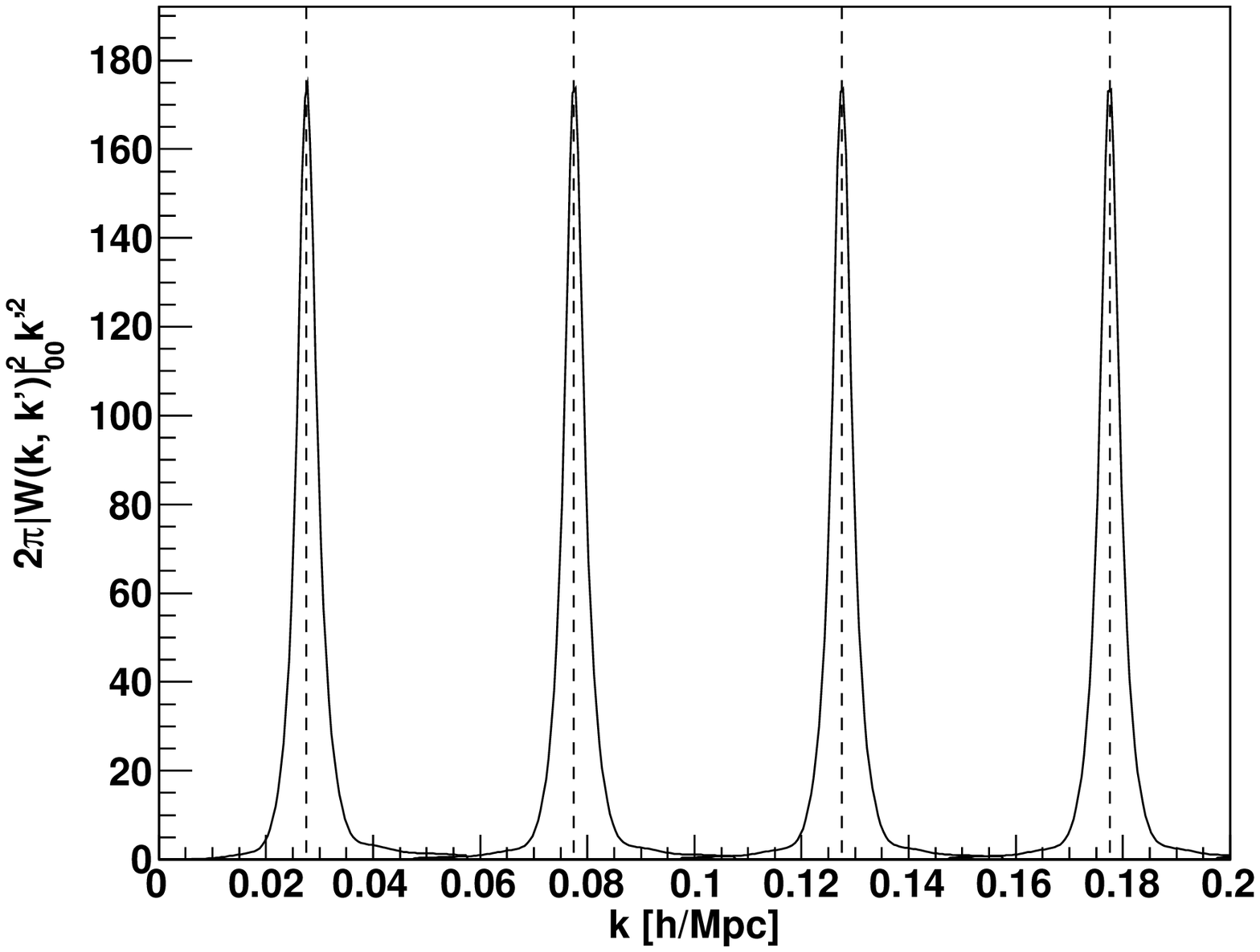,width=8cm}
\epsfig{file=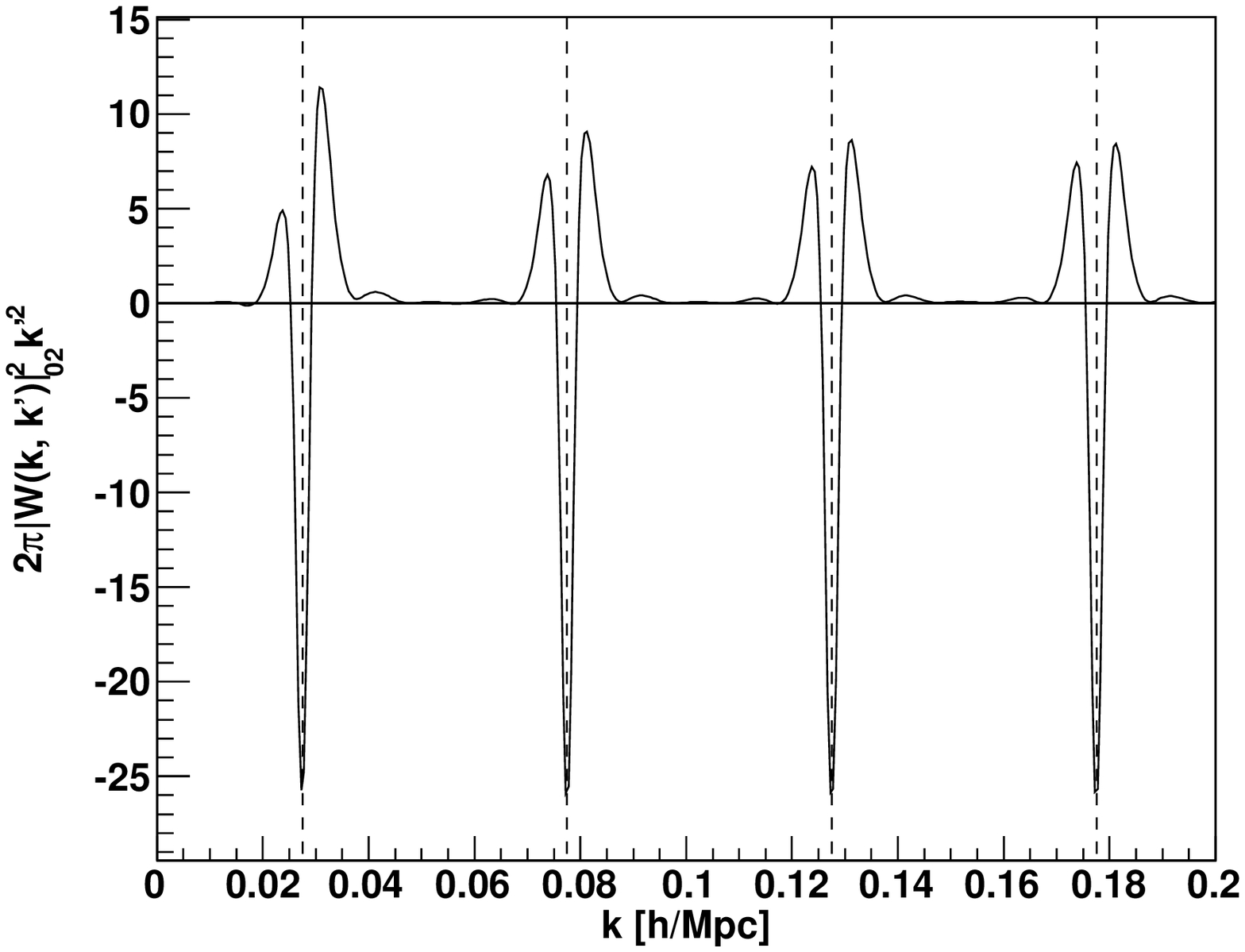,width=8cm}\\
\epsfig{file=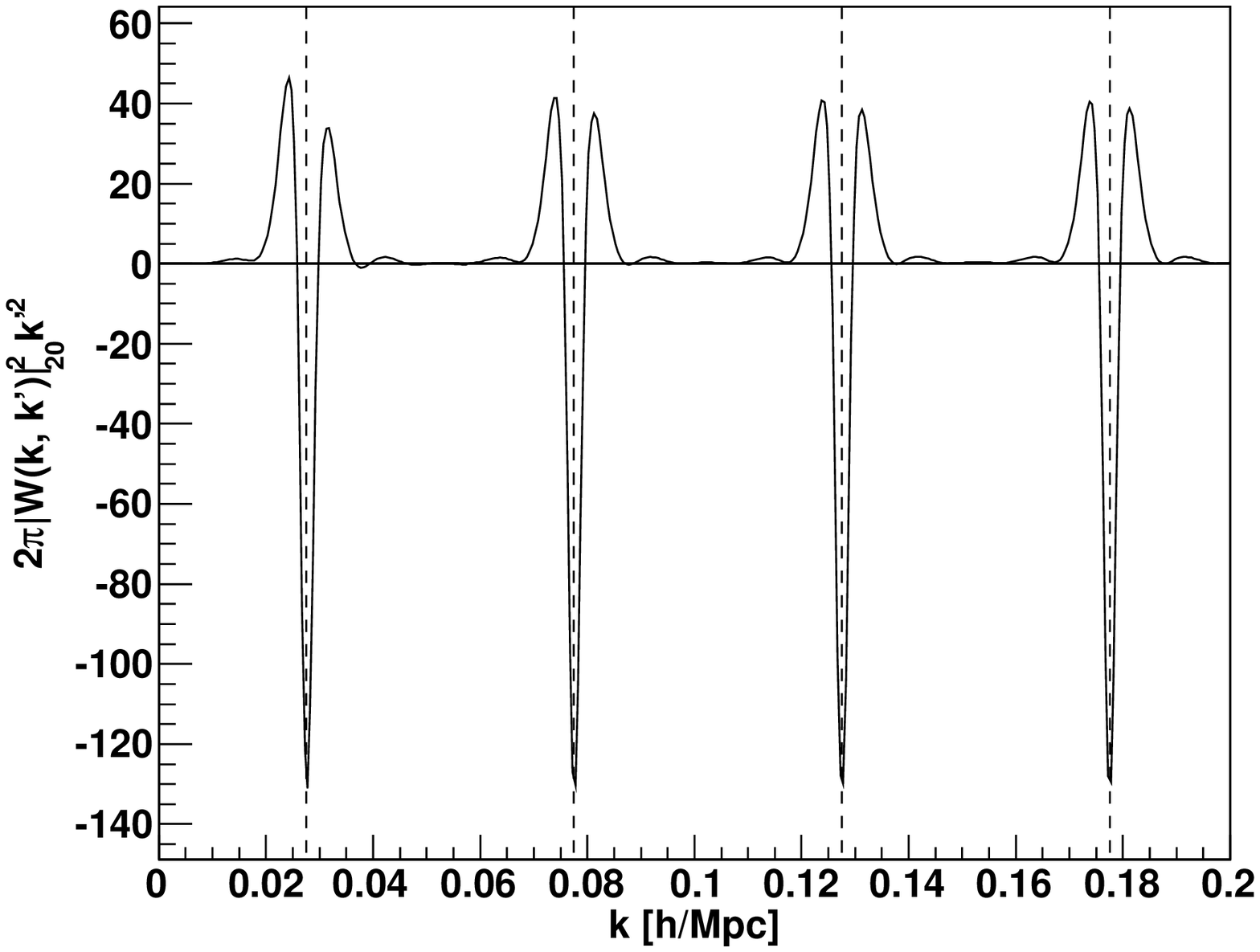,width=8cm}
\epsfig{file=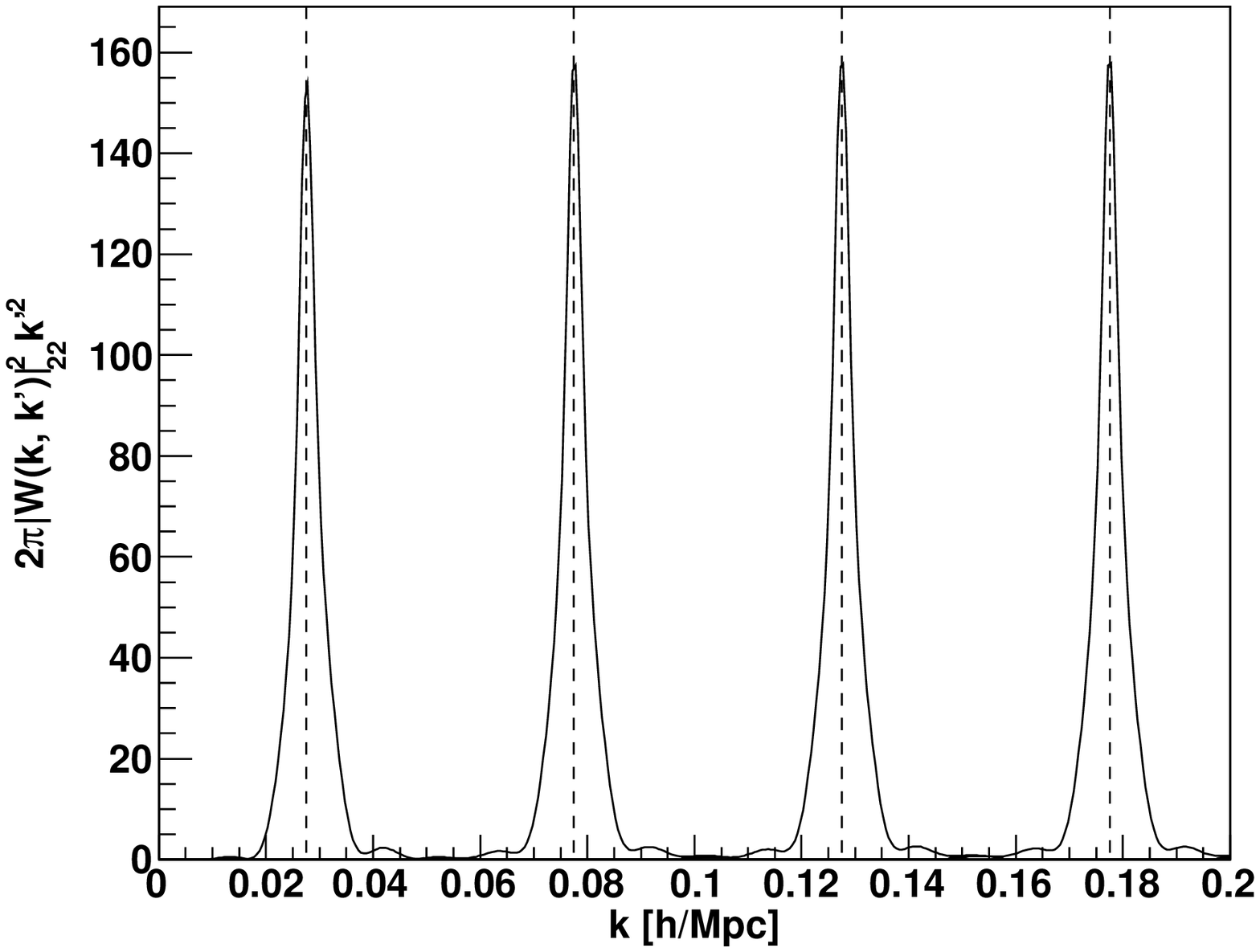,width=8cm}
\caption{The window function multipoles of the NGC of CMASS-DR11 required in eq.~\ref{eq:conv} and calculated using eq.~\ref{eq:conv2}. The window function multipoles are plotted as a function of $k$ for fixed values of $k' = (0.0275, 0.0775, 0.1275, 0.1775)$ (black dashed lines). Note that the window function multipoles are not symmetric under $\ell$ and $L$ (see eq.~\ref{eq:conv2}).}
\label{fig:Wk2}
\end{center}
\end{figure*}

Window function effects in the measured power spectrum do not necessarily represent a problem, since the survey window function is known in principle. One possible way to handle the window function is to deconvolve the measured power spectrum to get the true galaxy power spectrum~\citep{Baugh:1993,Lin:1996td, Sato:2010my,Sato:2013hea}. Here we follow the more common procedure to convolve each model power spectrum (i.e., $P^{\rm true}$) with the survey window function and derive a model $P^{\rm conv}$, which is then compared to the measured power spectrum. However, the straightforward implementation of eq.~\ref{eq:surveywindow} mode-by-mode would lead to a complexity of $\sim \mathcal{O}(N_c^2)$, where $N_c$ is the total number of modes. For most practical cases this is impossible to evaluate. Therefore, most studies in the past evaluated eq.~\ref{eq:surveywindow} as a convolution with the spherically averaged window function, $W_s$ (see e.g.~\citealt{deLaix:1997rr, Percival:2001hw, Cole:2005sx, Percival:2006gt,Ross:2012sx})
\begin{equation}
P^{\rm conv}(k) = \int d\vec{\epsilon}\,P^{\rm true}(\vec{k}+\vec{\epsilon})|W(\vec{\epsilon})|^2_s
\label{eq:sph}
\end{equation}
which assumes an isotropic power spectrum. The spherically averaged window function is defined as
\begin{equation}
|W(\vec{\epsilon})|_s^2 = \frac{1}{4\pi}\int d\Omega_{\epsilon'}|W(\vec{\epsilon}')|^2\delta(r_{\epsilon'} - r_{\epsilon}),
\end{equation}
with $r_{\epsilon} = |\vec{k} + \vec{\epsilon}|$. In our analysis we want to measure anisotropic signals in the power spectrum (AP effect and RSD), and hence the assumption of an isotropic power spectrum seems contradictory.

In a recent analysis,~\citet{Sato:2013hea} suggested splitting the survey into sub-regions (see also~\citealt{Hemantha:2013}), which are small enough that the plane parallel approximation can be applied. In this case the window function can be calculated using FFTs. However, the window function effect on the power spectrum in any sub-region will be larger than in the original survey, and there is a trade-off between keeping the window function(s) compact and making the plane parallel approximation work. These problems become especially prominent for the higher order multipoles. In addition to the enhanced window function effects, splitting the survey will discard large scale modes.

In this section we will present a treatment of the convolution of the power spectrum with the window function without any assumptions regarding isotropy and without the need to split the survey into sub-regions. Our approach has a complexity of only $\mathcal{O}(N^2_{\rm ran})$. We believe that our approach is more rigorous and allows a more efficient use of the available data, compared to the methods discussed above.

\begin{figure*}
\begin{center}
\epsfig{file=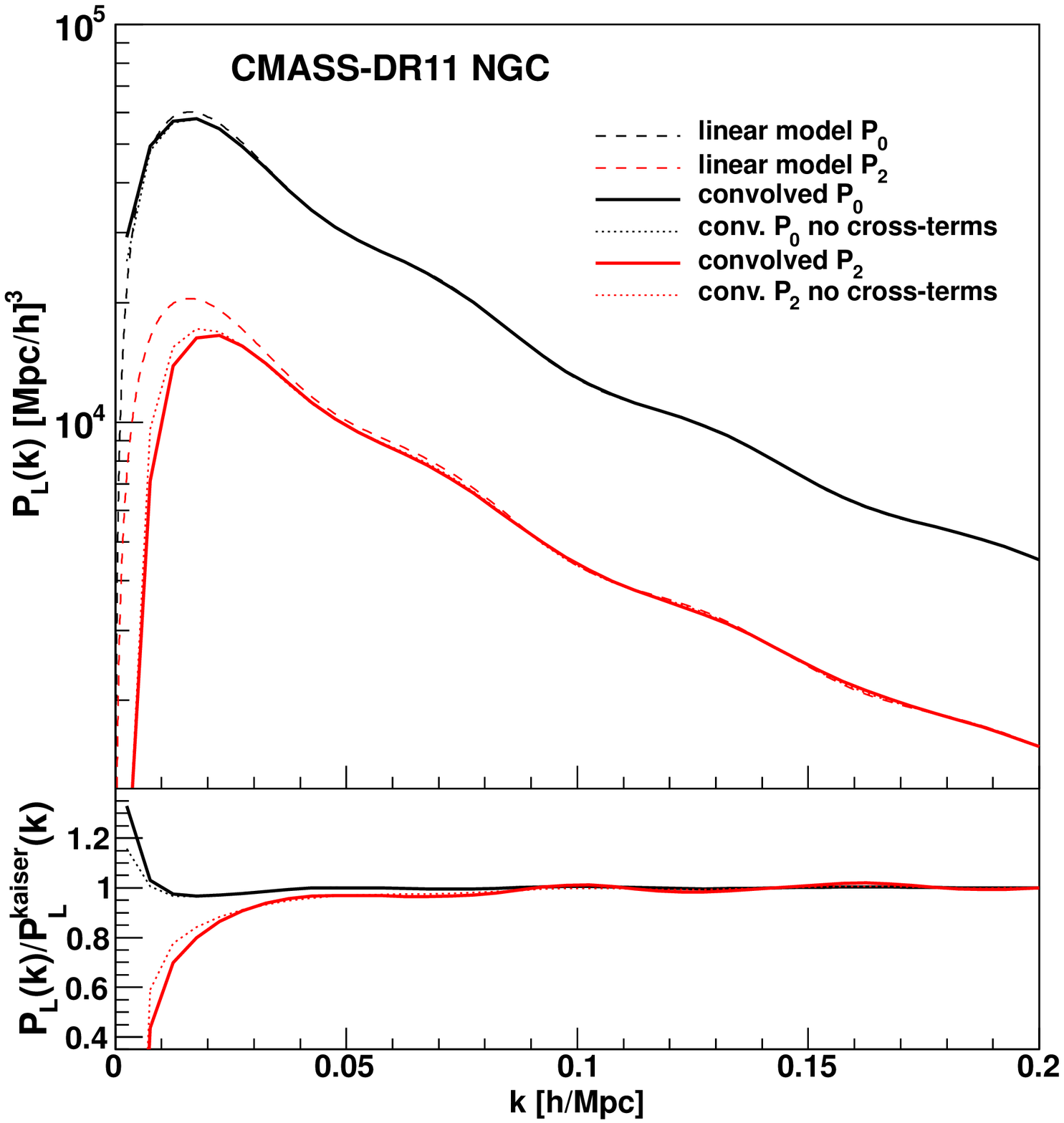,width=8.8cm}
\epsfig{file=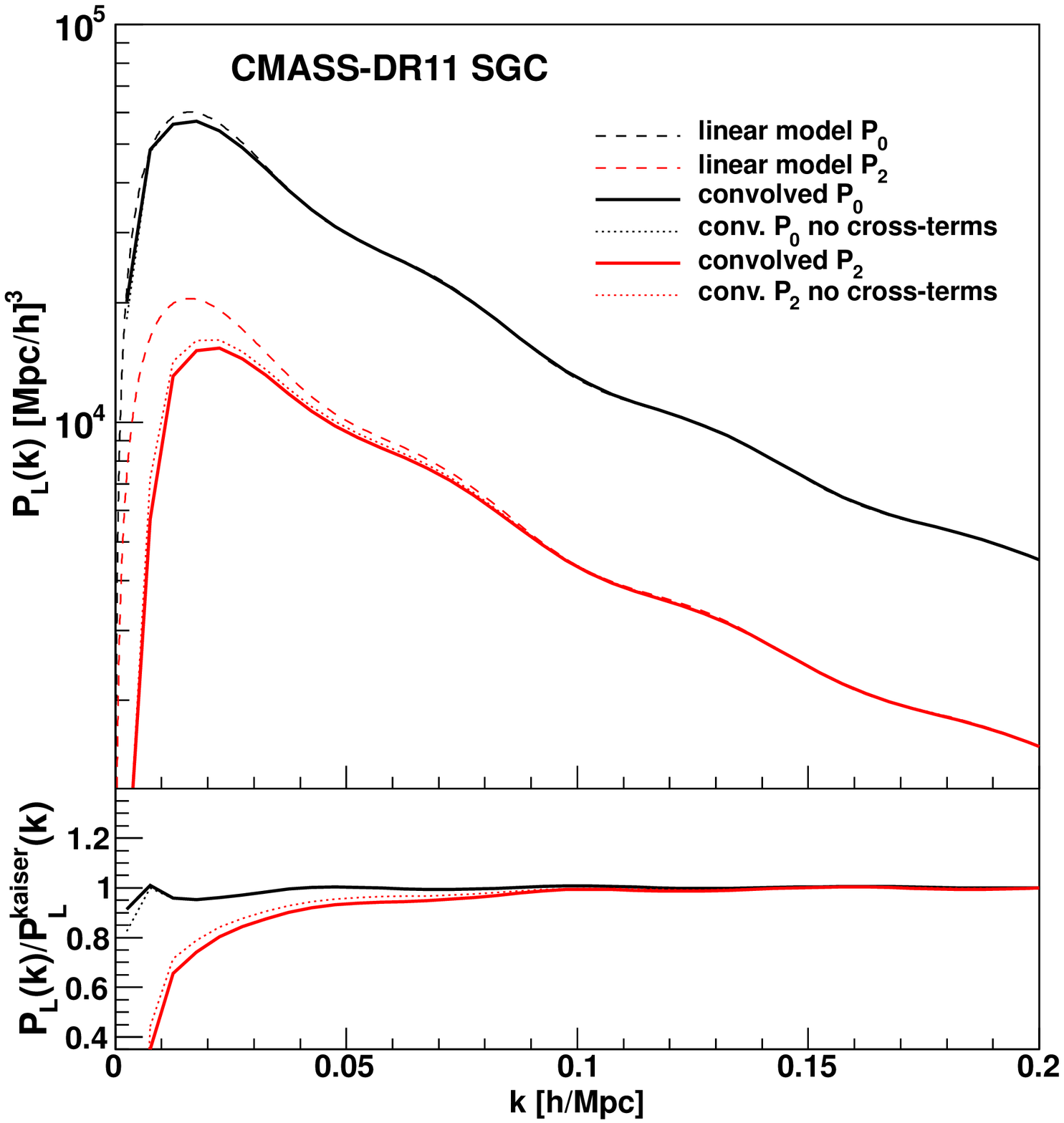,width=8.8cm}
\caption{A model monopole (black dashed lines) and quadrupole (red dashed lines) power spectra using Planck cosmological parameters and the linear Kaiser effect. The solid lines show the same models convolved with the CMASS-DR11 window functions for the NGC (left) and the SGC (right). The black dotted lines show the convolved monopole power spectra where the quadrupole contribution to the monopole has been ignored and the red dotted lines shows the convolved quadrupole power spectra where the monopole contribution to the quadrupole has been ignored (see eq.~\ref{eq:conv}). The bottom panels show the same power spectra relative to the original linear power spectra including the Kaiser effect (dashed lines in the top panels).}
\label{fig:ps_conv}
\end{center}
\end{figure*}

We can express eq.~\ref{eq:surveywindow} in terms of the wavevector amplitude $k = |\vec{k}|$, the cosine of the angle to the line-of-sight $\mu$ and the azimuthal angle $\phi$:
\begin{align}
P^{\rm conv}_{\ell}(k) &= \frac{2\ell + 1}{2}\int d\mu \int \frac{d\phi}{2\pi}\int d\vec{k}'P^{\rm true}(\vec{k}') |W(\vec{k} - \vec{k}')|^2\mathcal{L}_{\ell}(\mu)\notag\\
&= 2\pi\int dk' k'^2\sum_L P^{\rm true}_L(k')|W(k, k')|^2_{\ell L},
\label{eq:conv}
\end{align}
where the window function is now expanded into the Legendre multipole space, and analytical integration over the angles yields
\begin{equation}
\begin{split}
|W (k,k')|^2_{\ell L} &= 2i^{\ell}(-i)^L(2\ell + 1)\sum^{N_{\rm ran}}_{ij, i\neq j}w_{\text{\tiny{FKP}}}(\vec{x}_i)w_{\text{\tiny{FKP}}}(\vec{x}_j)\\
&\;\;\;\;\;j_{\ell}(k|\Delta\vec{x}|) j_L(k'|\Delta\vec{x}|)\mathcal{L}_{\ell}({\hat{\vec{x}}_h\cdot \Delta\hat{\vec{x}}})\mathcal{L}_{L}({\hat{\vec{x}}_h\cdot \Delta\hat{\vec{x}}}).
\end{split}
\label{eq:conv2}
\end{equation}
In this equation $j_{\ell}$ represents the spherical Bessel function of order $\ell$ and $\Delta\vec{x} = \vec{x}_i-\vec{x}_j$ (for a detailed derivation of this equation see appendix~\ref{ap:window}). We plot the different window function multipoles for CMASS-DR11 in Figure~\ref{fig:Wk2}. Eq.~\ref{eq:conv2} shows that there are cross terms between different multipoles, meaning that there is a contribution from e.g. the monopole to the convolved quadrupole. In other words, the survey window may induce an anisotropic signal in the convolved power spectrum even without the RSD or AP effect. These cross terms are neglected in the simplified treatment of eq.~\ref{eq:sph}.

The normalisation for the window function is given by
\begin{align}
\int d\vec{k}' |W(\vec{k} - \vec{k}')|^2 = 1.
\label{eq:norm2D}
\end{align}
In Figure~\ref{fig:ps_conv} we show linear model monopole and quadrupole power spectra before (dashed lines) and after (solid lines) the convolution with the CMASS-DR11 window functions. The dotted lines show the convolved monopole power spectra ignoring the quadrupole contribution in eq.~\ref{eq:conv} (black dotted line) and the convolved quadrupole power spectra ignoring the monopole contribution (red dotted line). While the quadrupole contribution to the monopole seems negligible, there is a small monopole contribution to the quadrupole. All window function effects seem quite small in CMASS-DR11, because of the very compact window function. Whether the full treatment of eq.~\ref{eq:conv} and eq.~\ref{eq:conv2} is needed, or whether one of the approximations discussed in the beginning of this section can be employed, needs to be tested for each galaxy survey. 

\subsection{The integral constraint}
\label{sec:ic}

\begin{figure}
\begin{center}
\epsfig{file=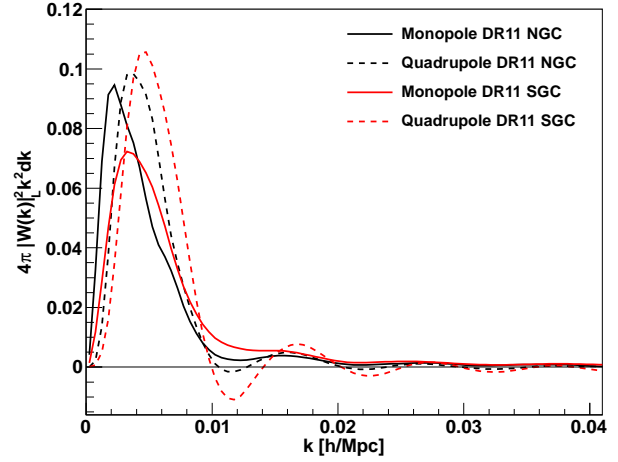,width=8.8cm}
\caption{The window function monopole (solid lines) and quadrupole (dashed lines) for the NGC (black lines) and SGC (red lines) calculated using eq.~\ref{eq:ic2}. The NGC multipoles of CMASS-DR11 peak at smaller wave-numbers $k$ and show weaker oscillations, which is a result of the larger sky coverage (see Figure~\ref{fig:footprint}). The window function multipoles shown in this figure are needed for the integral constraint calculation in eq.~\ref{eq:ic1}.}
\label{fig:Wk}
\end{center}
\end{figure}

Here we discuss the second term of eq.~\ref{eq:surveywindow}. If we go to our original power spectrum estimator (section~\ref{sec:Yamamoto}), we can see that for the mode at $k=0$ we have by design of the random catalogue:
\begin{equation}
\delta(k=0) = \sum^{N_{\rm gal}}_iw_c(\vec{x}_i)w_{\text{\tiny{FKP}}}(\vec{x}_i) - \alpha' \sum^{N_{\rm ran}}_iw_{\text{\tiny{FKP}}}(\vec{x}_i) = 0.
\end{equation}
By setting the $k=0$ mode to zero, we assume that the average density of our survey is equal to the average density of the Universe. The existence of sample variance tells us that this assumption must introduce a bias in our power spectrum estimate, which is known as integral constraint. The effect is that we underestimate the power in modes with wavelength approaching the size of our survey. So even neglecting the window function, we do not measure the true underlying power spectrum, but rather a power spectrum with the property $P(k) \rightarrow 0$ for $k\rightarrow 0$ (see e.g.~\citealt{Peacock:1991}). This is the reason for the second term in eq.~\ref{eq:surveywindow}. It represents the subtraction of the $P(0)$ component which spreads to larger $\vec{k}$, because of the convolution with the window function. Similar to what we did with the window function in the last section, we express the second term in eq.~\ref{eq:surveywindow} in terms of amplitude $k=|\vec{k}|$, the cosine of the angle to the line-of-sight $\mu$ and the azimuthal angle $\phi$:
\begin{equation}
\begin{split}
P^{\rm ic}_{\ell}(k) &= \frac{2\ell + 1}{2}\int d\mu \int \frac{d\phi}{2\pi}\,\frac{|W(\vec{k})|^2}{|W(0)|^2}\\
&\;\;\;\;\; \bigg[\int d\vec{k}' P^{\rm true}(\vec{k}')|W(\vec{k}')|^2\bigg]\mathcal{L}_{\ell}(\mu)\\
&= 2\pi\frac{|W(k)|^2_{\ell}}{|W(0)|^2_{0}} \int dk'k'^2 \sum_LP_L^{\rm true}(k')|W(k')|_L^2\frac{2}{2L+1}
\end{split}
\label{eq:ic1}
\end{equation}
with
\begin{equation}
\begin{split}
|W(k)|^2_{\ell} &= i^{\ell}(2\ell + 1) \sum^{N_{\rm ran}}_{ij, i\neq j}w_{\text{\tiny{FKP}}}(\vec{x}_i)w_{\text{\tiny{FKP}}}(\vec{x}_j)\\
&\;\;\;\;\; j_{\ell}(k|\Delta \vec{x}|)\mathcal{L}_{\ell}(\hat{\vec{x}}_h\cdot \Delta\hat{\vec{x}}).
\label{eq:ic2}
\end{split}
\end{equation} 
This window function is normalised to
\begin{align}
4\pi\int dk' k'^2 |W(k')|^2_{0} = 1,
\end{align}
which is equivalent to eq.~\ref{eq:norm2D}. In Figure~\ref{fig:Wk} we plot the window function multipoles for the NGC and SGC of CMASS-DR11. 
The NGC window function multipoles are more compact (concentrated to small $k$), which results in smaller window function effects in Figure~\ref{fig:ps_conv}. Later, when we fit the measured power spectrum multipoles, we calculate the integral constraint correction for each model multipole power spectrum and subtract it, following eq.~\ref{eq:surveywindow}. This allows a consistent comparison of model power spectra with our measurement. 

\section{Modelling the multipole power spectra}
\label{sec:model}

In this section we discuss our approach to modelling the multipole power spectra to be compared with 
the CMASS-DR11 measurement. 
In order to robustly extract information on RSD and AP from the anisotropic galaxy power 
spectrum in redshift space, it is crucial to prepare a theoretical template 
which takes account of the non-linear effects of 
gravitational evolution, galaxy bias, and RSD at a sufficiently accurate level. 
Particularly in terms of non-linear RSD, several different approaches to model the power 
spectrum or correlation function of the anisotropic galaxy clustering have been suggested 
in recent years ~\citep{Scoccimarro:2004tg,Matsubara:2007wj,Matsubara:2008wx, Carlson:2009it,
Taruya:2010mx,Reid:2011ar,Matsubara:2011ck,Seljak:2011tx,Vlah:2012ni,Wang:2013hwa,Matsubara:2013ofa,Taruya:2013my,Vlah:2013lia,Blazek:2013kfa}. 

We are going to use perturbation theory (PT) 
for such non-linear corrections, which is physically well motivated and widely applicable. 
We first introduce the model of the anisotropic power spectrum in two-dimensional space, 
and then explain how to incorporate the AP effect. 

\subsection{PT approach to model the galaxy power spectrum in redshift space}
\label{sec:taruya}

\begin{figure*}
\begin{center}
\epsfig{file=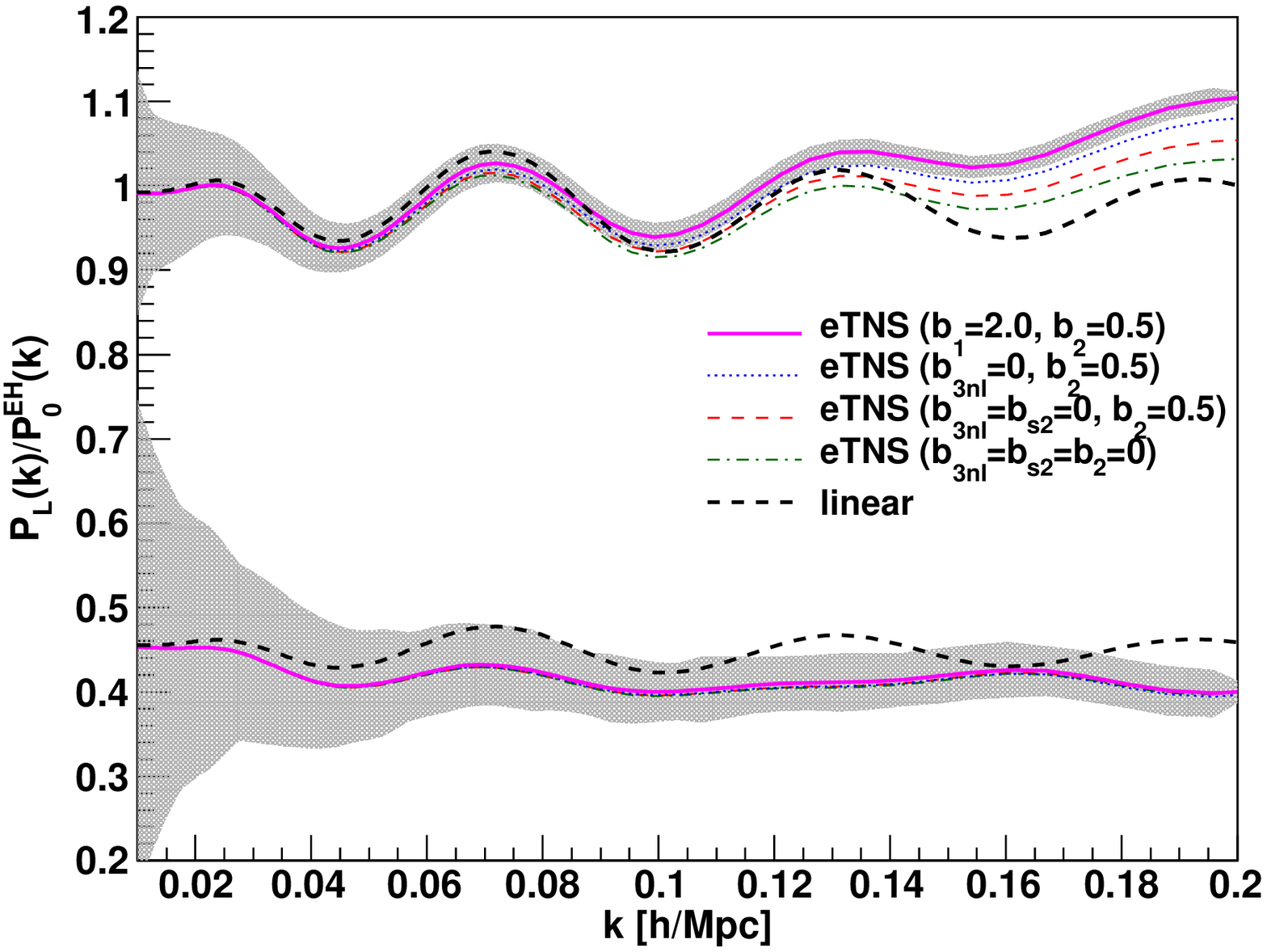,width=8.8cm}
\epsfig{file=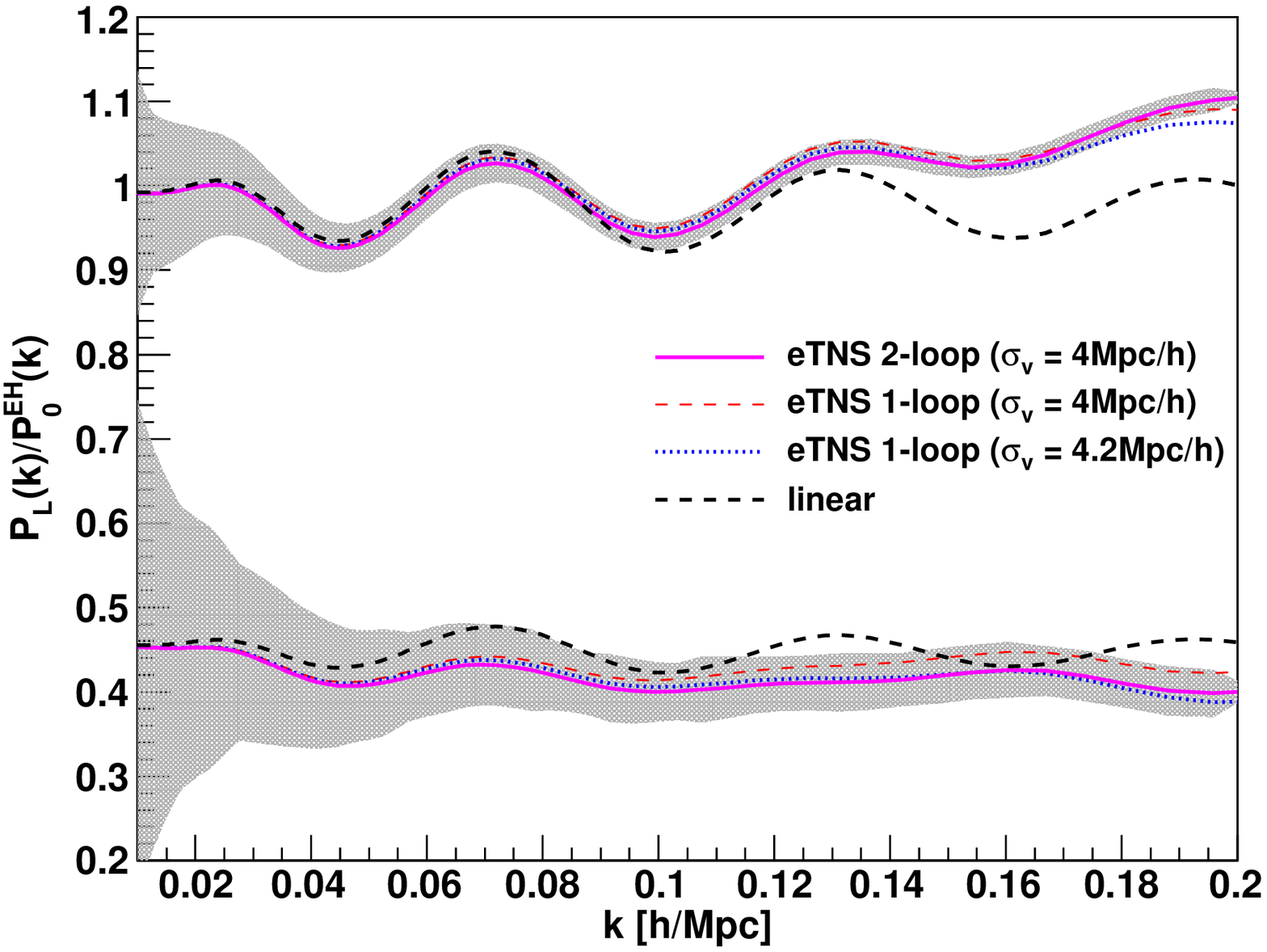,width=8.8cm}
\caption{These plots show the effect of different ingredients in the extended TNS (eTNS) model (see eq.~\ref{eq:taruya},~\ref{eq:40} and~\ref{eq:41}). All multipole power spectra are shown relative to an~\citet{Eisenstein:1997ik} no-BAO monopole power spectrum. The black dashed line is the linear CAMB power spectrum, the magenta line is the eTNS model including all correction terms of section~\ref{sec:taruya} and calculating $P_{\delta\delta}$, $P_{\delta\theta}$ and $P_{\theta\theta}$ using $2$-loop perturbation theory. The grey shaded area along the magenta line shows the $1\sigma$ power spectrum errors of CMASS-DR11 (NGC). (left) The dash-dot green line shows the eTNS model setting $b_2 = 0$, $b_{\rm 3nl} = 0$ and $b_{s2} = 0$. The dashed red line and the dotted blue line show the effect of the different bias terms. (right) The dashed red line shows the eTNS model using $1$-loop perturbation theory and the dotted blue line shows the same model with a different damping $\sigma_v$. Comparing the dotted blue line and the solid magenta line, one can see that the difference between $1$-loop and $2$-loop PT calculation can be absorbed by $\sigma_v$ to some extent.}
\label{fig:ps_red}
\end{center}
\end{figure*}

Our model for the anisotropic galaxy power spectrum is based on \citet{Taruya:2010mx} (TNS):
\begin{equation}
\begin{split}
P_{\rm g}(k,\mu) &= \exp\left\{-(fk\mu\sigma_v)^2\right\}\big[P_{{\rm g},\delta\delta}(k)\\
&\;\;\;\; + 2f\mu^2P_{{\rm g},\delta\theta}(k) + f^2\mu^4P_{\theta\theta}(k)\\
&\;\;\;\; + b_1^3A(k,\mu,\beta) + b_1^4B(k,\mu,\beta)\big], 
\label{eq:taruya}
\end{split}
\end{equation}
where $\mu$ denotes the cosine of the angle between the wavenumber vector and the line-of-sight direction. The overall exponential factor represents the suppression due to the Finger of God effect, and we treat $\sigma_{v}$ as a free parameter.

The first three terms in the square bracket in eq.~\ref{eq:taruya} describe an extension of the Kaiser factor. The density ($P_{\delta\delta}$), velocity divergence ($P_{\theta\theta}$) and their cross-power spectra ($P_{\delta\theta}$) are identical in linear theory,  while in the quasi non-linear regime, the density power spectrum increases and velocities are randomised on small scales which damps the velocity power spectrum~\citep{Scoccimarro:2004tg}. Besides this fact, we need to relate the density and velocity fields for (dark) matter to those of galaxies. Here we assume no velocity bias, i.e., $\theta_{\rm g}=\theta$, but include every possible galaxy bias term at next-to-leading order using symmetry arguments~\citep{McDonald:2009dh}: 
\begin{align}
\begin{split}
P_{{\rm g},\delta\delta}(k) &= b_1^2P_{\delta\delta}(k) + 2b_2b_1 P_{b2,\delta}(k) 
+ 2b_{s2}b_1P_{bs2,\delta}(k)\\
& + 2b_{\rm 3nl}b_1\sigma_3^2(k)P^{\rm L}_{\rm m}(k) + b^2_2P_{b22}(k)\\
& + 2b_2b_{s2}P_{b2s2}(k) + b^2_{s2}P_{bs22}(k) + N,
\label{eq:40}
\end{split}\\
\begin{split}
P_{{\rm g},\delta\theta}(k) &= b_1P_{\delta\theta}(k) + b_2P_{b2,\theta}(k) 
+ b_{s2}P_{bs2,\theta}(k)\\
& + b_{\rm 3nl}\sigma_3^2(k)P^{\rm lin}_{\rm m}(k),
\label{eq:41}
\end{split}
\end{align}
where $P^{\rm lin}_{\rm m}$ is the linear matter power spectrum. Here we introduce five galaxy bias parameters: the renormalised linear bias, $b_{1}$, $2$nd-order local bias, $b_2$, $2$nd-order non-local bias, $b_{s2}$, $3$rd-order non-local bias, $b_{\rm 3nl}$, and the constant stochasticity term, $N$. From now we will call the model in eq.~\ref{eq:taruya},~\ref{eq:40} and~\ref{eq:41} extended TNS (eTNS) model. We evaluate the non-linear matter power spectra, $P_{\delta\delta}$, $P_{\delta\theta}$, $P_{\theta\theta}$ with the RegPT scheme at $2$-loop order~\citep{Taruya:2012ut}. The other bias terms are given by 
\begin{align}
\begin{split}
P_{b2,\delta}(k) &= \int \frac{d^3q}{(2\pi)^3}P^{\rm lin}_{\rm m}(q)P^{\rm lin}_{\rm m}(|k-q|)\\
&\;\;\;\;\;\times F^{(2)}_{\rm S}(q,k-q),
\end{split}\\
\begin{split}
P_{b2,\theta}(k) &= \int \frac{d^3q}{(2\pi)^3}P^{\rm lin}_{\rm m}(q)P^{\rm lin}_{\rm m}(|k-q|)\\
&\;\;\;\;\;\times G^{(2)}_{\rm S}(q,k-q),
\end{split}\\
\begin{split}
P_{bs2,\delta}(k) &= \int \frac{d^3q}{(2\pi)^3}P^{\rm lin}_{\rm m}(q)P^{\rm lin}_{\rm m}(|k-q|)\\
&\;\;\;\;\;\times F^{(2)}_{\rm S}(q,k-q)S^{(2)}(q,k-q),
\end{split}\\
\begin{split}
P_{bs2,\theta}(k) &= \int \frac{d^3q}{(2\pi)^3}P^{\rm lin}_{\rm m}(q)P^{\rm lin}_{\rm m}(|k-q|)\\
&\;\;\;\;\;\times G^{(2)}_{\rm S}(q,k-q)S^{(2)}(q,k-q),
\end{split}\\
\begin{split}
P_{b22}(k) &= \frac{1}{2}\int \frac{d^3q}{(2\pi)^3}P^{\rm lin}_{\rm m}(q)\Big[P^{\rm lin}_{\rm m}(|k-q|)\\
&\;\;\;\;\; - P^{\rm lin}_{\rm m}(q)\Big],
\end{split}\\
\begin{split}
P_{b2s2}(k) &= -\frac{1}{2}\int \frac{d^3q}{(2\pi)^3}P^{\rm lin}_{\rm m}(q)\Big[\frac{2}{3}P^{\rm lin}_{\rm m}(q)\\
&\;\;\;\;\; - P^{\rm lin}_{\rm m}(|k-q|)S^{(2)}(q,k-q)\Big],
\end{split}\\
\begin{split}
P_{bs22}(k) &= -\frac{1}{2}\int \frac{d^3q}{(2\pi)^3}P^{\rm lin}_{\rm m}(q)\Big[\frac{4}{9}P^{\rm lin}_{\rm m}(q)\\
&\;\;\;\;\; - P^{\rm lin}_{\rm m}(|k-q|)S^{(2)}(q,k-q)^2\Big],
\end{split}
\label{eq:nlterms}
\end{align}
where the symmetrised $2$nd-order PT kernels, $F^{(2)}_{S}$, $G^{(2)}_{S}$, 
and $S^{(2)}$ are given by
\begin{align}
F^{(2)}_{\rm S}(q_1,q_2) &= \frac{5}{7} + \frac{q_1\cdot q_2}{2q_1q_2}
\left(\frac{q_1}{q_2} + \frac{q_2}{q_1}\right) + \frac{2}{7}\left(\frac{q_1\cdot q_2}{q_1q_2}\right)^2,\\
G^{(2)}_{\rm S}(q_1,q_2) &= \frac{3}{7} + \frac{q_1\cdot q_2}{2q_1q_2}
\left(\frac{q_1}{q_2} + \frac{q_1}{q_2}\right) + \frac{4}{7}\left(\frac{q_1\cdot q_2}{q_1q_2}\right)^2,\\
S^{(2)}(q_1,q_2) & = \left(\frac{q_1\cdot q_2}{q_1q_2}\right)^2 - \frac{1}{3}.
\end{align}
If we additionally define 
\begin{equation}
D^{(2)}(q_1,q_2) = \frac{2}{7}\left[S^{(2)}(q_1,q_2) - \frac{2}{3}\right],
\end{equation}
we can write down $\sigma^2_3(k)$ of eq.~\ref{eq:41} as
\begin{equation}
                \sigma_{3}^{2}(k) = \frac{105}{16}\int \frac{d^3 q}{(2\pi)^{3}}P^{\rm lin}_{\rm m}(q)
                \left[  D^{(2)}(-q,k)S^{(2)}(q, k-q) + \frac{8}{63}  \right].
\end{equation}
As shown in~\citet{Chan:2012jj} non-linear gravitational evolution naturally induces such non-local bias terms even 
starting from purely local bias at an initial time. \citet{Baldauf:2012hs} shows that the $2$nd-order bias is important to explain the large-scale bispectrum in simulations, while the $3$rd-order non-local bias terms play a more important role in the power spectrum~\citep{Saito2013}. 
In the case of the local Lagrangian bias picture in which the initial non-local bias 
is neglected, we can predict the amplitude of the non-local bias as~\citep{Chan:2012jj, Baldauf:2012hs, Saito2013}
\begin{align}
b_{s2} &= -\frac{4}{7}(b_1 - 1),\\
b_{\rm 3nl} &= \frac{32}{315}(b_1 - 1), 
\end{align}
which are in good agreement with the values measured in simulations. 
In this work, we adopt these relations for simplicity, 
while we float $b_{1}$, $b_{2}$ and $N$ as free\footnote{We actually vary $b_1\sigma_8$, $b_2\sigma_8$ and $N$, see section~\ref{sec:para}.}. 
The impact of the $2$nd-order bias terms on the power spectrum 
is somewhat small. Figure~\ref{fig:ps_red} (left) shows the power spectrum multipoles when all higher order bias terms are set to zero (dash-dot green line). The solid magenta line uses $b_2 = 0.5$ and $b_1 = 2.0$. We can see that the higher order bias terms mainly affect the monopole and while the effect is small, it is significant when compared to the measurement errors (grey shaded area). 

We should also mention that the stochastic term, $N$, can in general depend 
on scale~\citep{Dekel:1998eq,Baldauf:2012hs}, while we treat it as a constant and free parameter.
The final ingredients in our model of eq.~\ref{eq:taruya} are the correction 
terms, $A$ and $B$, which originate from the higher-order correlation between 
Kaiser terms and velocity fields in mapping to redshift space~\citep{Taruya:2010mx}. 
We refer the reader to~\citet{Taruya:2010mx} for the definitions of the $A$ and $B$ terms. 
Note that these terms are in fact proportional to $b_{1}^{2}$ as physically expected 
if one takes account of $\beta=f/b_{1}$. Also notice that we drop the $2$nd-order 
bias terms in the $A$ and $B$ correction terms. 

\subsection{The Alcock-Paczynski effect}
\label{sec:alcock}

If our fiducial cosmological parameters that we use to convert galaxy redshifts into distances deviate from the true cosmology, we artificially introduce an anisotropy in our clustering measurement, which is known as Alcock-Paczynski distortion~\citep{Alcock:1979mp}. This effect can be used to measure cosmological parameters~\citep{Matsubara:1996nf,Ballinger:1996cd}. To account for the Alcock-Paczynski effect and its different scaling along and perpendicular to the line-of-sight direction, we can introduce the scaling factors 
\begin{align}
\alpha_{\parallel} &= \frac{H^{\rm fid}(z)r^{\rm fid}_s(z_d)}{H(z)r_s(z_d)},\\
\alpha_{\perp} &= \frac{D_A(z)r^{\rm fid}_s(z_d)}{D^{\rm fid}_A(z)r_s(z_d)},
\end{align}
where $H^{\rm fid}(z)$ and $D^{\rm fid}_A(z)$ are the fiducial values for the Hubble constant and angular diameter distance at $z = 0.57$ and $r^{\rm fid}_s(z_d)$ is the fiducial sound horizon assumed in the power spectrum template. The true wave-numbers $k_{\parallel}'$ and $k_{\perp}'$ are then related to the observed wave-numbers by $k_{\parallel}' = k_{\parallel}/\alpha_{\parallel}$ and $k_{\perp}' = k_{\perp}/\alpha_{\perp}$. Transferring this into scalings for the absolute wavenumber $k = \sqrt{k^2_{\parallel} + k^2_{\perp}}$ and the cosine of the angle to the line-of-sight $\mu$ we can relate the true and observed values by~\citep{Ballinger:1996cd}
\begin{align}
k' &= \frac{k}{\alpha_{\perp}}\left[1 + \mu^2\left(\frac{1}{F^2} - 1\right)\right]^{1/2},
\label{eq:scaling1}\\
\mu' &= \frac{\mu}{F}\left[1 + \mu^2\left(\frac{1}{F^2} - 1\right)\right]^{-1/2}
\label{eq:scaling2}
\end{align}
with $F = \alpha_{\parallel}/\alpha_{\perp}$. The multipole power spectrum including the Alcock-Paczynski effect can then be written as
\begin{align}
P_{\rm \ell}(k) &= \frac{(2\ell + 1)}{2\alpha^2_{\perp}\alpha_{\parallel}}\int^1_{-1}d\mu\; P_{\rm g}\left(k', \mu'\right)\mathcal{L}_{\ell}(\mu),
\label{eq:multi}
\end{align}
where we use the extended TNS model for $P_{\rm g}(k',\mu')$. The AP effect constrains the parameter combination $F_{\rm AP}(z) = (1+z)D_A(z)H(z)/c$, while the BAO feature constrains the combination $D_V(z)/r_s(z_d) \propto \left[D^2_A(z)/H(z)\right]^{1/3}$. Together these two signals allow us to break the degeneracy between $D_A(z)$ and $H(z)$. We will include the scaling parameters $\alpha_{\parallel}$ and $\alpha_{\perp}$ in our model parametrisation, which will be discussed in the next section.

\subsection{Model parameterisation}
\label{sec:para}

We parametrize our model using the scaling parameters $\alpha_{\parallel}$ and $\alpha_{\perp}$ introduced in the last section. Using these parameters we can derive 
\begin{equation}
\frac{D_{V}(z_{\rm eff})}{r_s(z_d)} = \frac{\left(\alpha^2_{\perp}\alpha_{\parallel}\left[(1 + z_{\rm eff})D_A^{\rm fid}(z_{\rm eff})\right]^2\frac{cz_{\rm eff}}{H^{\rm fid}(z_{\rm eff})}\right)^{\frac{1}{3}}}{r_s^{\rm fid}(z_d)}
\label{eq:DVrs}
\end{equation} 
and 
\begin{equation}
\begin{split}
F_{\rm AP}(z_{\rm eff}) &= \frac{\alpha_{\parallel}}{\alpha_{\perp}}(1+z_{\rm eff})D^{\rm fid}_A(z_{\rm eff})H^{\rm fid}(z_{\rm eff})/c\\
&= (1+z_{\rm eff})D_A(z_{\rm eff})H(z_{\rm eff})/c.
\end{split}
\end{equation} 
The parameter combination $D_{V}(z)/r_s(z_d)$ represents the actual quantity which is constrained by the BAO signal, while $F_{\rm AP}(z)$ is the parameter combination which the AP effect is sensitive to~\citep{Padmanabhan:2009yr}. Once such geometric parameters are constrained, the relative amplitude of the monopole and quadrupole constrains the growth rate $f(z)\sigma_8(z)$. Beside the three main parameters above ($\alpha_{\parallel}$, $\alpha_{\perp}$ and $f\sigma_8$) we also include four nuisance parameters in our power spectrum model: The power spectrum amplitudes, $b_1\sigma_8(z_{\rm eff})$ and $b_2\sigma_8(z_{\rm eff})$, the velocity dispersion $\sigma_v$ and the shot noise component $N$.

Any use of the parameter constraints from this analysis should take into account the underlying assumption of our analysis. We assume that the measured Planck cosmology at very high redshift can be used to build the $``$initial condition$"$ for the linear clustering amplitude on which our power spectrum model, including all non-linear corrections, is based. 

\subsection{Effective wave-number}
\label{sec:keff}

In our introduction we advertised RSD as one probe which is able to test GR on very large scales. So what is the scale of our measurement? The information covariance, $C^{-1}_{ij, \rm info}$ can be calculated as
\begin{equation}
\begin{split}
C^{-1}_{ij, \rm info} = \sum_{\ell \ell'} \frac{d\ln P_{\ell}(k_i)}{df\sigma_8}C^{-1}_{ij,\rm Hartlap}\frac{d\ln P_{\ell'}(k_j)}{df\sigma_8},
\end{split}
\end{equation}
where $P_{\ell}$ is the extended TNS model power spectrum we introduced in section~\ref{sec:taruya} and $C^{-1}_{ij,\rm Hartlap}$ is the covariance matrix we derived in section~\ref{sec:cov}. We now can calculate the effective wave-number as 
\begin{equation}
k_{\rm eff} = \sqrt{\frac{1}{A}\sum_{i,j} k_i C^{-1}_{ij, \rm info}k_j}.
\end{equation}
Here the normalisation $A$ is given by $A=\sum_{ij} C^{-1}_{ij, \rm info}$. Using $k_{\rm max} = 0.20h/$Mpc we get $k_{\rm eff} = 0.178h/$Mpc, which can be related to a real-space scale by $s = 1.15\pi/k_{\rm eff} \approx 20.3\,$Mpc$/h$~\citep{Reid:2011ar}.
The effective wave-number of our measurement using $k_{\rm max} = 0.15h/$Mpc is $k_{\rm eff} = 0.132h/$Mpc.

\section{Testing for systematic uncertainties and determining the maximum wavenumber, \lowercase{$k_{\rm max}$}}
\label{sec:sys}

The question of the maximum wavenumber, $k_{\rm max}$ up to which we can trust our power spectrum model, is directly linked to the question of possible systematic uncertainties. We would like to make use of as much data as possible, but there are significant power spectrum modelling issues given the small error bars of our measurement. 

\begin{table*}
\begin{center}
\caption{Summary of systematic uncertainties of $\alpha_{\parallel}$, $\alpha_{\perp}$ and $f(z_{\rm eff})\sigma_8(z_{\rm eff})$. The shift parameters $\alpha_{\parallel}$ and $\alpha_{\perp}$ are closely related to $H(z_{\rm eff})$ and $D_A(z_{\rm eff})$, respectively. The different lines in this table are: Comparison to N-body simulations (see section~\ref{sec:Nbody}),  comparison between 1-loop and 2-loop perturbation theory (PT) (see section~\ref{sec:PT}) and varying the underlying HOD (see section~\ref{sec:HOD}). In the case of the HOD test we include the result for $M_{\rm sat}-1\sigma = 5\times 10^{13}M_{\odot}/h$, which represents the largest variation compared to the CMASS HOD. We find significant systematic uncertainties only for $f(z_{\rm eff})\sigma_8(z_{\rm eff})$. Based on these uncertainties we chose $k_{\rm max}=0.20h/$Mpc, since this is where the error on $f(z_{\rm eff})\sigma_8(z_{\rm eff})$ is minimised (using the quadrature sum of the statistical and the largest systematic error). For comparison in the last row we included the expected statistical uncertainty for each parameter with different $k_{\rm max}$, which we obtained by fitting the mean of the $999$ mock catalogues using the data covariance matrix.
}
	\begin{tabular}{llllllllll}
     		\hline
		 source & \multicolumn{2}{c}{$\alpha_{\parallel}$ $\left[H(z_{\rm eff})\right]$}\hspace{0.6cm} & \multicolumn{2}{c}{$\alpha_{\perp}$ $\left[D_A(z_{\rm eff})\right]$}\hspace{0.6cm} & \multicolumn{2}{c}{$f(z_{\rm eff})\sigma_8(z_{\rm eff})$}\\
		 $k_{\rm max}$ [$h$/Mpc] & $\;\;\;0.15$ & $\;\;\;0.20$\hspace{1cm} & $\;\;\;0.15$ & $0.20$\hspace{1cm} & $\;\;\;0.15$ & $\;\;\;0.20$\\
		\hline
		model test & $\;\;\;0.11\pm0.13\%$ & $\;\;\;0.00\pm0.10\%$ & $\;\;\;0.352\pm0.061\%$ & $0.052\pm0.049\%$ & $-0.66\pm0.29\%$ &$-3.08\pm0.26\%$\\
		PT test & $\;\;\;0.04\pm0.14\%$ & $-0.32\pm0.12\%$ & $-0.075\pm0.074\%$ & $0.168\pm0.060\%$ & $-0.65\pm0.33\%$ & $-1.01\pm0.30\%$\\
		HOD test & $-1.07\pm0.89\%$ & $\;\;\;0.21\pm0.67\%$ & $-0.09\pm0.42\%$ & $0.50\pm0.38\%$ & $\;\;\;2.6\pm2.4\%$ & $\;\;\;1.5\pm2.1\%$\\
		\hline
		statistical error & $\;\;\;4.0\%$ & $\;\;\;3.1\%$ & $\;\;\;1.9\%$ & $1.6\%$ & $\;\;\;9.1\%$ & $\;\;\;8.3\%$\\
	  \end{tabular}
	  \label{tab:sys}
\end{center}
\end{table*}

\subsection{Test with N-body simulation}
\label{sec:Nbody}

\begin{figure}
\begin{center}
\epsfig{file=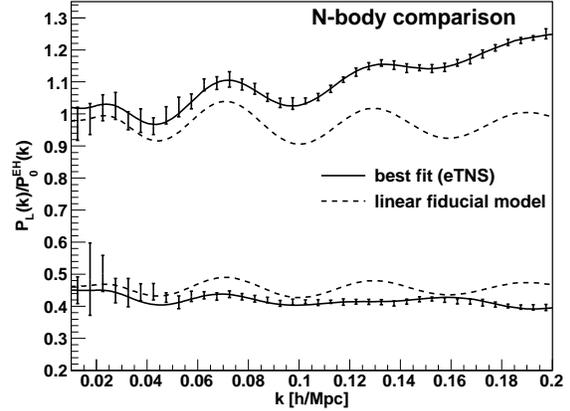,width=8.cm}
\caption{The power spectrum monopole (top) and quadrupole (bottom) measured in a set of N-body simulations (black data points) plotted relative to the fiducial~\citet{Eisenstein:1997ik} no-BAO monopole power spectrum. The solid black line represents the best fitting model. The fitting range is $k = 0.01$ - $0.20h/$Mpc. The error at each data point is the variation between the $20$ simulation boxes covering a total volume of $67.5\,[$Gpc$/h]^{3}$.}
\label{fig:runA}
\end{center}
\end{figure}

\begin{figure}
\begin{center}
\epsfig{file=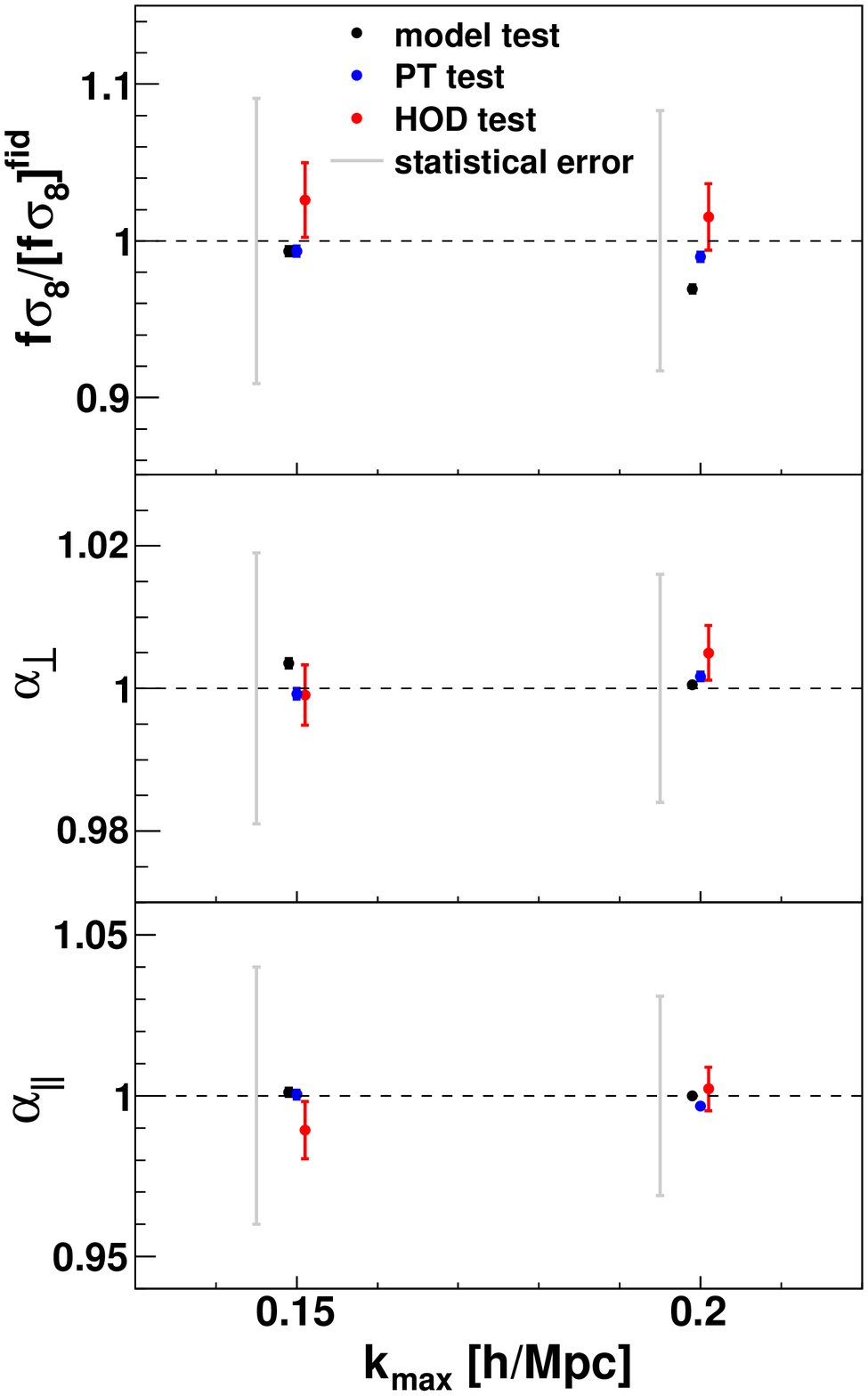,width=8.cm}
\caption{The best fitting values for $\alpha_{\parallel}$, $\alpha_{\perp}$ and $f\sigma_8/[f\sigma_8]^{\rm fid}$ for the different systematics tests performed in this analysis using the fitting range $k=0.01$ - $0.15\,h/$Mpc and $k=0.01$ - $0.20\,h/$Mpc. The data points have been shifted away slightly from $k_{\rm max} = 0.15\,h/$Mpc  and $k_{\rm max} = 0.20\,h/$Mpc  for clarity. The black data points are obtained from the comparison with N-body simulations (see section~\ref{sec:Nbody}), the blue data points show the result when using $1$-loop perturbation theory (see section~\ref{sec:PT}) and the red data points show the result when varying the underlying HOD (see section~\ref{sec:HOD}). For this plot we restrict ourselves to the case $M_{\rm sat}-1\sigma = 5\times 10^{13}M_{\odot}/h$, which has the largest deviation from the CMASS HOD. The PT test used the mean of the $999$ QPM mocks and has error-bars a factor of $\sim\sqrt{999}$ smaller than the plotted statistical error (grey line). The HOD tests have been performed on the mean of $20$ mock catalogues and hence have errors $\sim\sqrt{20}$ smaller than the statistical errors.}
\label{fig:sys}
\end{center}
\end{figure}

To test whether our power spectrum model can extract the correct cosmological parameters from a power spectrum measurement, we use a set of $20$ N-body simulations described in~\citet{White:2010ed} that were generated using a TreePM code. The simulations cover a total volume of $67.5\,[$Gpc$/h]^{3}$. Note, that we use these N-body simulations only for this sub-section and use the QPM simulations for the rest of this paper. We calculate the monopole and quadrupole power spectrum for these simulations and perform a fit using our power spectrum model. When using the fitting range $k = 0.01$ - $0.20h/$Mpc, the best fitting value of $f(z_{\rm eff})\sigma_8(z_{\rm eff})$ deviates from the fiducial value of the simulation by $3.1\%$, while we cannot find any significant deviation for $\alpha_{\parallel}$ and $\alpha_{\perp}$. A comparison between the model and the measured power spectrum in these N-body simulations can be seen in~Figure~\ref{fig:runA}. Using $k_{\rm max} = 0.15h/$Mpc we find deviations of $-0.1\%$, $-0.1\%$ and $-0.7\%$ for $\alpha_{\parallel}$, $\alpha_{\perp}$ and $f(z_{\rm eff})\sigma_8(z_{\rm eff})$, respectively. We include these values in Table~\ref{tab:sys} and Figure~\ref{fig:sys}.

Several authors have recently performed similar studies to what we have done here~\citep{Nishimichi:2011jm,delaTorre:2012dg,Ishikawa:2013aea,Oka:2013cba}. They studied the systematic uncertainty against halos (or sub-halos) in N-body simulations using the TNS model. Although some of these studies ignore the Alcock-Paczynski effect, which is degenerate with $f\sigma_8$ and use a phenomenological treatment of the galaxy/halo bias, they reach very similar conclusions.

\subsection{Uncertainties from perturbation theory}
\label{sec:PT}

Because we want to make use of the power spectrum beyond $k=0.10h/$Mpc we cannot rely on standard perturbation theory (SPT) which seems to break down at low redshift for $k > 0.10h/$Mpc, where the $2$-loop term turns out to be larger than the $1$-loop term~\citep{Crocce:2005xz,Taruya:2009ir,Carlson:2009it}. We therefore use re-normalised perturbation theory to calculate $P_{\delta\delta}$, $P_{\theta\theta}$ and $P_{\delta\theta}$~\citep{Taruya:2007xy,Taruya:2009ir,Taruya:2013my} and include corrections up to 2-loop order. We make use of the publicly available RegPT code~\citep{Taruya:2012ut}. 

The authors of this code suggested a phenomenological rule for the maximum wavenumber up to which the model is numerically stable, which they call $k_{\rm crit}$ given by
\begin{equation}
\frac{k^2_{\rm crit}}{6\pi^2} \int^{k_{\rm crit}}_0 dk\,P^{\rm lin}_{\rm m}(k) = 0.7.
\end{equation}
This rule is roughly based on percent level accuracy. At redshift $z_{\rm eff}=0.57$ with a Planck cosmological model we get $k_{\rm crit} = 0.28h/$Mpc. 

To get a rough upper limit on the effect of ignoring terms higher than second order, we estimate the effect of ignoring the second order term since we expect that the effect of the former is smaller than the latter. We therefore calculate the power spectra at $1$-loop order and measure the amplitude differences of the power spectra at different wave-numbers. We find $\Delta P_{\delta\delta}$ of ($0.5$, $0.2$, $3.2$)$\%$ at $k = (0.10$, $0.15$, $0.20)h/$Mpc. The corresponding values for $\Delta P_{\delta\theta}$ are ($3.4$, $5.2 $, $4.8 $)$\%$ and for $\Delta P_{\theta\theta}$ we find ($6.3$, $10.3$, $12.2$)$\%$. While these differences seem very significant, we are actually only interested in the bias these uncertainties introduce in our cosmological parameters. We use the $1$-loop power spectra calculated from RegPT instead of the $2$-loop power spectra and build our model following section~\ref{sec:taruya}. We then fit this model to the mean of the $999$ QPM mock power spectra. The shifts in the cosmological parameters are shown in Table~\ref{tab:sys} and Figure~\ref{fig:ps_red} (right). We see a shift of $1.0\%$ in $f(z_{\rm eff})\sigma_8(z_{\rm eff})$ when using the fitting range $k=0.01$ - $0.20h/$Mpc, while the shifts in $\alpha_{\parallel}$ and $\alpha_{\perp}$ are much smaller.

Figure~\ref{fig:ps_red} (right) shows the extended TNS model using $2$-loop and $1$-loop perturbation theory. The $1$-loop case has a larger amplitude in the quadrupole, while the monopole is much less affected. The differences in the quadrupole are caused mainly by the big changes in $P_{\theta\theta}$ going from the $2$-loop to $1$-loop calculation. Most of the difference can be absorbed by nuisance parameters like $\sigma_v$. This is also included in Figure~\ref{fig:ps_red} (right) as the dotted blue line, where we use the $1$-loop calculations, but changed $\sigma_v$ from $4.0\,$Mpc$/h$ to $4.2\,$Mpc$/h$ bringing the model in good agreement with the $2$-loop calculation (solid magenta line). Therefore, $\sigma_v$ can absorb the difference between $1$-loop and $2$-loop calculation to a great extent, which is the reason why the large difference in the power spectrum amplitude does not transfer into large differences in the actual parameter constraints.

\subsection{The impact of different HODs}
\label{sec:HOD}

\begin{figure}
\begin{center}
\epsfig{file=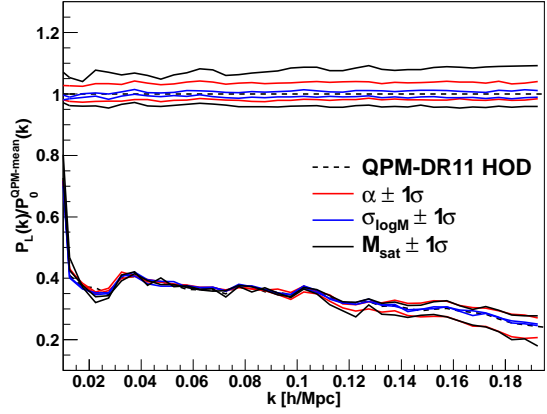,width=8.cm}
\caption{We plot the mean of the power spectrum monopole and quadrupole measured from $20$ CMASS mock catalogues with varying HOD relative to the power spectrum monopole using the fiducial HOD parametrisation of section~\ref{sec:cov}. The red lines show the power spectrum multipoles where we varied $\alpha$ (see section~\ref{sec:HOD} for details) while the blue  and black lines show variations in $\sigma_{\log M}$ and $M_{\rm sat}$, respectively.}
\label{fig:hod}
\end{center}
\end{figure}

Here we want to test how sensitive our power spectrum model is to the underlying HOD. Ideally one would want to constrain the HOD parameters together with the cosmological parameters, by using all information in the galaxy clustering, down to very small scales. However, current model uncertainties do not allow such studies. 

The CMASS-DR11 mock catalogues which we introduced in section~\ref{sec:cov} are populated with a specific HOD model. The question is, whether our ability to extract the correct cosmological parameters does depend on this HOD?

To test this, we create CMASS-DR11 catalogues, based on the same original simulation box as the mock catalogues used in section~\ref{sec:cov}, but populated with different HODs. We vary the three HOD parameters ($\sigma_{\log M}$, $\alpha$ and $M_{sat}$) by the $1\sigma$ uncertainties reported in~\citet{White:2010ed}. The explicit variations are $\sigma_{\sigma_{\log M}} = 0.04$, $\sigma_{\alpha} = 0.2$ and $\sigma_{M_{\rm sat}} = 1.3\times 10^{13}M_{\odot}/h$, meaning we generate six different HOD models. We choose $M_{\rm min}$ so that the number density is kept fixed. Because~\citet{White:2010ed} used a dataset about $10$ times smaller than CMASS-DR11, the real uncertainties on the HOD parameters should be significantly smaller. For each new set of HOD parameters we create $20$ mock catalogues. We calculate the mean of the $20$ power spectra and fit our model to it. We show the power spectrum monopole and quadrupole for the different HODs in Figure~\ref{fig:hod}. As expected, different HODs mainly affect the amplitude of the monopole, but do not cause significant changes in the shape even at $k = 0.20h/$Mpc. 

All parameter fits result in constraints on $\alpha_{\parallel}$, $\alpha_{\perp}$ and $f(z)\sigma_8(z)$ in good agreement with the original HOD parameterisation (black dashed line in Figure~\ref{fig:hod}). Since we are only fitting the mean of $20$ mock catalogues for each HOD model, we are only sensitive to shifts $\sim 5$ times smaller than our measurement uncertainties\footnote{Since we are using the same cosmic volume as in the original mock catalogues our sensitivity is a little bit better than just a factor of $5$.}. However, we consider this level of accuracy to be sufficient for the purpose of this analysis. We include the result for $M_{\rm sat} -1\sigma = 5\times 10^{13}M_{\odot}/h$ in Table~\ref{tab:sys} and Figure~\ref{fig:sys}, since this is where we find the largest deviation from the CMASS HOD.

RSDs are induced by the peculiar velocities which are assumed to follow the underlying dark matter field. Violations of this assumption are usually called velocity bias. In our analysis we do not consider the issues related to the velocity bias, which could have a non-negligible impact. We here simply assume that the galaxies follow the velocity field of dark matter halos. There are various scenarios that could affect the galaxy peculiar velocity field, such as the velocity bias related to the peak formation~\citep{Bardeen:1985tr,Desjacques:2009kt}, the offset of the central galaxies~\citep{Hikage:2011ut,Hikage:2012zk,Hikage:2013yja} and the kinematical features of the satellite galaxies~\citep{Masaki:2012gh,Nishimichi:2013aba}. 
These issues are beyond the scope of this paper and should be addressed using the galaxy clustering or the galaxy-galaxy lensing signal at somewhat smaller scales where the $1$-halo term is more dominant (for CMASS see~\citealt{Miyatake:2013,Reid:2013}). Nevertheless, we believe that our results should be fairly robust against such effects, since we do not confirm any significant differences when changing the fitting range (see Table~\ref{tab:para} and the discussion in section~\ref{sec:results}). 

\subsection{Uncertainty in the underlying linear matter power spectrum}
\label{sec:PLANCK}

The BOSS dataset, like all galaxy redshift survey datasets, cannot constrain all $\Lambda$CDM parameters just by itself, due to parameter degeneracies. Our analysis therefore makes use of the information coming from the analysis of the CMB, in a sense that we take the cosmological parameters found in Planck and use them as initial conditions. We then test whether such initial conditions lead to the clustering signal measured with our dataset.
In our model we are using a power spectrum with fixed cosmological parameters. The assumption here is that the Planck uncertainty in most of the parameters which define the shape of the power spectrum is much smaller than the uncertainty of our measurement and hence can be neglected. This assumption has been found to be reasonable for the CMASS-DR9 dataset combined with WMAP7~\citep{Reid:2012sw}. We repeat the test of~\citet{Reid:2012sw}, where we only consider the Planck uncertainty in $\omega_c=\Omega_ch^2$, representing the least well constrained parameter important for our analysis. We then calculate the quantity
\begin{equation}
s = \frac{\Delta p}{\Delta \omega_c}\frac{\sigma_{\omega_c}}{\sigma_p},
\end{equation}
where $\Delta p$ stands for the change in our parameter constraint when changing $\omega_c$ by $\Delta \omega_c$ and $\sigma_p$ is the uncertainty in the parameter $p$ at fixed $\omega_c$. The uncertainty in $p$ when marginalised over $\omega_c$ is increased by $\sqrt{1 + s^2}$ assuming Gaussian probability distribution functions. By fitting the mean of the $999$ mock catalogues and using the fitting range $k=0.01$ - $0.20h/$Mpc we find $\sigma_p = (0.031, 0.016, 0.038)$ for $\alpha_{\parallel}$, $\alpha_{\perp}$ and $f(z_{\rm eff})\sigma_8(z_{\rm eff})$, respectively. For $\Delta\omega_c = 0.02$ we find $\Delta \alpha_{\parallel} = 0.015\,$, $\Delta \alpha_{\perp}  = 0.016$ and $\Delta f(z_{\rm eff})\sigma_8(z_{\rm eff}) = 0.008$ leading to $s = 0.07$, $s = 0.14$ and $s = 0.03$, respectively. These results imply that the error in $\alpha_{\perp}$ would increase by only $1.0\%$ if the Planck errors are propagated to our results while the effect on $\alpha_{\parallel}$ and $f(z_{\rm eff})\sigma_8(z_{\rm eff})$ is even smaller. These uncertainties are negligible and justify our choice to fix these parameters in our analysis. As a further test we changed the power spectrum template from the fiducial cosmology to a different one, varying the cosmological parameters within the WMAP9 uncertainties and find that the best fitting values changed by $< 0.1\%$.

\subsection{Summary of the study of possible systematics}

Table~\ref{tab:sys} and Figure~\ref{fig:sys} summarise the results of our systematics test. Since the systematic errors we found are related, we use only the largest systematic error and combine it with the statistical error (in quadrature). We only find significant systematic bias for $f\sigma_8$ when using the larger fitting range of $k=0.01$ - $0.20\,h/$Mpc, given by $3.1\%$. For all other parameters as well as for the smaller fitting range of $k=0.01$ - $0.15h/$Mpc, we did not find any significant systematic errors.
Since $f\sigma_8$ is the parameter of interest for this analysis, we chose the maximum wave-number according to where the total error of $f\sigma_8$ is minimised. This is the case at $k_{\rm max} = 0.20h/$Mpc. We did not test wave-numbers beyond $k= 0.20h/$Mpc. Note, that the geometric parameters $\alpha_{\parallel}$ and $\alpha_{\perp}$ are more robust against systematic errors and could go to larger wave-numbers when marginalising over $f(z_{\rm eff})\sigma_8(z_{\rm eff})$. Such an analysis can be found in~\citet{Anderson2.0} and if only the geometric information is needed, we recommend useing the constraints quoted in this analysis. Note however, that the extra information contained in the growth rate can lead to substantially improved constraints even for geometric parameters, like the dark energy equation of state $w$~\citep{Rapetti:2012bu,Reid:2012sw,Chuang:2013hya}.

There are other aspects of galaxy clustering which we did not investigate here, which could also introduce systematic biases into our measurement. Naturally our analysis has to be interpreted with respect to the tests made in this section. 

\section{Analysis}
\label{sec:analysis}

This section is devoted to presenting our main results. First we will discuss the setup of our fitting procedure, before discussing the results of the parameter fits.

\subsection{Fitting preparation}
\label{sec:fit}

In recent years different areas of cosmology haven been pushing for blinded analysis techniques to avoid any possible (confirmation) bias. We are using a blinded analysis with the following setup: (1) All tests of the power spectrum model, its parameterisations and possible systematic uncertainties have been done using mock catalogues only, (2) the conditions of the fit, like the maximum wavenumber, $k_{\max}$ and the binning of the power spectrum, have been set before the data is analysed, (3) the data has been fit only once for each fitting range.  

We decided to bin the power spectrum in bins of $\Delta k = 5\times 10^{-3}h/$Mpc~\citep{Percival:2013} and to use the fitting range $k = 0.01$ - $0.20h/$Mpc as the main result of this paper. The choice of our maximum wavenumber, $k_{\rm max} = 0.20h/$Mpc is based on the systematics analysis in the previous section. We will also provide the results using $k_{\rm max} = 0.15h/$Mpc for two reasons: (1) Some people might be concerned about systematic uncertainties not considered in our analysis and (2) the results with the two different fitting ranges can be used to test the scale dependence of $f\sigma_8$, since the two cases have different effective wave-numbers. Such scale-dependence is a property of many modified gravity theories. We emphasise here however, that our assumption of the scale-independent $f\sigma_8$ is to some extent only a consistency check of GR. In order to constrain a modified gravity theory, it is desirable to prepare a new theoretical template in a theory-dependent manner (e.g. for $f(R)$, see~\citealt{Taruya:2013quf}).

We also have to define the effective redshift of the CMASS-DR11 dataset. We calculate the effective redshift by
\begin{equation}
z_{\rm eff} = \frac{\sum_i^{N_{\rm gal}}w_{\text{\tiny{FKP}}}(\vec{x}_i)z_i}{\sum_i^{N_{\rm gal}}w_{\text{\tiny{FKP}}}(\vec{x}_i)},
\end{equation}
where we find $z_{\rm eff} \approx 0.57$. This is the same effective redshift as used in the CMASS-DR9 analysis and the accompanying papers of CMASS-DR11.

Using the covariance matrix derived in section~\ref{sec:cov} we perform a $\chi^2$ minimisation to find the best fitting parameters.
In addition to the scaling of the covariance matrix of eq.~\ref{eq:covhartlap} we have to propagate the error in the covariance matrix to the error on the estimated parameters. We can do this by scaling the variance for each parameter by~\citep{Percival:2013}
\begin{equation}
M = \sqrt{\frac{1 + B(n_b - n_p)}{1 + A + B(n_p + 1)}},
\end{equation}
where $n_p$ is the number of parameters and
\begin{align}
A &= \frac{2}{(N_s - n_b - 1)(N_s - n_b - 4)},\\
B &= \frac{N_s - n_b - 2}{(N_s - n_b - 1)(N_s - n_b - 4)}.
\end{align}
Taking the quantities which apply in our case ($N_s = 999$, $n_b=76$, $n_p =7$) results in a very modest correction of $M\approx 1.03$.

\subsection{Results}
\label{sec:results}

\begin{table*}
\begin{center}
\caption{The maximum likelihood and mean together with the $1\sigma$ error for the main cosmological parameters (first $3$ rows), the $4$ nuisance parameters (middle $4$ rows) as well as several derived parameters (last $7$ rows). While we report the results for two different fitting ranges, we regard the results for the fitting range $k = 0.01$ - $0.20\,h/$Mpc as the main results of this work. Our measurements have an effective redshift of $z_{\rm eff} = 0.57$. The effective wave-number is $k_{\rm eff} = 0.132\,h/$Mpc when using $k_{\rm max} = 0.15\,h/$Mpc and $k_{\rm eff} = 0.178\,h/$Mpc when using $k_{\rm max} = 0.20\,h/$Mpc (see section~\ref{sec:keff}). The best fitting $\chi^2/\rm d.o.f.$ is $90.3/(112-7)$ and $140.5/(152-7)$ when using the smaller and larger fitting range, respectively. We include the systematic error on $f\sigma_8$ for the larger fitting range (note that the systematic error has to be added in quadrature, resulting in $f(z_{\rm eff})\sigma_8(z_{\rm eff}) = 0.419\pm0.044$). The last three rows of the table contain the derived parameter $\beta = f(z_{\rm eff})\sigma_8(z_{\rm eff})/[b_1\sigma_8(z_{\rm eff})]$, as well as the bias parameters $b_1$ and $b_2$. To derive the bias parameters we assumed a fiducial $\sigma^{\rm fid}_8(z=0) = 0.80$. Since the cosmological parameters included in this table are correlated, we recommend useing the multivariate Gaussian likelihood presented in section~\ref{sec:use}.
}
	\begin{tabular}{cllll}
     		\hline
		fitting range & \multicolumn{2}{c}{$0.01$ - $0.15h/$Mpc} & \multicolumn{2}{c}{$0.01$ - $0.20h/$Mpc}\\
		& best fit & mean $\pm 1\sigma$ & best fit & mean $\pm 1\sigma$\\
		\hline
		$\alpha_{\parallel}$ & $1.008$ & $1.005\pm 0.057$ & $1.014$ & $1.018\pm 0.036$\\
		$\alpha_{\perp}$ & $1.026$ & $1.029\pm 0.023$ & $1.029$ & $1.029\pm 0.015$\\
		$f(z_{\rm eff})\sigma_8(z_{\rm eff})$ & $0.420$ & $0.423\pm 0.052$ & $0.422$ & $0.419\pm (\overset{\rm stat}{0.042}+\overset{\rm sys}{0.014})$\\
		\hline
		$b_1\sigma_8(z_{\rm eff})$ & $1.221$ & $1.222\pm 0.044$ & $1.221$ & $1.224\pm 0.031$\\
		$b_2\sigma_8(z_{\rm eff})$ & $1.7$ & $0.7\pm 1.2$ & $-0.21$ & $-0.09\pm 0.62$\\
		$\sigma_{v}$ & $4.6\,$Mpc$/h$ & $4.3\pm 1.3\,$Mpc$/h$ & $4.63\,$Mpc$/h$ & $4.65\pm 0.81\,$Mpc$/h$\\
		$N$ & $1030\,$[Mpc$/h]^3$ & $1080\pm 620\,$[Mpc$/h]^3$ & $1890\,$[Mpc$/h]^3$ & $1690\pm 600\,$[Mpc$/h]^3$\\		
		\hline
		$D_V(z_{\rm eff})/r_s(z_d)$ & $13.83$ & $13.85\pm 0.27$ & $13.88$ & $13.89\pm 0.18$\\
		$F_{\rm AP}(z_{\rm eff})$ & $0.684$ & $0.686\pm 0.046$ & $0.683$ & $0.679\pm 0.031$\\
		$H(z_{\rm eff})r_s(z_d)/r^{\rm fid}_s(z_d)$ & $94.0\,$km/s/Mpc & $94.1\pm 5.4\,$km/s/Mpc & $93.5\,$km/s/Mpc & $93.1\pm 3.3\,$km/s/Mpc\\
		$D_A(z_{\rm eff})r_s^{\rm fid}(z_d)/r_s(z_d)$ & $1385\,$Mpc & $1389\pm 31\,$Mpc & $1389\,$Mpc & $1388\pm 22\,$Mpc\\
		$\beta$ & $0.344$& $0.346\pm 0.043$& $0.346$& $0.342\pm 0.037$\\
		$b_1\times(0.8/\sigma_8)$ & $2.035$& $2.037\pm 0.073$& $2.035$ & $2.040\pm 0.052$\\
		$b_2\times(0.8/\sigma_8)$ & $2.8$& $1.2\pm 2.0$& $-0.4$& $-0.2\pm 1.0$\\
		\hline
	  \end{tabular}
	  \label{tab:para}
\end{center}
\end{table*}

\begin{figure*}
\begin{center}
\epsfig{file=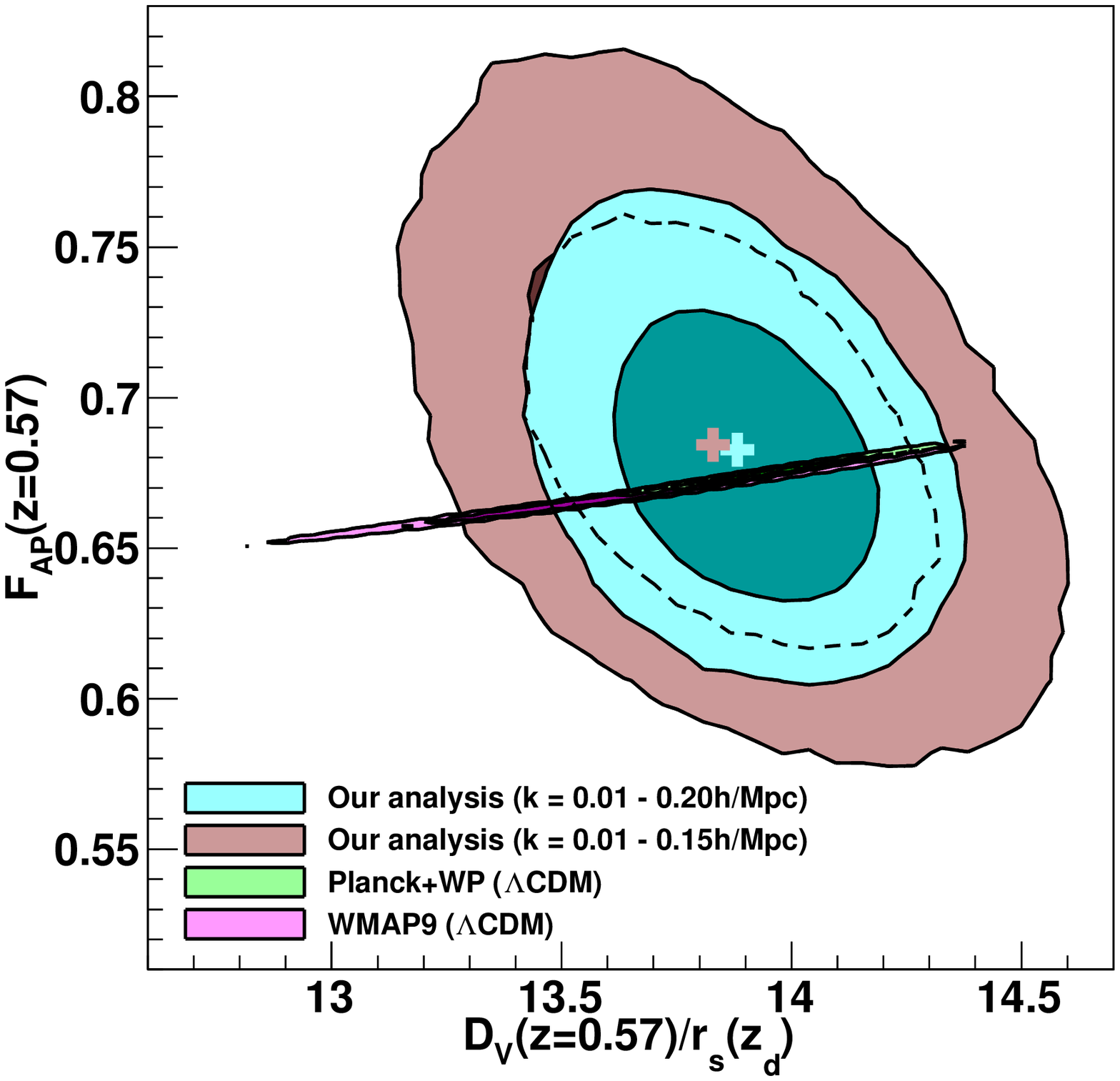,width=8.8cm}
\epsfig{file=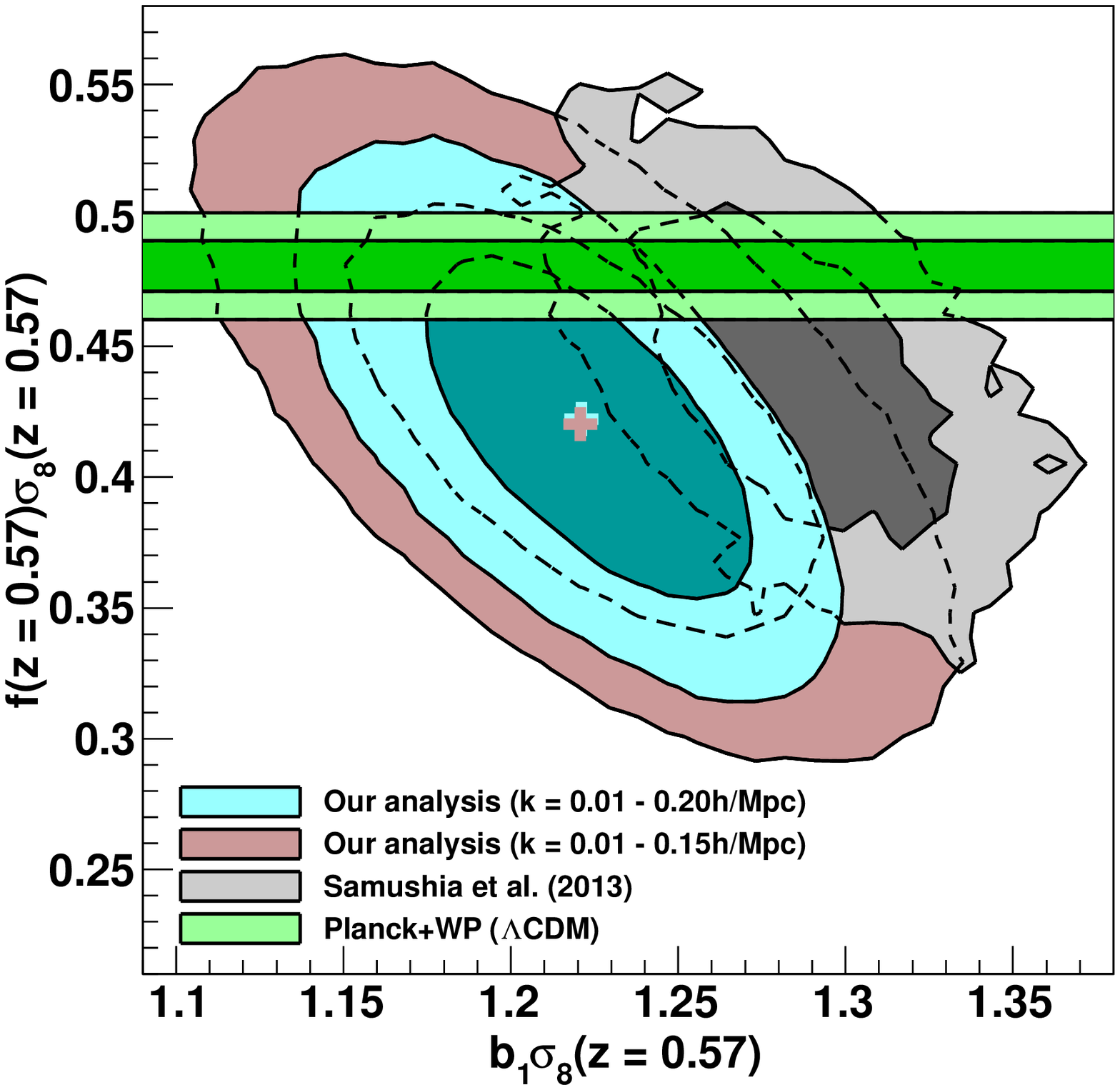,width=8.8cm}\\
\epsfig{file=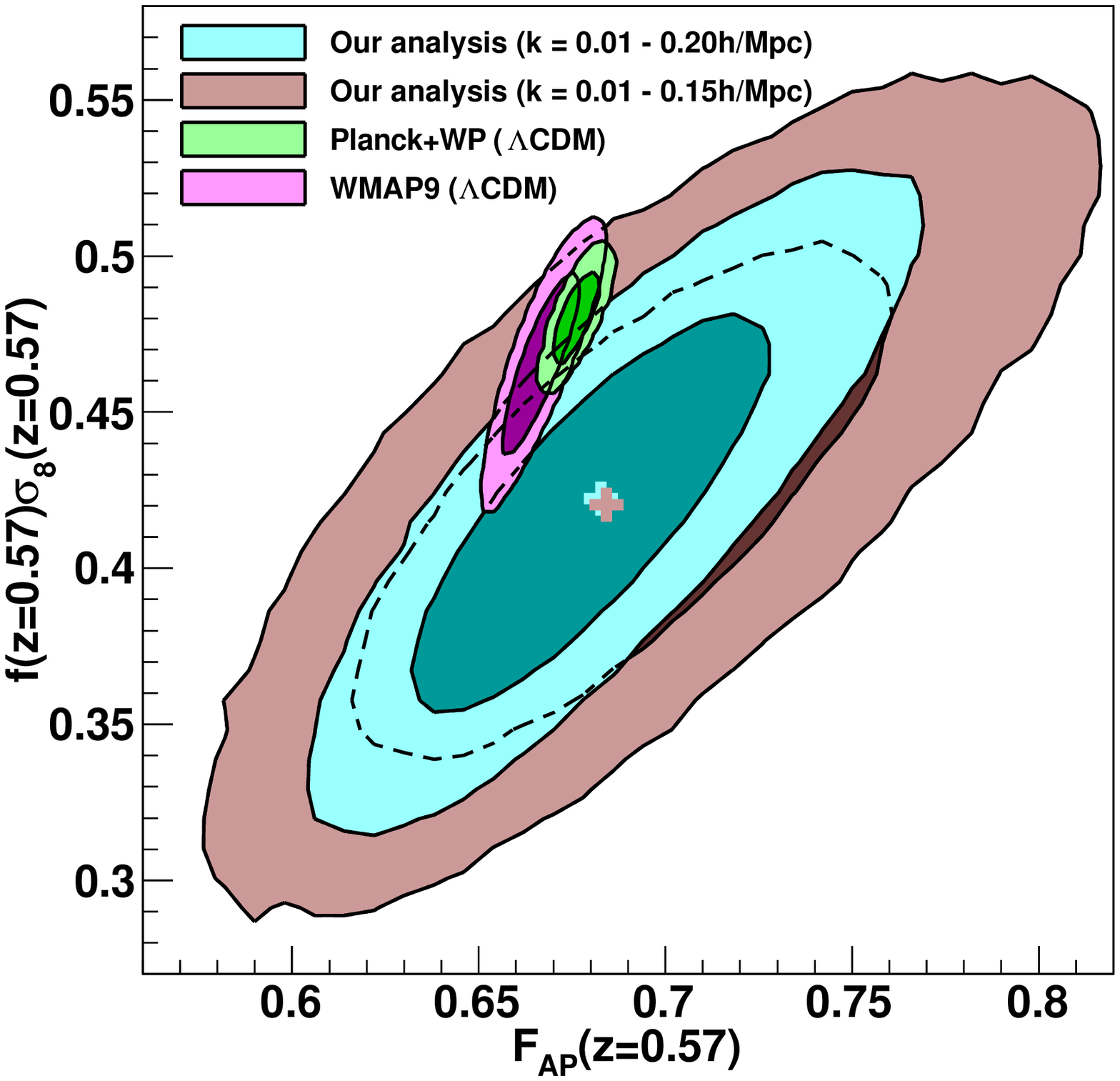,width=8.8cm}
\epsfig{file=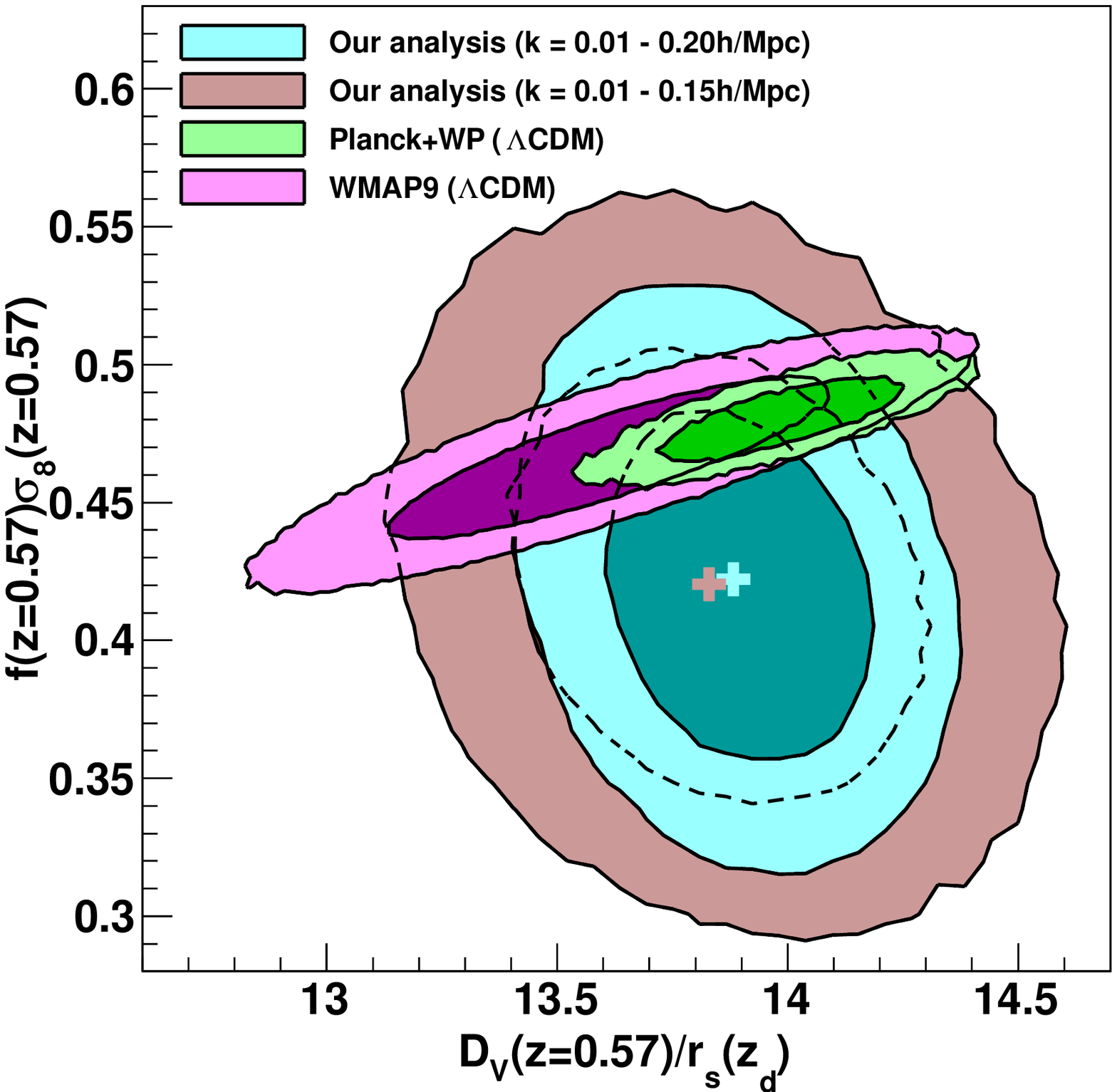,width=8.8cm}
\caption{Two dimensional likelihood distribution of $D_V(z_{\rm eff})/r_s(z_d)$ and $F_{\rm AP}(z_{\rm eff})$ (top left), $b_1\sigma_8(z_{\rm eff})$ and $f(z_{\rm eff})\sigma_8(z_{\rm eff})$ (top right), $F_{\rm AP}(z_{\rm eff})$ and $f(z_{\rm eff})\sigma_8(z_{\rm eff})$ (bottom left), $D_V(z_{\rm eff})/r_s(z_d)$ and $f(z_{\rm eff})\sigma_8(z_{\rm eff})$ (bottom right). We show the $68\%$ and $95\%$ confidence regions. The plot on the top right also includes the result of~\citet{Samushia:2013}. All contours are directly derived from the MCMC chains and do not include the systematic uncertainties. The crosses mark the maximum likelihood values with colours corresponding to the contours. In all plots we also compare to Planck within $\Lambda$CDM (green contours) and WMAP9 within $\Lambda$CDM (magenta contours).}
\label{fig:ps_red4}
\end{center}
\end{figure*}

\begin{figure*}
\begin{center}
\epsfig{file=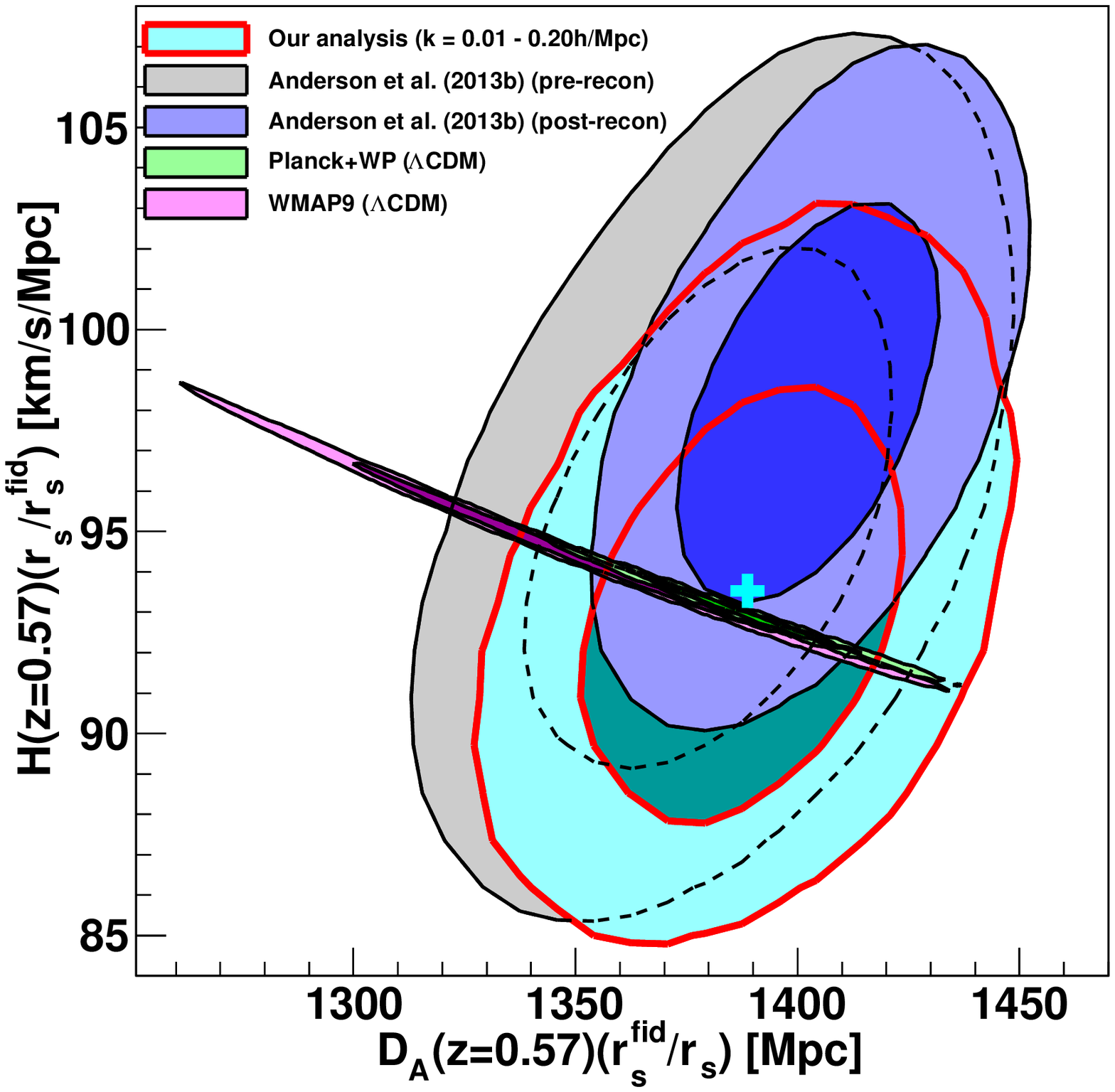,width=8.8cm}
\epsfig{file=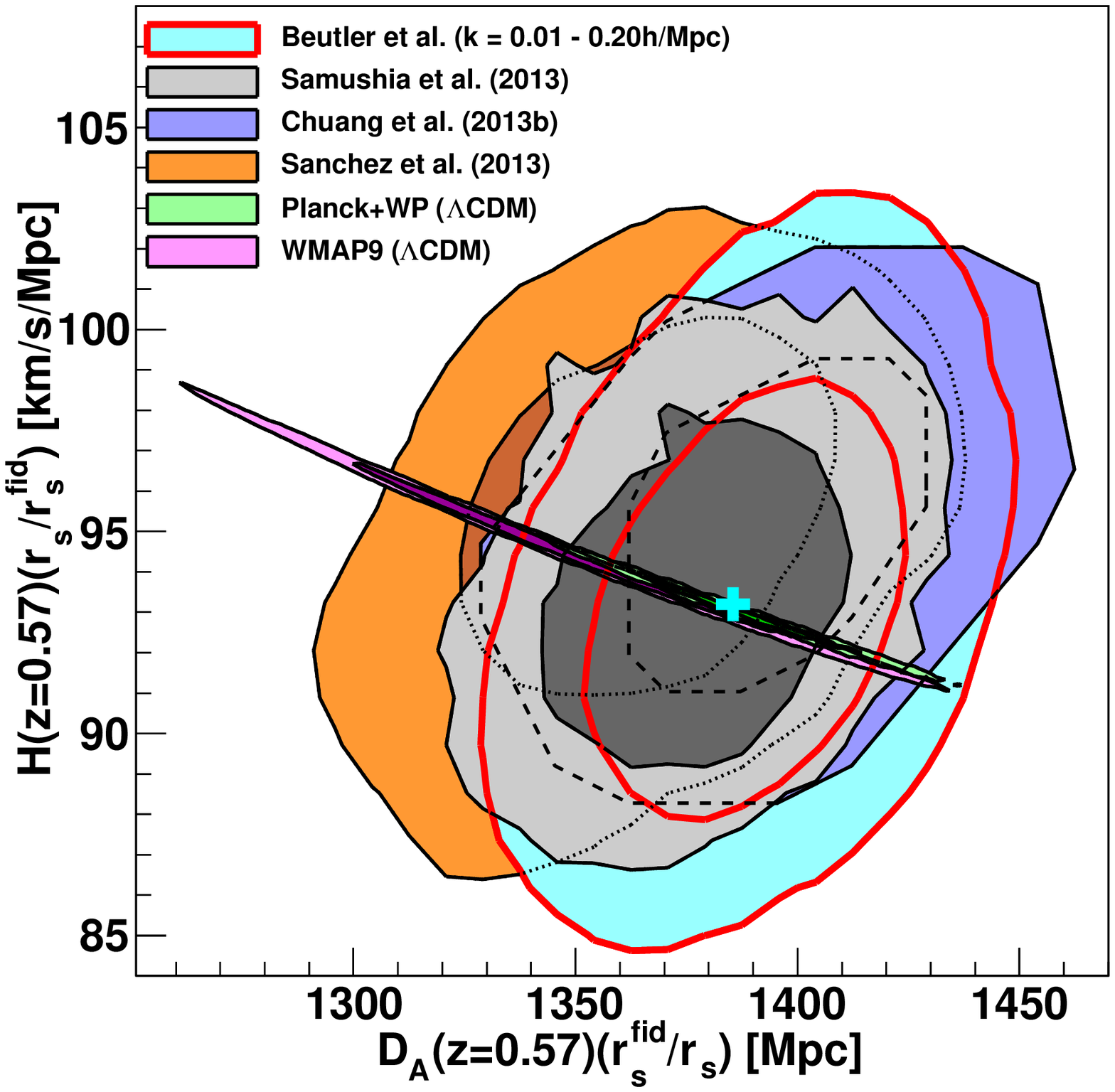,width=8.8cm}
\caption{Comparison of the two dimensional likelihood distribution of $D_A(z_{\rm eff})r_s^{\rm fid}(z_d)/r_s(z_d)$ and $H(z_{\rm eff})r_s(z_d)/r_s^{\rm fid}(z_d)$. We show the $68\%$ and $95\%$ confidence regions. The plot on the left compares our analysis (cyan contours) to the analysis by~\citet{Anderson2.0} before applying density field reconstruction (grey contours) and after applying density field reconstruction (blue contours). The plot on the right compares our analysis (cyan contours) to the analysis by~\citet{Samushia:2013} (grey contours), the analysis by~\citet{Chuang:2013wga} (blue contours) and~\citet{Sanchez:2013} (orange contours). In both plots we also compare to Planck within $\Lambda$CDM (green contours) and WMAP9 within $\Lambda$CDM (magenta contours).}
\label{fig:Hz_DA}
\end{center}
\end{figure*}

We are using a Markov-Chain Monte-Carlo (MCMC) method to find the best fitting values for the measurement of the CMASS-DR11 monopole and quadrupole. 
The seven free parameters of this fit are: $\alpha_{\parallel}$, $\alpha_{\perp}$, the growth rate $f(z_{\rm eff})\sigma_8(z_{\rm eff})$, the power spectrum amplitudes, $b_1\sigma_8(z_{\rm eff})$ and $b_2\sigma_8(z_{\rm eff})$, the velocity dispersion, $\sigma_v$ and the shot noise component $N$.

We summarise our best-fitting results with marginalised errors for each free parameter in Table~\ref{tab:para} and we show 2D contour plots in Figures~\ref{fig:ps_red4} and Figure~\ref{fig:Hz_DA}. Using the fitting range $k = 0.01$ - $0.20h/$Mpc we find $\alpha_{\parallel} = 1.018\pm 0.036\,$, $\alpha_{\perp} = 1.029\pm 0.015$ and $f(z_{\rm eff})\sigma_8(z_{\rm eff}) = 0.419\pm 0.042$. The constraints on $\alpha_{\parallel}$ and $\alpha_{\perp}$ can be expressed as $D_V(z_{\rm eff})/r_s(z_d) = 13.89\pm 0.18$ and $F_{\rm AP} = 0.679\pm 0.031$. Another alternative is to express the geometric constraints as the expansion rate $H(z_{\rm eff})r_s(z_d)/r^{\rm fid}_s(z_d) = 93.1\pm 3.3\,$km/s/Mpc and the angular diameter distance $D_A(z_{\rm eff})r_s^{\rm fid}(z_d)/r_s(z_d) = 1388\pm 22\,$Mpc where $r_s(z_d)$ is the sound horizon at the drag redshift $z_d$. For the nuisance parameters we find $b_1\sigma_8(z_{\rm eff}) = 1.224\pm 0.031$, $b_2\sigma_8(z_{\rm eff}) = -0.09\pm 0.62$, $\sigma_v = 4.65\pm 0.81\,$Mpc$/h$ and $N = 1690\pm 600$. The $\chi^2$ of our best fit is $140.5$ with $152$ bins and $7$ free parameters. The best fitting $\chi^2$ has a contribution of $66.6$ from the NGC of CMASS-DR11 and $73.9$ from the SGC with $76$ bins each. Splitting between the monopole and quadrupole, we find that the monopole contribution to $\chi^2$ is $79.8$, while the quadrupole contribution is $68.7$, again with $76$ bins each\footnote{The sum of the monopole and quadrupole contributions does not add up to the best fitting $\chi^2$ of $140.5$, because of the cross-correlation between the monopole and quadrupole.}. Overall we find a better fit for the NGC than for the SGC and a better fit for the quadrupole than for the monopole. 

Using the fitting range $k = 0.01$ - $0.15h/$Mpc we find $\alpha_{\parallel} = 1.005\pm 0.057$, $\alpha_{\perp} = 1.029\pm 0.023$ and $f(z_{\rm eff})\sigma_8(z_{\rm eff}) = 0.423\pm 0.052$. The constraints on $\alpha_{\parallel}$ and $\alpha_{\perp}$ can again be expressed as $D_V(z_{\rm eff})/r_s(z_d) = 13.85\pm 0.27$ and $F_{\rm AP} = 0.686\pm 0.046$ or alternatively $H(z_{\rm eff})r_s(z_d)/r^{\rm fid}_s(z_d) = 94.1\pm 5.4\,$km/s/Mpc and $D_A(z_{\rm eff})r_s^{\rm fid}(z_d)/r_s(z_d) = 1389\pm 31\,$Mpc. For the nuisance parameters we find $b_1\sigma_8(z_{\rm eff}) = 1.222\pm 0.044$, $b_2\sigma_8(z_{\rm eff}) = 0.7\pm 1.2$, $\sigma_v = 4.3\pm 1.3\,$Mpc$/h$ and $N = 1080\pm 620$. The $\chi^2/\rm d.o.f$ of our best fit is $90.3/105$.

In Figure~\ref{fig:ps_red4} we show the constraints on $D_V/r_s$, $F_{\rm AP}$ and $f\sigma_8$ comparing our results using the fitting range $k = 0.01$ - $0.20h/$Mpc in cyan and $k = 0.01$ - $0.15h/$Mpc in brown. While the constraints weaken for the brown contours due to the smaller number of modes, the two fits give very similar best fitting values.

\subsection{To use our results}
\label{sec:use}

In this subsection, we present our main results for future use, i.e. best-fitting values of the two geometric constraints ($D_V/r_s(z_d)$ and $F_{\rm AP}$) and the RSD parameter together with the covariance matrix. If readers are interested in using our constraints to test cosmological models or modifications of GR, they should be aware of the assumptions underlying our constraints given in section~\ref{sec:para}.

Since we present our result in a different base compared to the base we used for the study of systematics in section~\ref{sec:sys}, we made sure that the negligible systematic uncertainties in $\alpha_{\parallel}$ and $\alpha_{\perp}$ transfer into negligible shifts in $D_V/r_s$ and $F_{\rm AP}$. 
For most purposes our results can be well approximated by a multivariate Gaussian likelihood with
\begin{equation}
V^{\rm data}_{k_{\rm max}=0.20} = \begin{pmatrix}D_V(z_{\rm eff})/r_s(z_d)\\ F_{\rm AP}(z_{\rm eff})\\ f(z_{\rm eff})\sigma_8(z_{\rm eff})\end{pmatrix} = \begin{pmatrix}13.88\\ 0.683\\ 0.422\end{pmatrix}
\label{eq:likefinal0}
\end{equation} 
and the symmetric covariance matrix is given by
\begin{equation}
10^3C_{k_{\rm max}=0.20} = \begin{pmatrix} 36.400 & -2.0636 & -1.8398\\
			        & 1.0773 & 1.1755\\
			        &  & 1.8478 + 0.196\end{pmatrix}
\label{eq:likefinal2}
\end{equation} 
leading to 
\begin{equation}
C^{-1}_{k_{\rm max}=0.20} = \begin{pmatrix} 31.032 & 77.773 & -16.796\\
			        & 2687.7 & -1475.9\\
			        &  & 1323.0\end{pmatrix}.
\label{eq:likefinal22}
\end{equation} 
For $f\sigma_8$ we included the systematic error of $3.1\%$ (see section~\ref{sec:sys}), where we assumed uncorrelated systematic errors. The sound horizon scale used in our analysis is given by $r_s(z_d) = 147.36\,$Mpc.
The diagonal elements of the inverse covariance matrix represent the error on the different parameters when not marginalising over the other parameters. For example for the growth rate we find $f(z_{\rm eff})\sigma_8(z_{\rm eff}) = 0.422\pm0.027$. Note, that this constraint assumes that we know the geometry of the Universe exactly and neglects the large correlation between $f\sigma_8$ and $F_{\rm AP}$. We recommend using the full multivariate Gaussian for any cosmological model constraints.

We encourage the use of our results for $k_{\rm max} = 0.20h/$Mpc, but we also provide the results using $k_{\rm max} = 0.15h/$Mpc. The maximum likelihood values for the fitting range $k = 0.01$ - $0.15h/$Mpc are
\begin{equation}
V^{\rm data}_{k_{\rm max}=0.15} = \begin{pmatrix}D_V(z_{\rm eff})/r_s(z_d)\\ F(z_{\rm eff})\\ f(z_{\rm eff})\sigma_8(z_{\rm eff})\end{pmatrix} = \begin{pmatrix} 13.83\\ 0.684\\ 0.420\end{pmatrix}
\end{equation} 
and the symmetric covariance matrix is given by
\begin{equation}
10^3C_{k_{\rm max}=0.15} = \begin{pmatrix} 84.732 & -5.7656 & -3.0985\\
			        & 2.2777 & 1.9755\\
			        &  & 2.9532\end{pmatrix}
\end{equation} 
leading to 
\begin{equation}
C^{-1}_{k_{\rm max}=0.15} = \begin{pmatrix} 14.877 & 57.455 & -22.825\\
			        & 1267.7 & -787.74\\
			        &  & 841.62\end{pmatrix},
\end{equation} 
where no systematic error is included. Note that the values above are based on the sound horizon, $r_s$, calculated from CAMB~\citep{Lewis:1999bs}, while the equivalent values using $r_s$ calculated from~\citet{Eisenstein:1997ik} are given in appendix~\ref{app:EH}. The likelihood for any cosmological model using our constraints can then be calculated as 
\begin{equation}
\mathcal{L} \propto \exp\left[-(V^{\rm data} - V^{\rm m})^TC^{-1}(V^{\rm data} - V^{\rm m})/2\right],
\end{equation}
where $V^{\rm m}$ is a vector with model predictions for the three cosmological parameters.

\subsection{Comparison to other measurements}
\label{sec:compare}

In Figure~\ref{fig:Hz_DA} we show the constraints on $H(z_{\rm eff})r_s/r_s^{\rm fid}$ and $D_A(z_{\rm eff})r_s^{\rm fid}/r_s$ from different CMASS analyses as well as the Planck prediction within $\Lambda$CDM. Our analysis using the fitting range $k=0.01$ - $0.20h/$Mpc is included as the cyan contours. \citet{Anderson2.0} updated the CMASS-DR9 analysis published in~\citet{Anderson:2013oza}, where only the BAO information is exploited, while the RSD signal and broadband shape is marginalised out. The BAO constraint can be improved substantially by using density field reconstruction. We compare our results with~\citet{Anderson2.0} in Figure~\ref{fig:Hz_DA} (left), before reconstruction (grey contours) and after reconstruction (blue contours). Figure~\ref{fig:Hz_DA} (right) shows our results compared to other CMASS-DR11 studies, namely~\citet{Samushia:2013} (grey contours),~\citet{Chuang:2013wga} (blue contours) and~\citet{Sanchez:2013} (orange contours). While our analysis is in Fourier-space, all companion BOSS-DR11 papers we compare with in Figure~\ref{fig:Hz_DA} do their analysis in configuration space. The different CMASS-DR11 studies shown in Figure~\ref{fig:Hz_DA} use the same dataset, but use (1) different information from this dataset, (2) different fitting regions and (3) different clustering models. From Figure~\ref{fig:Hz_DA} we can see that all CMASS studies show agreement within $1\sigma$.

In Figure~\ref{fig:ps_red4} (top right) we show another comparison between our result and~\citet{Samushia:2013}, this time using the 2D likelihood of $f\sigma_8$ together with the power spectrum normalisation, $b_1\sigma_8$. We can see that the two results agree well on $f\sigma_8$ but find different clustering amplitudes. Using the fitting range $k=0.01$ - $0.20h/$Mpc we find $b_1\sigma_8(z_{\rm eff}) = 1.227\pm 0.030$ while~\citet{Samushia:2013} finds $b_1\sigma_8(z_{\rm eff}) = 1.289\pm0.032$. Using the fiducial values for $\sigma_8(z_{\rm eff})$ the bias obtained in our analysis is $b_1= 2.040\pm0.052$, while~\citet{Samushia:2013} finds $b_1=2.096\pm0.052$. The reason for this difference could be (1) the different scales which are used in the two different studies or (2) the details of the modelling, i.e., we include higher-order bias terms: $b_2$, $b_{s2}$, $b_{3nl}$ and $N$, while~\citet{Samushia:2013} only includes linear bias. Since the clustering amplitude is just considered a nuisance parameter in our analysis, this difference does not represent a problem for our main cosmological results. Comparing the constraints on $f\sigma_8$ with the prediction of Planck (green contours) we find that our best fitting value is below the Planck prediction ($\left[f(z=0.57)\sigma_8(z=0.57)\right]_{\rm Planck} = 0.481\pm0.010$) at $1.4\sigma$ significance level, when marginalising over all other parameters.

\citet{Reid:2012sw} and~\citet{Chuang:2013hya} analysed the power spectrum multipoles in CMASS-DR9 finding $f(z=0.57)\sigma_8(z=0.57) = 0.427^{+0.069}_{-0.063}$~\citep{Reid:2012sw} and $f(z=0.57)\sigma_8(z=0.57) = 0.428\pm 0.066$~\citep{Chuang:2013hya} in good agreement with our results. The error decreased from DR9 to DR11 by roughly a factor of $1.6$, which agrees with  the expectation due to the survey volume increase.

We should also mention other RSD measurements in the literature. \citet{Blake:2011ep} analysed the WiggleZ power spectrum simultaneously fitting for $F_{\rm AP}(z)$ and $f(z)\sigma_8(z)$. Because of the small sky coverage of the different patches of the WiggleZ survey, it is possible to measure the power spectrum multipoles in WiggleZ using the FKP estimator~\citep{Blake:2011rj}. They found constraints on $f(z)\sigma_8(z)$ between $21\%$ and $32\%$ for four redshift bins ($0.22$, $0.41$, $0.6$ and $0.78$). Their constraint at $z=0.6$ is $f\sigma_8 = 0.37\pm0.08$, which is statistically consistent with our result. 
Within the Luminous Red Galaxy (LRG) sample in SDSS-II DR7,~\citet{Samushia:2011cs} reported growth of structure measurements in two redshift bins, finding $f(z=0.25)\sigma_8(z=0.25) = 0.351\pm0.058$ and $f(z=0.37)\sigma_8(z=0.37) = 0.460\pm0.038$. While~\citet{Samushia:2011cs} fixed the AP effect, \citet{Oka:2013cba} put a simultaneous constraint on the RSD and the AP effect using the power spectrum multipoles finding $f(z=0.3)\sigma_8(z=0.3) = 0.49\pm0.08$. The 6dFGS team recently reported a growth of structure measurement of $f(z)\sigma(z) = 0.423\pm0.055$~\citep{Beutler:2012px} at $z=0.067$.

We note that like the $f\sigma_8$ constraint reported in our paper, most growth of structure constraints obtained in other galaxy surveys lie below the Planck $\Lambda$CDM-GR prediction. 

\section{Cosmological implications}
\label{sec:cos}

This section contains two simple applications of the constraints we obtained with CMASS-DR11. First we perform a $\Lambda$CDM consistency check by combining the CMASS constraints with the Planck data to test GR. The second application assumes $\Lambda$CDM and GR and constrains $\sigma_8$ using only the CMASS dataset.

\subsection{Consistency check using CMASS-DR11 and Planck 2013}
\label{sec:consistency}

\begin{table*}
\begin{center}
\caption{This table summarises cosmological parameter constraints obtained in section~\ref{sec:cos} using CMASS-DR11. The first four rows contain constraints on the growth index $\gamma$ and $\Omega_m$ when combining CMASS with Planck and WMAP9 (see Figure~\ref{fig:gamma}). The fifth and sixth row contains constraints on $\sigma_8$ and $\Omega_m$ using only the growth rate and the AP effect ($f\sigma_8$ and $F_{\rm AP}$) of the CMASS dataset. The last four rows contain constraints on $\sigma_8$ and $\Omega_m$ using all CMASS-DR11 constraints ($D_V/r_s$, $F_{\rm AP}$ and $f\sigma_8$) and assuming the sound horizon of Planck or WMAP9 (see Figure~\ref{fig:om_sig8}) in co-moving units. In this case the constraint on $\Omega_m$ is dependent on the CMB experiment used to calibrate the standard ruler, while the constraint on $\sigma_8$ is fairly independent of this choice.}
	\begin{tabular}{ccll}
		\hline
		\multicolumn{2}{c}{parameter constraint} & based on & assumptions\\
     		\hline
		section~\ref{sec:consistency}& & &\\
		$\gamma$ & $0.772^{+0.124}_{-0.097}$ & CMASS-($D_V/r_s$, $F_{\rm AP}$, $f\sigma_8$) + Planck & $\Lambda$CDM, $\Omega_m^{\gamma}(z)$\\
		$\Omega_m$ & $0.308\pm 0.011$ & CMASS-($D_V/r_s$, $F_{\rm AP}$, $f\sigma_8$) + Planck & $\Lambda$CDM, $\Omega_m^{\gamma}(z)$\\
		$\gamma$ & $0.76\pm0.11$ & CMASS-($D_V/r_s$, $F_{\rm AP}$, $f\sigma_8$) + WMAP9 & $\Lambda$CDM, $\Omega_m^{\gamma}(z)$\\
		$\Omega_m$ & $0.298\pm 0.013$ & CMASS-($D_V/r_s$, $F_{\rm AP}$, $f\sigma_8$) + WMAP9 & $\Lambda$CDM, $\Omega_m^{\gamma}(z)$\\
		\hline
		section~\ref{sec:sig8}& & &\\
		$\sigma_8$ & $0.731\pm 0.052$ & CMASS-($F_{\rm AP}$, $f\sigma_8$) & $\Lambda$CDM, $\Omega_m^{0.55}(z)$\\
		$\Omega_m$ & $0.33^{+0.15}_{-0.12}$ & CMASS-($F_{\rm AP}$, $f\sigma_8$) & $\Lambda$CDM, $\Omega_m^{0.55}(z)$\\
		$\sigma_8$ & $0.719\pm 0.047$ & CMASS-($D_V/r_s$, $F_{\rm AP}$, $f\sigma_8$) & $\Lambda$CDM, $\Omega_m^{0.55}(z)$, $r^{\rm Planck}_s(z_d) = 98.79\,$Mpc$/h$\\
		$\Omega_m$ & $0.341\pm 0.028$ & CMASS-($D_V/r_s$, $F_{\rm AP}$, $f\sigma_8$) & $\Lambda$CDM, $\Omega_m^{0.55}(z)$, $r^{\rm Planck}_s(z_d) = 98.79\,$Mpc$/h$\\
		$\sigma_8$ & $0.713\pm 0.047$ & CMASS-($D_V/r_s$, $F_{\rm AP}$, $f\sigma_8$) & $\Lambda$CDM, $\Omega_m^{0.55}(z)$, $r^{\rm WMAP9}_s(z_d) = 102.06\,$Mpc$/h$\\
		$\Omega_m$ & $0.274\pm 0.023$ & CMASS-($D_V/r_s$, $F_{\rm AP}$, $f\sigma_8$) & $\Lambda$CDM, $\Omega_m^{0.55}(z)$, $r^{\rm WMAP9}_s(z_d) = 102.06\,$Mpc$/h$\\
		\hline
	  \end{tabular}
	  \label{tab:para2}
\end{center}
\end{table*}

\begin{figure*}
\begin{center}
\epsfig{file=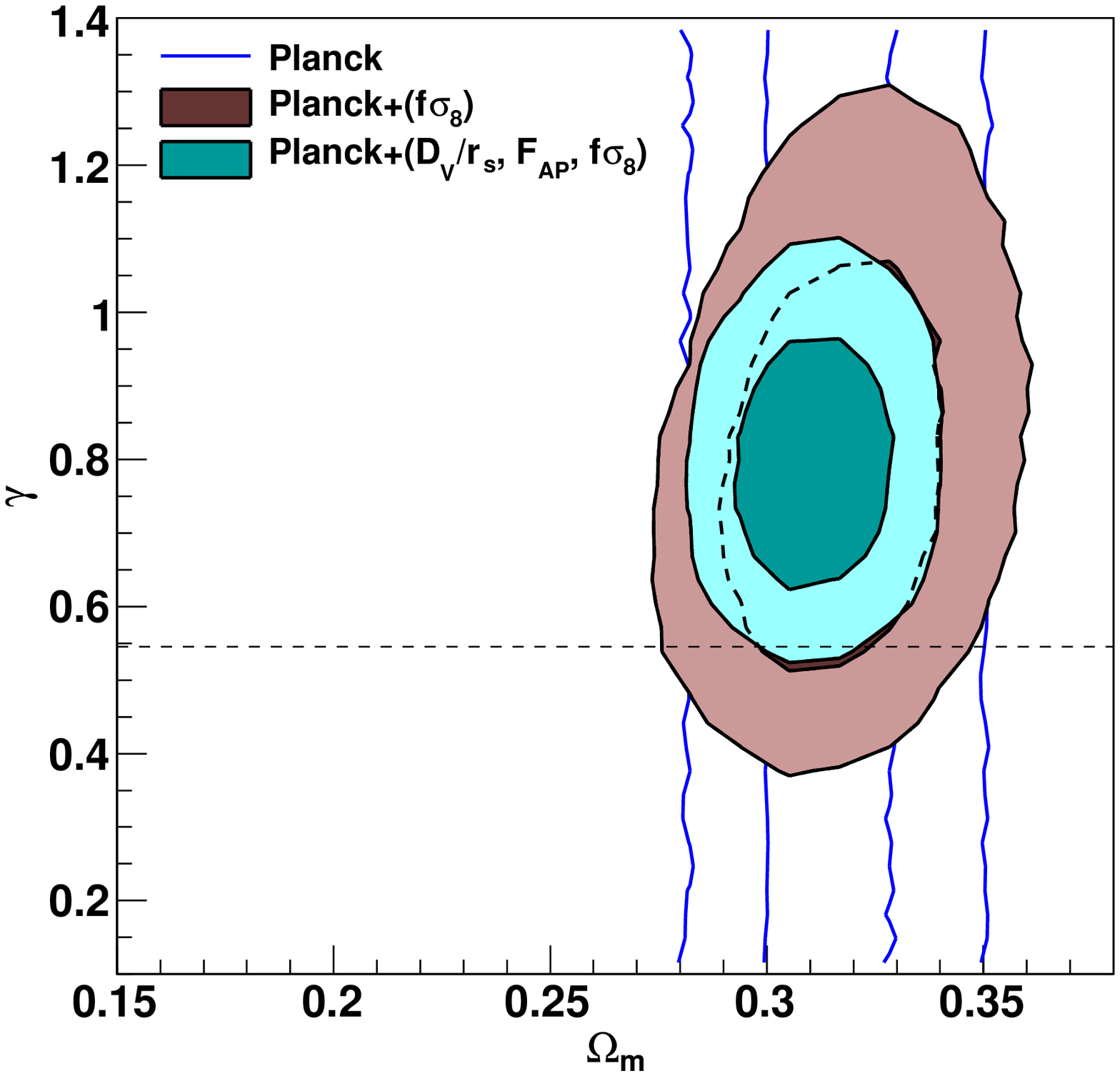,width=8.cm}
\epsfig{file=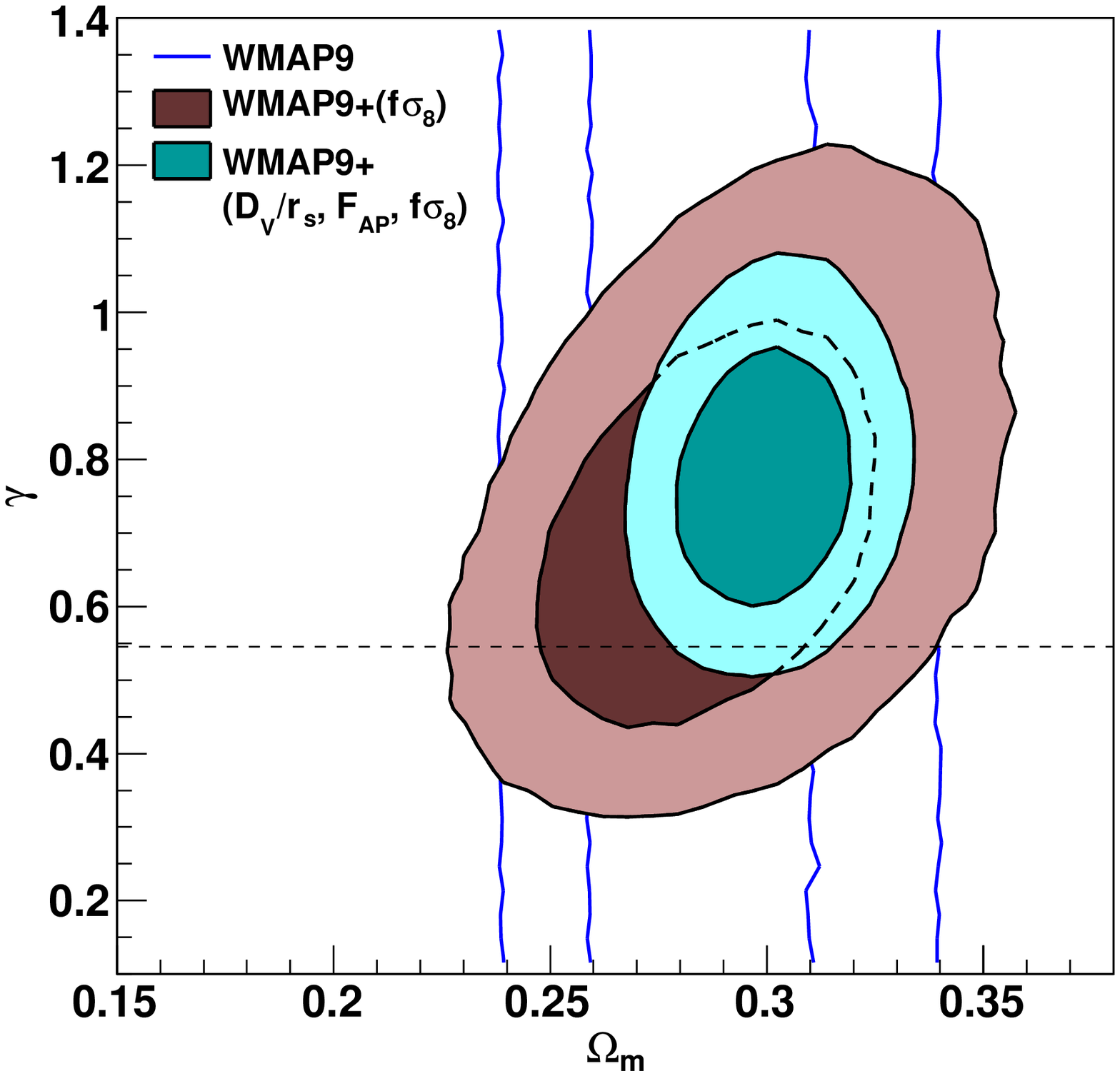,width=8.cm}
\caption{The 2D likelihood distribution for $\gamma$ and $\Omega_m$ from Planck+CMASS (left) and WMAP9+CMASS (right). We show the $68\%$ and $95\%$ confidence regions. The different contours are for the CMB constraints alone (blue lines), CMB + $f\sigma_8$ from CMASS-DR11 (brown contours) and CMB + ($D_V/r_s$, $F_{\rm AP}$, $f\sigma_8$) from eq.~\ref{eq:likefinal0} and~\ref{eq:likefinal22} (cyan contours). Since we do not exploit the Integrated Sachs-Wolfe (ISW) effect for this test, the CMB datasets cannot set constraints on $\gamma$. The CMB data are needed for tight constraints on $\Omega_m$ and for the normalisation of the power spectrum, $\sigma_8(z)$.}
\label{fig:gamma}
\end{center}
\end{figure*}

Within $\Lambda$CDM-GR it has been shown that the growth rate can be parameterised as $f(z) = \Omega^{\gamma}_m(z)$, where $\gamma$ is the growth index, predicted to be $\gamma\approx 0.55$ in GR~\citep{Linder:2005in}. As a consistency check for $\Lambda$CDM-GR within the Planck cosmology we use our constraint on $f(z_{\rm eff})\sigma_8(z_{\rm eff})$ obtained using the fitting range $k=0.01$ - $0.20h/$Mpc, to set constraints on $\gamma$. To do this, we download the Planck MCMC chain for $\Lambda$CDM\footnote{http://irsa.ipac.caltech.edu/data/Planck/release\_1/ancillary-data/} and importance sample this chain. The analysis method is described in the following three steps:
\begin{enumerate}
\item For each MCMC chain element, we randomly choose a value of $\gamma$ with the flat prior $0 < \gamma < 2$. Since the value of $\sigma_8(z_{\rm eff})$ depends on $\gamma$ we have to re-calculate this value for each chain element. First we calculate the growth factor
\begin{equation}
D(a_{\rm eff}) = \exp\left[ -\int_{a_{\rm eff}}^1da'\,f(a')/a'\right],
\end{equation}
where $a_{\rm eff}$ is the scale factor at the effective redshift $a_{\rm eff} = 1/(1 + z_{\rm eff})$. In order to derive $\sigma_{8,\gamma}(z_{\rm eff})$ we have to extrapolate from the matter dominated epoch to the effective redshift,
\begin{equation}
\sigma_{8,\gamma}(z_{\rm eff}) = \frac{D_{\gamma}(z_{\rm eff})}{D(z_{hi})}\sigma_8(z_{hi}),
\end{equation} 
where we calculate $\sigma_8(z_{hi})$ at $z_{hi} = 50$, well in the matter-dominated regime, where $f(z) \approx 1$.
\item Now we calculate the growth rate using $f_{\gamma}(z_{\rm eff}) \simeq \Omega_m^{\gamma}(z_{\rm eff})$. This gives us all the ingredients to construct the parameter combination $f_{\gamma}(z_{\rm eff})\sigma_{8,\gamma}(z_{\rm eff})$. 
\item We also calculate $D_V/r_s(z_d)$ and $F_{\rm AP}$ for each chain element. We then use the maximum likelihood values and inverse covariance matrix of eq.~\ref{eq:likefinal0} and \ref{eq:likefinal22} to calculate a CMASS-DR11 likelihood and combine this with the Planck likelihood. 
\end{enumerate}
The result is shown in Figure~\ref{fig:gamma} (left). Marginalising over the remaining parameters we get $\gamma = 0.772^{+0.124}_{-0.097}$ (Planck+CMASS), while the prediction of $\Lambda$CDM+GR is $\gamma \approx 0.55$. Only $1.7\%$ of the likelihood can be found below $\gamma = 0.55$ and therefore GR lies outside the $96.6\%$ confidence level. We can now ask, whether this situation changes if we use WMAP9 instead of Planck. WMAP9 measured a smaller value of $\Omega_m$ and therefore predicts a smaller value of $f\sigma_8$. If we use only the measured $f\sigma_8$, ignoring the geometric information (brown contours in Figure~\ref{fig:gamma}, right) we find better agreement with $\gamma=0.55$ compared to the same situation for Planck. When we include the geometric information (cyan contours) the errors become smaller and the preferred value of gamma changes from $\gamma = 0.65^{+0.22}_{-0.14}$ (WMAP9+$f\sigma_8$) to $\gamma = 0.76\pm0.11$ (WMAP9+$D_V/r_s$, $F_{\rm AP}$, $f\sigma_8$), very similar to the value we find in Planck+CMASS. The shift of $\gamma$ towards larger values when including geometric information is caused by the slight tension between WMAP9 and our geometric parameters. In both cases we see that the constraints improve considerably, when including the geometric information. Since the geometric parameters are not sensitive to $\gamma$, this improvement comes through the improvement on $\Omega_m$ and $\sigma_8$. We regard our measurement of $\gamma$ using the Planck chain as the final result of this consistency check and include it in Figure~\ref{fig:keff} at the scale of $\sim 30\,$Mpc (see section~\ref{sec:keff}). 

From the theoretical side it is difficult to find models of modified gravity which suppress the growth of structure. Most models actually predict a stronger structure growth (see e.g.~\citealt{Mortonson:2008qy,Dodelson:2013sma}). One example of a model which does predict smaller structure growth is the DGP model~\citep{Dvali:2000hr}, which however has theoretical issues~\citep{Gorbunov:2005zk} and also seems to predict the wrong expansion history (e.g.~\citealt{Davis:2007na,Fang:2008kc}).

\begin{figure}
\begin{center}
\epsfig{file=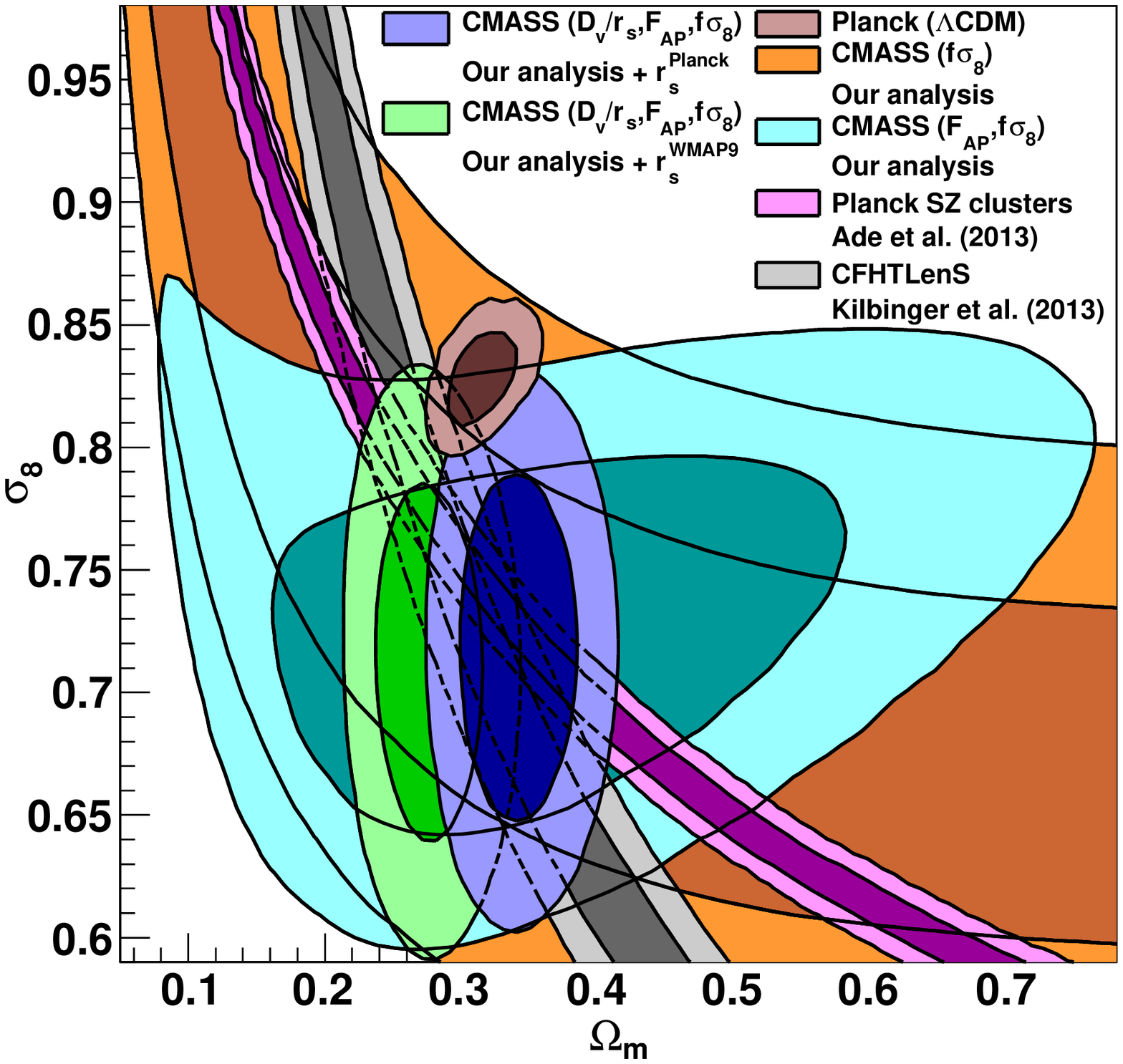,width=8.8cm}
\caption{Comparison between Planck~\citep{Ade:2013zuv}, Planck SZ clusters~\citep{Ade:2013lmv}, CFHTLenS lensing~\citep{Kilbinger:2012qz} and our results in the $\sigma_8$-$\Omega_m$ plane. When using only the $f\sigma_8$ constraint from our analysis (orange contours), there is a degeneracy, similar to the cluster and lensing datasets. The geometric information can break this degeneracy. While the AP effect is only depending on $\Omega_m$, our $D_V/r_s$ constraint does require calibration of the sound horizon. We show the results, where we fix the sound horizon to the value of Planck (blue contours) and the value reported by WMAP9 (green contours). The results are summarised in Table~\ref{tab:para2}. To turn our $f\sigma_8$ constraint into a constraint on $\sigma_8$ we assume GR ($\gamma = 0.55$) and $\Lambda$CDM similar to the Planck contours (brown contours). The tension in $\sigma_8$ between our measurement and Planck is directly related to the large $\gamma$ we find in our $\Lambda$CDM consistency check in section~\ref{sec:consistency}.}
\label{fig:om_sig8}
\end{center}
\end{figure}

There are many ways in which one could reduce the predicted structure growth of Planck, e.g. massive neutrinos, $w < -1$ or $\Omega_k > 0$. We should also mention that there are several other datasets in tension with the Planck inferred structure growth. Figure~\ref{fig:om_sig8} shows our result in the $\sigma_8$-$\Omega_m$ plane compared to Planck~\citep{Ade:2013zuv}, Planck SZ clusters~\citep{Ade:2013lmv} and CHFTLS lensing~\citep{Kilbinger:2012qz}. Using the CMASS $f\sigma_8$ measurement alone, there is a degeneracy between $\sigma_8$ and $\Omega_m$ similar to the lensing and cluster constraints. This degeneracy can be broken when including the geometric information ($F_{\rm AP}$ and $D_V/r_s$). We can see that Planck predicts a large $\sigma_8$ in tension with the other datasets included in this comparison (see also~\citealt{Mandelbaum:2012ay}). The large normalisation $\sigma_8$ of Planck directly leads to the large $\gamma$ we found in our consistency check above. Therefore Figure~\ref{fig:om_sig8} shows that we can relax the tension between our measurement and GR by using the normalisation from one of the other datasets shown in this Figure.

\subsection{Constraining $\sigma_8$ with CMASS-DR11}
\label{sec:sig8}

Assuming $\Lambda$CDM and GR in the form $\Omega_m^{0.55}(z)$ we can use our constraint on the growth of structure ($f\sigma_8$) and the AP effect ($F_{\rm AP}$) to set the constraint $\sigma_8 = 0.731\pm 0.052$ (cyan contours in Figure~\ref{fig:om_sig8}). Our dataset is therefore one of the few low redshift datasets, which is powerful enough to constrain $\sigma_8$ independently. We can also get a fairly weak constraint on the matter density of $\Omega_m = 0.33^{+ 0.15}_{-0.12}$.

Additionally we can include the BAO information ($D_V/r_s$), where we however have to fix the sound horizon size $r_s$. In Figure~\ref{fig:om_sig8} we show the constraint using the sound horizon of Planck (blue contours) and WMAP9 (green contours). We use the sound horizon in co-moving units $r_s^{\rm Planck}(z_d) = 98.79\,$Mpc$/h$ and $r^{\rm WMAP9}_s(z_d) = 102.06\,$Mpc$/h$, which includes information about the Hubble constant. Our constraint on $D_V/r_s$ together with the sound horizon from the CMB allows tight constraints on $\Omega_m$, while the constraint on $\sigma_8$ does not improve significantly (see Table~\ref{tab:para2} for details).

\section{Conclusion}
\label{sec:conclusion}

This paper analyses the BOSS CMASS-DR11 dataset employing a power spectrum estimator suggested by~\citet{Yamamoto:2005dz}, which allows us to measure the power spectrum monopole and quadrupole in a wide-angle survey like BOSS. We use Quick-Particle-Mesh (QPM) simulations to produce $999$ mock catalogues to derive a covariance matrix. The covariance matrix shows little correlation between the different bins in the power spectrum, which is very different to similar studies using the correlation function. 

Our model of the multipole power spectrum accounts for nonlinear evolution on the basis of perturbation theory. We adopt the modelling of non-linear redshift-space distortion by~\citet{Taruya:2010mx} and extend this approach to include the local and non-local galaxy bias with its stochasticity. 

The parameter fits using the fitting range $k = 0.01$ - $0.20h/$Mpc are considered the main results of this paper. We provide a multivariate Gaussian likelihood to use our results for cosmological constraints. 

Our analysis has been performed blind, meaning that all systematics checks and the set-up of the fitting procedure has been done on mock catalogues and only at the last stage did we analyse the actual CMASS-DR11 power spectrum measurement. The results of our analysis can be summarised in the following five points:
\begin{enumerate}
\item We provide a set of equations (eq.~\ref{eq:conv},~\ref{eq:conv2},~\ref{eq:ic1},~\ref{eq:ic2}), which allows us to incorporate the window function and the integral constraint into our analysis in a self-consistent manner, without using any simplifying assumptions and without the need to split the survey into sub-regions.
\item Our study of systematic uncertainties lead to a maximum wavenumber of $k_{\rm max} = 0.20h/$Mpc for our analysis, where the total error of $f(z_{\rm eff})\sigma_8(z_{\rm eff})$ is minimised. Our final systematic uncertainty for $f(z_{\rm eff})\sigma_8(z_{\rm eff})$ is $3.1\%$ when using the fitting range $k = 0.01$ - $0.20h/$Mpc. The geometric parameters $\alpha_{\parallel}$ and $\alpha_{\perp}$ ($D_V/r_s$ and $F_{\rm AP}$) do not show any significant systematic uncertainties.
\item Our power spectrum model includes $7$ free parameters: the two geometric parameters, $\alpha_{\parallel}$ and $\alpha_{\perp}$, the growth rate $f(z_{\rm eff})\sigma_8(z_{\rm eff})$ and $4$ nuisance parameters. 
We find $\alpha_{\parallel} = 1.018\pm 0.036$, $\alpha_{\perp} = 1.029\pm 0.015\,$ and $f(z_{\rm eff})\sigma_8(z_{\rm eff}) = 0.419\pm0.044$, where we included the systematic uncertainty of $3.1\%$. The geometric parameters $\alpha_{\parallel}$ and $\alpha_{\perp}$ can be expressed as $D_V(z_{\rm eff})/r_s(z_d) = 13.89\pm 0.18$ and $F_{\rm AP}(z_{\rm eff}) = (1+z_{\rm eff})D_A(z_{\rm eff})H(z_{\rm eff})/c = 0.679\pm0.031$. While the geometric parameters found in our analysis agree very well with the Planck prediction within $\Lambda$CDM, the growth rate is about $1.4\sigma$ below the Planck prediction. 
We provide a multivariate Gaussian likelihood to use our results (see section~\ref{sec:use}). All results are summarised in Table~\ref{tab:para}, where we also provide the parameter constraints using the more conservative fitting range $k =0.01$ - $0.15h/$Mpc. We also provide the power spectrum measurements itself, together with the covariance matrices and the window functions online at \url{https://sdss3.org/science/boss_publications.php}.
\item We performed a $\Lambda$CDM-GR consistency check within the Planck cosmology, which results in a measurement of the growth index $\gamma = 0.772^{+0.124}_{-0.097}$. This value excludes the GR prediction of $\gamma \approx 0.55$ by more than $2\sigma$. When replacing Planck with WMAP9 we find a very similar result of $\gamma = 0.76\pm0.11$. We conclude that there is tension between our result combined with Planck (WMAP9) and the prediction by GR. This tension could be (1) a statistical fluctuation, (2) an indication for unaccounted systematic uncertainties in CMASS and/or Planck (WMAP9) or (3) ask for modifications in $\Lambda$CDM or GR.
\item Assuming $\Lambda$CDM and GR we can use our measurement of the growth rate ($f\sigma_8$) together with the information from the Alcock-Paczynski effect ($F_{\rm AP}$) to constrain $\sigma_8 = 0.731\pm 0.052$. The low value of $\sigma_8$ is directly connected to the high value of the growth index $\gamma$ obtained from our dataset. While galaxy datasets in the past only constrained a degenerate combination of $\sigma_8$ and $\Omega_m$, our data is now good enough to break this degeneracy. This represents one of the best independent $\sigma_8$ constraints at low redshift.
\end{enumerate}
Finally we should also mention that separate studies within the BOSS collaboration are currently working on measurements of CMASS clustering combined with lensing, as well as measurements of the CMASS bispectrum, which should provide additional information about the bias parameters $b_1$ and $b_2$, respectively. This will help us to go from $f\sigma_8$ directly to the growth rate $f$ and test gravity models without using the CMB normalisation.  

\section*{Acknowledgments}

FB would like to thank Chris Blake, Uros Seljak, Eric Linder, Beth Reid, Martin White, Morag Scrimgeour and Julien Guy for helpful discussion. SS is supported by a Grant-in-Aid for Young Scientists (Start-up) from the Japan Society for the Promotion of Science (JSPS) (No. 25887012).

Funding for SDSS-III has been provided by the Alfred P. Sloan Foundation, the Participating Institutions, the National Science Foundation, and the U.S. Department of Energy Office of Science. The SDSS-III web site is http://www.sdss3.org/.

SDSS-III is managed by the Astrophysical Research Consortium for the Participating Institutions of the SDSS-III Collaboration including the University of Arizona, the Brazilian Participation Group, Brookhaven National Laboratory, Carnegie Mellon University, University of Florida, the French Participation Group, the German Participation Group, Harvard University, the Instituto de Astrofisica de Canarias, the Michigan State/Notre Dame/JINA Participation Group, Johns Hopkins University, Lawrence Berkeley National Laboratory, Max Planck Institute for Astrophysics, Max Planck Institute for Extraterrestrial Physics, New Mexico State University, New York University, Ohio State University, Pennsylvania State University, University of Portsmouth, Princeton University, the Spanish Participation Group, University of Tokyo, University of Utah, Vanderbilt University, University of Virginia, University of Washington, and Yale University.

This research used resources of the National Energy Research Scientific Computing Center, which is supported by the Office of Science of the U.S. Department of Energy under Contract No. DE-AC02-05CH11231.

\setlength{\bibhang}{2em}
\setlength{\labelwidth}{0pt}

\newpage

\appendix
\numberwithin{equation}{section}

\section{Derivation of the minimum variance weight, $\lowercase{w}_{\rm FKP}$ in the presence of systematic weights}
\label{ap:minvar}

This derivation follows the original derivation in~\citet{Feldman:1993ky}, with the addition of a systematic weight, $w_{\rm sys}$. Under the assumption that the width $\Delta k$ of the spherical shell (or bin size) is larger than the coherence length ($\sim 1/D$, with $D$ being the size of the survey) we can write the error in the power spectrum as
\begin{equation}
\sigma_P^2(k) \simeq \frac{1}{V_k}\int d\vec{k}' |P(\vec{k})Q(\vec{k}') + S(\vec{k}')|^2
\end{equation}
with
\begin{align}
\begin{split}
Q(\vec{k}) &= \frac{1}{A}\int d\vec{x}\; {n'}_g^2(\vec{x})w^2_{\text{\tiny{FKP}}}(\vec{x})e^{-i \vec{k}\cdot \vec{x}}
\end{split}\\
\begin{split}
S(\vec{k}) &= \frac{1}{A}\bigg(\int d\vec{x}\; n'_g(\vec{x})w_{\rm sys}(\vec{x})w^2_{\text{\tiny{FKP}}}(\vec{x})e^{-i\vec{k}\cdot \vec{x}}\\
&\;\;\;\;\;+ \alpha\int d\vec{x}\; n_g'(\vec{x})w^2_{\text{\tiny{FKP}}}(\vec{x})e^{-i\vec{k}\cdot \vec{x}}\bigg)
\end{split}
\end{align}
and the normalisation $A = \int d\vec{x}\; {n'}_g^2(\vec{x})w^2_{\text{\tiny{FKP}}}(\vec{x})$. The fractional variance of the power can be written as
\begin{align}
\begin{split}
\left(\frac{\sigma_P(k)}{P(k)}\right)^2 &= \frac{1}{V_k}\int d\vec{k}' \left|Q(\vec{k}') + \frac{S(\vec{k}')}{P(\vec{k})}\right|^2
\end{split}\\
& = \frac{1}{V_kA^2}\int d\vec{k}' \bigg|\int d\vec{x}\; w^2_{\text{\tiny{FKP}}}(\vec{x})e^{-i\vec{k}'\cdot \vec{x}}\\
&\;\;\;\;\;\left({n'}_g^2(\vec{x}) + \frac{n'_g(\vec{x})w_{\rm sys}(\vec{x}) + \alpha n'_g(\vec{x})}{P(\vec{k})}\right)\bigg|^2.\notag
\end{align}
Using Parseval's theorem in the form
\begin{equation}
\int d\vec{k}' \left|\int d\vec{x} F(\vec{x})e^{-i\vec{k}'\cdot \vec{x}}\right|^2 = (2\pi)^3\int d\vec{x}\; F^2(\vec{x})
\end{equation}
the equation can be further simplified to
\begin{align}
\left(\frac{\sigma_P(k)}{P(k)}\right)^2 &= \frac{(2\pi)^3}{V_kA^2}\int d\vec{x}\; w^4_{\text{\tiny{FKP}}}(\vec{x})\\
&\;\;\;\;\;\left({n'}_g^2(\vec{x}) + \frac{n'_g(\vec{x})w_{\rm sys}(\vec{x}) + \alpha n'_g(\vec{x})}{P(k)}\right)^2.\notag
\label{eq:opweight}
\end{align}
Introducing the functions
\begin{align}
f(\vec{x}) &= \left({n'}^2(\vec{x}) + \frac{n'_g(\vec{x})w_{\rm sys}(\vec{x}) + \alpha n'_g(\vec{x})}{P(k)}\right)^2,\\
g(\vec{x}) &= {n'}^2(\vec{x})
\end{align}
we can write
\begin{equation}
\left(\frac{\sigma_P(k)}{P(k)}\right)^2 = \frac{\int d\vec{x} w^4_{\text{\tiny{FKP}}}(\vec{x})f(\vec{x})}{\left[\int d\vec{x}\; w^2_{\text{\tiny{FKP}}}(\vec{x})g(\vec{x})\right]^2}.
\end{equation}
Now we perturb the weight $w_{\text{\tiny{FKP}}}(\vec{x}) \rightarrow w_{\text{\tiny{FKP}}}(\vec{x}) + \Delta w(\vec{x})$, which leads to
\begin{align}
\left(\frac{\sigma_P(k)}{P(k)}\right)^2 &= \frac{\int d\vec{x}\; [w_{\text{\tiny{FKP}}}(\vec{x}) + \Delta w(\vec{x})]^4f(\vec{x})}{\left(\int d\vec{x}\; \left[w_{\text{\tiny{FKP}}}(\vec{x}) + \Delta w(\vec{x})\right]^2g(\vec{x})\right)^2}\\
&\approx \frac{\int d\vec{x}\; w^4_{\text{\tiny{FKP}}}(\vec{x})\left[1 + 4\frac{\Delta w(\vec{x})}{w_{\text{\tiny{FKP}}}(\vec{x})}\right]f(\vec{x})}{\left[\int d\vec{x}\; w^2_{\text{\tiny{FKP}}}(\vec{x}) \left[1 + 2\frac{\Delta w(\vec{x})}{w_{\text{\tiny{FKP}}}(\vec{x}_i)} \right]g(\vec{x})\right]^2}.
\end{align}
Using Taylor expansion up to second order around $\Delta w = 0$ we get
\begin{equation}
\begin{split}
\left(\frac{\sigma_P(k)}{P(k)}\right)^2 &= \frac{\int d^3 x\; w^4_{\text{\tiny{FKP}}}(\vec{x})f(\vec{x})}{\left[\int d^3 x\; w^2_{\text{\tiny{FKP}}}(\vec{x})g(\vec{x})\right]^2}\times\\
&\bigg( 1 + 4\bigg[\frac{\int d^3x\; w^3_{\text{\tiny{FKP}}}(\vec{x})f(\vec{x})\Delta w(\vec{x})}{\int d^3x\;w^4_{\text{\tiny{FKP}}}(\vec{x})f(\vec{x})}\\
&- \frac{\int d^3x\; w_{\text{\tiny{FKP}}}(\vec{x})g(\vec{x})\Delta w(\vec{x})}{\int d^3x\; w^2_{\text{\tiny{FKP}}}(\vec{x})g(\vec{x})}\bigg]\bigg) + ...
\end{split}
\end{equation}
Therefore, the optimal weighting function has to satisfy
\begin{equation}
\frac{\int d^3x\; w^3_{\text{\tiny{FKP}}}(\vec{x})f(\vec{x})\Delta w(\vec{x})}{\int d^3x\; w^4_{\text{\tiny{FKP}}}(\vec{x})f(\vec{x})} = \frac{\int d^3x\; w_{\text{\tiny{FKP}}}(\vec{x})g(\vec{x})\Delta w(\vec{x})}{\int d^3 x\; w^2_{\text{\tiny{FKP}}}(\vec{x})g(\vec{x})}.
\end{equation}
The solution of this equation is given by
\begin{align}
w_{\text{\tiny{FKP}}}(\vec{x}) \propto \sqrt{\frac{g(\vec{x})}{f(\vec{x})}} &= \frac{n'(\vec{x})}{{n'}^2(\vec{x}) + \frac{n'_g(\vec{x})w_{\rm sys}(\vec{x}) + \alpha n'_g(\vec{x})}{P(k)}}\\
&= \frac{1}{n'(\vec{x}) + \frac{w_{\rm sys}(\vec{x}) + \alpha}{P(k)}}.
\end{align}
Finally, the dimensionless optimal weighting function is
\begin{equation}
w_{\text{\tiny{FKP}}}(\vec{x}) = \frac{1}{\frac{P(k)n'(\vec{x})}{w_{\rm sys}(\vec{x})} + 1 + \frac{\alpha}{w_{\rm sys}(\vec{x})}}.
\end{equation}
When we choose a large number of random galaxies ($\alpha \ll 1$), we get
\begin{equation}
w_{\text{\tiny{FKP}}}(\vec{x}) = \frac{1}{1 + \frac{P(k)n'(\vec{x})}{w_{\rm sys}(\vec{x})}}.
\end{equation}
which in the case of $w_{\rm sys} = 1$ recovers the original minimum variance weight reported in~\citet{Feldman:1993ky}.

\section{Window function}
\label{ap:window}

\subsection{The survey window function: Derivation of eq.~{\protect\ref{eq:conv2}}}

We simplify the convolution integral of eq.~\ref{eq:conv} to
\begin{align}
\begin{split}
P^{\rm conv}_{\ell}(k) &= \frac{2\ell + 1}{2}\int d\mu \int \frac{d\phi}{2\pi}\int d\vec{k}'P^{\rm true}(\vec{k}')|W(\vec{k}-\vec{k}')|^2\mathcal{L}_{\ell}(\mu)\\
&= \frac{2\ell + 1}{2}\int d\mu \int \frac{d\phi}{2\pi}\int d\mu'\int d\phi'\int dk' k'^2 P^{\rm true}(k',\mu')\\
&\;\;\;\;\; \sum^{N_{\rm ran}}_{ij,i\neq j} w_{\rm FKP}(\vec{x}_i) w_{\rm FKP}(\vec{x}_j)e^{i\vec{k}\cdot\Delta\vec{x}}e^{-i\vec{k}'\cdot\Delta\vec{x}} \mathcal{L}_{\ell}(\mu),
\end{split}
\end{align}
where $\Delta \vec{x} = \vec{x}_i - \vec{x}_j$. We now expand the power spectrum into multipoles $P(k',\mu') = \sum_L P_L(k')\mathcal{L}_L(\mu')$ where $\mu' = \hat{\vec{k}}'\cdot\hat{\vec{x}}_h$. We also apply the relation 
\begin{equation}
e^{i|\vec{k}||\vec{x}|\mu} = \sum_s i^s (2s + 1) j_s(|\vec{k}||\vec{x}|)\mathcal{L}_s(\mu),
\label{eq:exp}
\end{equation}
as well as the identity 
\begin{equation}
\begin{split}
&\frac{2\ell + 1}{2}\int d\mu \int \frac{d\phi}{2\pi}\, \mathcal{L}_{\ell}(\hat{\vec{k}}\cdot\Delta\hat{\vec{x}})\mathcal{L}_{\ell'}(\hat{\vec{k}}\cdot\hat{\vec{x}}_h)\\
&=\mathcal{L}_{\ell}({\hat{\vec{x}}_h\cdot \Delta\hat{\vec{x}}})\delta_{\ell\ell'}.
\end{split}
\label{eq:hee2}
\end{equation}
See appendix~\ref{sec:hee} for a proof of this identity. We can now re-write the convolution as
\begin{align}
\begin{split}
P^{\rm conv}_{\ell}(k) &= 2\pi\int dk'k'^2 \sum_L P^{\rm true}_L(k')(-i)^L\\
&\;\;\;\;\;2 i^{\ell}(2\ell + 1)\sum^{N_{\rm ran}}_{ij, i\neq j} w_{\rm FKP}(\vec{x}_i)w_{\rm FKP}(\vec{x}_j)\\
&\;\;\;\;\;j_{\ell}(k|\Delta\vec{x}|) j_L(k'|\Delta\vec{x}|)\mathcal{L}_{\ell}({\hat{\vec{x}}_h\cdot \Delta\hat{\vec{x}}})\mathcal{L}_{L}({\hat{\vec{x}}_h\cdot \Delta\hat{\vec{x}}}).\\
&= 2\pi\int dk'k'^2\sum_L P^{\rm true}_L(k')|W(k, k')|^2_{\ell L}
\end{split}
\end{align}
with the window function defined as 
\begin{align}
\begin{split}
|W (k,k')|^2_{\ell L} &= 2 i^{\ell}(-i)^L(2\ell + 1)\sum^{N_{\rm ran}}_{ij, i\neq j}w_{\rm FKP}(\vec{x}_i)w_{\rm FKP}(\vec{x}_j)\\
&\;\;\;\;\;j_{\ell}(k|\Delta\vec{x}|) j_L(k'|\Delta\vec{x}|)\mathcal{L}_{\ell}({\hat{\vec{x}}_h\cdot \Delta\hat{\vec{x}}})\mathcal{L}_{L}({\hat{\vec{x}}_h\cdot \Delta\hat{\vec{x}}}).
\end{split}
\end{align}
This equation does not depend on the vector $\vec{k}$ anymore (but only its amplitude $k$) and hence does not scale with the number of modes $N_c$. 

\subsection{Integral constraint: Derivation of eq.~{\protect\ref{eq:ic1}}}

We start with the observed density field~\citep{Peacock:1991}
\begin{equation}
\delta'(\vec{x})W(\vec{x}) = W(\vec{x})\left[\delta(\vec{x}) - \int d\vec{x} \delta(\vec{x})W(\vec{x})\right],
\label{eq:peacock1}
\end{equation}
where the second term on the right comes from the assumption that the mean density of the survey is equal to the mean density of the Universe. The density field measured with a galaxy survey has the survey window function $W(\vec{x})$ imprinted. In Fourier space this equation becomes
\begin{align}
\begin{split}
\int d\vec{k}'\delta'(\vec{k}')W(\vec{k}-\vec{k}') &= \int d\vec{k}'\delta(\vec{k}')W(\vec{k}-\vec{k}')\\
&\;\;\;\;\;- \frac{W(\vec{k})}{W(0)}\int d\vec{k}' \delta(\vec{k}')W(\vec{k}').
\end{split}
\end{align}
Taking $\langle \delta'\delta'^*\rangle$ we get equation~\ref{eq:surveywindow}. Focusing on the integral constraint for multipoles, we can write
\begin{align}
\begin{split}
P^{\rm ic}_{\ell}(k) &= \frac{2\ell + 1}{2}\int d\mu\int \frac{d\phi}{2\pi} \frac{|W(\vec{k})|^2}{|W(0)|^2_{0}}\\
&\;\;\;\;\;\bigg[\int d\vec{k}' P^{\rm true}(\vec{k}')|W(\vec{k}')|^2\bigg]\mathcal{L}_{\ell}(\mu)
\end{split}\\
&= 2\pi\frac{|W(k)|^2_{\ell}}{|W(0)|^2_{0}}\int dk'k'^2 \sum_LP_L^{\rm true}(k')|W(k')|_L^2\frac{2}{2L+1}
\end{align}
with the window function
\begin{align}
|W(k)|^2_{\ell} &= \frac{2\ell + 1}{2}\int d\mu\int\frac{d\phi}{2\pi} \,W(\vec{k})W^*(\vec{k})\mathcal{L}_{\ell}(\mu)\\
\begin{split}
&= \frac{2\ell + 1}{2}\int d\mu\int\frac{d\phi}{2\pi} \sum^{N_{\rm ran}}_{ij, i\neq j}w_{\rm FKP}(\vec{x}_i)w_{\rm FKP}(\vec{x}_j)\\
&\;\;\;\;\; e^{i\vec{k}\cdot\vec{x}_i}e^{-i\vec{k}\cdot\vec{x}_j}\mathcal{L}_{\ell}(\mu)\\
\end{split}\\
\begin{split}
&= i^{\ell}(2\ell + 1)\sum^{N_{\rm ran}}_{ij, i\neq j}w_{\rm FKP}(\vec{x}_i)w_{\rm FKP}(\vec{x}_j)\\
&\;\;\;\;\;j_{\ell}(k|\Delta\vec{x}|)\mathcal{L}_{\ell}({\vec{x}_h\cdot \Delta\hat{\vec{x}}}).
\end{split}
\end{align}
For this derivation we used eq.~\ref{eq:exp} and the identity relation~\ref{eq:hee2} in the same way as we did in the last section.

\subsection{Proof of the identity relation in eq.~\ref{eq:hee2}}
\label{sec:hee}

We want to proof:
\begin{equation}
\begin{split}
&\frac{2\ell + 1}{2}\int d\mu\int \frac{d\phi}{2\pi}\,\mathcal{L}_{\ell}(\hat{\vec{k}}\cdot\Delta\hat{\vec{x}})\mathcal{L}_{\ell'}(\hat{\vec{k}}\cdot\hat{\vec{x}}_h)\\
&= \mathcal{L}_{\ell}({\hat{\vec{x}}_h\cdot\Delta\hat{\vec{x}}})\delta_{\ell\ell'}.
\end{split}
\label{eq:hee3}
\end{equation}
To do this we are going to use
\begin{equation}
\mathcal{L}_{\ell}(\hat{\vec{k}}\cdot\hat{\vec{k}}') = \frac{4\pi}{2\ell + 1}\sum^{\ell}_{m=-\ell}Y_{\ell m}(\hat{\vec{k}})Y^*_{\ell m}(\hat{\vec{k}}')
\end{equation}
and 
\begin{equation}
\int d\mu\int d\phi\,Y_{m\ell}(\hat{\vec{k}})Y^*_{m'\ell'}(\hat{\vec{k}}) = \delta_{mm'}\delta_{\ell\ell'},
\end{equation}
where the spherical harmonics are given by
\begin{equation}
Y^m_{\ell}(\hat{\vec{k}}) = (-1)^m\sqrt{\frac{(2\ell + 1)(\ell - m)!}{4\pi (\ell + m)!}}\mathcal{L}^m_{\ell}(\mu)e^{im\phi}
\end{equation}
with $\hat{\vec{k}} = (\theta, \phi)$ and $\mu = \cos(\theta)$. We start with the left side of eq.~\ref{eq:hee3}:
\begin{align}
&\;\;\;\;\;\frac{2\ell + 1}{2}\int d\mu\int \frac{d\phi}{2\pi}\, \mathcal{L}_{\ell}(\hat{\vec{k}}\cdot\Delta\hat{\vec{x}})\mathcal{L}_{\ell'}(\hat{\vec{k}}\cdot\hat{\vec{x}}_h)\\
\begin{split}
& = \frac{2\ell + 1}{2}\int d\mu\int \frac{d\phi}{2\pi}\; \frac{4\pi}{2\ell + 1}\sum^{\ell}_{m=-\ell}Y_{\ell m}(\hat{\vec{k}})Y^*_{\ell m}(\Delta\hat{\vec{x}})\\
&\;\;\;\;\;\;\;\;\;\;\;\;\;\;\;\;\;\;\;\;\;\;\;\;\; \frac{4\pi}{2\ell' + 1}\sum^{\ell'}_{m'=-\ell'}Y^*_{\ell' m'}(\hat{\vec{k}})Y_{\ell' m'}(\hat{\vec{x}}_h) 
\end{split}\\
& = \frac{2\ell + 1}{4\pi} \left(\frac{4\pi}{2\ell + 1}\right)^2\sum^{\ell}_{m=-\ell}Y^*_{\ell m}(\Delta\hat{\vec{x}})Y_{\ell m}(\hat{\vec{x}}_h)\delta_{\ell\ell'}\\
& = \mathcal{L}_{\ell}(\hat{\vec{x}}_h\cdot \Delta\hat{\vec{x}})\delta_{\ell\ell'}
\end{align}

\section{Multivariate Gaussian with the sound horizon from E\lowercase{isenstein} \& H\lowercase{u} (1998)}
\label{app:EH}

Here we provide the multivariate Gaussian likelihood using the sound horizon calculated from the approximate equation in~\citet{Eisenstein:1997ik} while in section~\ref{sec:use} we used the sound horizon calculated with CAMB. The ratio of the two calculations is roughly $1.026$ and when treated consistently both methods should lead to the same conclusions (see~\citealt{Mehta:2011xf} for details). For the fitting range $k=0.01$ - $0.20h/$Mpc we have
\begin{equation}
V^{\rm data}_{k_{\rm max}=0.20} = \begin{pmatrix}D_V(z_{\rm eff})/r^{\rm EH}_s(z_d)\\ F(z_{\rm eff})\\ f(z_{\rm eff})\sigma_8(z_{\rm eff})\end{pmatrix} = \begin{pmatrix}13.53\\ 0.683\\ 0.422\end{pmatrix}
\end{equation} 
and the symmetric covariance matrix is given by
\begin{equation}
10^3C_{k_{\rm max}=0.20} = \begin{pmatrix} 34.576 & -2.0110 & -1.7926\\
			        & 1.0776 & 1.1757\\
			        &  & 1.8475 +0.196\end{pmatrix}
\end{equation} 
leading to 
\begin{equation}
C^{-1}_{k_{\rm max}=0.20} = \begin{pmatrix} 32.668 & 79.761 & -17.231\\
			        & 2686.8 & -1475.8\\
			        &  & 1323.3\end{pmatrix}.
\end{equation} 
For $f\sigma_8$ we included the systematic error of $3.1\%$ (see section~\ref{sec:sys}), where we assumed uncorrelated systematic errors. The sound horizon scale derived with the approximate equation in~\citet{Eisenstein:1997ik} is $r_s(z_d) = 151.28\,$Mpc.

The maximum likelihood values for the fitting range $k = 0.01$ - $0.15h/$Mpc are
\begin{equation}
V^{\rm data}_{k_{\rm max}=0.15} = \begin{pmatrix}D_V(z_{\rm eff})/r^{\rm EH}_s(z_d)\\ F(z_{\rm eff})\\ f(z_{\rm eff})\sigma_8(z_{\rm eff})\end{pmatrix} = \begin{pmatrix} 13.48\\ 0.684\\ 0.420\end{pmatrix}
\end{equation} 
and the symmetric covariance matrix is given by
\begin{equation}
10^3C_{k_{\rm max}=0.15} = \begin{pmatrix} 82.202 & -5.7200 & -3.0557\\
			        & 2.2844 & 1.9768\\
			        &  & 2.9510\end{pmatrix}
\end{equation} 
leading to
\begin{equation}
C^{-1}_{k_{\rm max}=0.15} = \begin{pmatrix} 15.388 & 58.870 & -23.503\\
			        & 1266.8 & -787.65\\
			        &  & 842.17\end{pmatrix},
\end{equation} 
where no systematic error is included.

\newpage

\label{lastpage}

\end{document}